\documentclass{emulateapj}

\def\kmsec{\mbox{km~s$^{\rm -1}$}}
\def\logg{\mbox{log~{\it g}}}

\def\teff{\mbox{$T_{\rm eff}$}}
\def\vt{\mbox{$v_{\rm t}$}}

\def\rpro{\mbox{$r$-process}}
\def\spro{\mbox{$s$-process}}

\def\loggf{$\log gf$}
\def\logeps{$\log \epsilon$}
\def\bd{\object[BD+440493]{BD$+$44~493}}
\def\csa{\object[BPS CS 22891-200]{CS~22891--200}}
\def\csb{\object[BPS CS 22893-010]{CS~22893--010}}
\def\csc{\object[BPS CS 22949-037]{CS~22949--037}}
\def\csd{\object[BPS CS 22957-027]{CS~22957--027}}
\def\cse{\object[BPS CS 22958-042]{CS~22958--042}}
\def\csf{\object[BPS CS 29498-043]{CS~29498--043}}
\def\csg{\object[BPS CS 30314-067]{CS~30314--067}}
\def\csh{\object[BPS CS 22877-001]{CS~22877--001}}
\def\csi{\object[BPS CS 22878-101]{CS~22878--101}}
\def\csk{\object[BPS CS 22943-201]{CS~22943--201}}
\def\csl{\object[BPS CS 22948-066]{CS~22948--066}}
\def\csm{\object[BPS CS 22960-064]{CS~22960--064}}
\def\csn{\object[BPS CS 29502-092]{CS~29502--092}}
\def\cso{\object[BPS CS 30492-001]{CS~30492--001}}
\def\he{\object[HE 1012-1540]{HE~1012$-$1540}}

\shorttitle{Early Neutron-Capture Element Production}
\shortauthors{Roederer et al.}
\slugcomment{Accepted for publication in the Astrophysical Journal}

\begin{document}

\title{Neutron-Capture Nucleosynthesis in the First Stars\footnotemark[1]}

\footnotetext[1]{
This paper includes data gathered with the 6.5 meter 
Magellan Telescopes located at Las Campanas Observatory, Chile,
and The McDonald Observatory of The University of Texas at Austin.
}

\author{Ian U.\ Roederer,\altaffilmark{2}
George W.\ Preston,\altaffilmark{3}
Ian B.\ Thompson,\altaffilmark{3}
Stephen A. Shectman,\altaffilmark{3}
Christopher Sneden\altaffilmark{4}
}

\altaffiltext{2}{Department of Astronomy, University of Michigan,
500 Church Street, Ann Arbor, MI 48109, USA;
iur@umich.edu}
\altaffiltext{3}{Carnegie Observatories, 
813 Santa Barbara St., Pasadena, CA 91101, USA}
\altaffiltext{4}{Department of Astronomy, University of Texas at Austin,
1 University Station C1400, Austin, TX 78712, USA}

% after all the footnotes...
\addtocounter{footnote}{4}

\begin{abstract}

Recent studies suggest that 
metal-poor stars enhanced in carbon
but containing low levels of neutron-capture elements
may have been among the first to incorporate
the nucleosynthesis products of the first generation of stars.
We have observed
16~stars with enhanced carbon or nitrogen
using the MIKE Spectrograph on the
Magellan Telescopes at Las Campanas Observatory
and the Tull Spectrograph 
on the Smith Telescope at McDonald Observatory.
We present radial velocities, stellar parameters, and
detailed abundance patterns for these stars.
Strontium, yttrium, zirconium, barium, europium, ytterbium,
and other heavy elements are detected.
In four stars, these heavy elements appear to have originated
in some form of \rpro\ nucleosynthesis.
In one star, a partial \spro\ origin is possible.
The origin of the heavy elements in the rest of the sample
cannot be determined unambiguously.
The presence of elements heavier than the iron group
offers further evidence 
that zero-metallicity rapidly-rotating massive stars
and pair instability supernovae
did not contribute substantial amounts of neutron-capture elements
to the regions where the stars in our sample formed.
If the carbon- or nitrogen-enhanced 
metal-poor stars with low levels
of neutron-capture elements were 
enriched by products of zero-metallicity supernovae only,
then the presence of these heavy elements
indicates that at least one form of neutron-capture reaction
operated in some of the first stars.

\end{abstract}


\section{Introduction}
\label{introduction}

The first stars formed from
only the products of Big Bang nucleosynthesis,
yet the metals they produced 
and distributed into the interstellar medium
forever changed the fundamental 
methods by which molecular gas clouds 
cool, collapse, and form stars.
Cooling by fine structure line emission of 
ionized carbon (C~\textsc{ii}),
neutral oxygen (O~\textsc{i}), and
thermal emission from collisionally-excited dust grains
is suspected to have enabled 
long-lived low-mass stars to form.
Many of the most iron-poor stars known,
with Fe/H ratios less than 10$^{-4}$ times the solar ratio,
show C/H and O/H ratios
of only 10$^{-3}$ to 10$^{-1}$.
This may indicate that substantial enrichment 
in carbon or oxygen was
closely linked with formation of the long-lived low-mass
stars that are observable in the solar neighborhood.
Theoretical studies of the 
nucleosynthesis reactions that 
may have occurred in the first stars have,
understandably, focused their effort on
production of metals from carbon
through the iron group.
Elements heavier than the
iron group, here considered to be
those with $Z >$~32,
are difficult to detect observationally in the most iron-poor stars and
almost certainly had no effect on subsequent star formation
due to their low abundances of 10$^{-13}$ or less per hydrogen atom.

Carbon-enhanced metal-poor stars with 
no enhancement of neutron-capture elements are commonly 
referred to as members of the ``CEMP-no'' class,
using the classification scheme proposed
by \citet{ryan05} and \citet{beers05}.
\citeauthor{ryan05}\ proposed that 
these stars formed from gas clouds 
pre-enriched with high levels of carbon
by previous generations of supernovae,
rather than acquiring the carbon enrichment
by mass transfer from an evolved companion.
The CEMP-no stars include three of the four most iron-poor stars known
\citep{christlieb02,christlieb04,bessell04,
frebel05,frebel06,frebel08,aoki06,norris07,norris12}.
CEMP-no stars are frequently enhanced in 
nitrogen (N, $Z =$~7), 
oxygen (O, $Z =$~8), 
sodium (Na, $Z =$~11), 
magnesium (Mg, $Z =$~12), 
aluminum (Al, $Z =$~13), and
silicon (Si, $Z =$~14).
A few CEMP-no stars are found in binary systems,
but the binary nature of the ensemble of CEMP-no stars is clearly 
unlike that of the
CEMP stars enhanced in material
produced by the slow neutron-capture process 
(the \spro)
in a companion that passed through the
thermally-pulsing asymptotic giant branch (TP-AGB) 
phase of evolution.

\citet{norris13} summarize these properties
and weigh the evidence that stars in the CEMP-no class
were among the first stars to incorporate metals produced by
zero-metallicity stars.
This evidence includes 
the chemically-primitive stars in ultra-faint dwarf galaxies
(e.g., \citealt{frebel10,norris10a,norris10b,simon10,koch13}),
including the discovery of a CEMP-no star in each of
Segue~1 \citep{norris10a} and 
Bo\"{o}tes~I \citep{lai11,gilmore13};
the greater chemical inhomogeneity, on average, of
field stars that are kinematically associated
with the outer halo
(e.g., \citealt{fulbright02,stephens02,gratton03,
roederer09a,ishigaki10,ishigaki13,nissen10});
the increasing fraction of 
carbon-enhanced stars with decreasing metallicity 
and increasing distance above the Galactic plane
\citep{frebel06,carollo12};
and the low frequency of carbon-enhanced
metal-poor damped Lyman-$\alpha$ (DLA) systems
at redshifts 2~$< z <$~6.3,
hinting that an epoch dominated by carbon-enhanced systems, 
if one existed at all,
must have occurred at even higher redshifts
(e.g., \citealt{cooke11a,cooke11b,becker12};
but note that \citeauthor{becker12}\
express reservations about the 
high level of carbon enhancement
in the DLA reported by \citeauthor{cooke11a}).
These links between carbon enhancement, low metallicity,
and remote environments
hint that carbon-enhanced stars formed in 
chemically-primitive regions
that experienced relatively few enrichment events.

What kind of stars were responsible
for prodigious carbon production in the early Universe?
The abundance patterns found in the CEMP-no stars
are consistent with model
predictions for zero-metallicity stars that 
were massive and rotating \citep{fryer01,meynet06,meynet10}, 
underwent faint ``mixing and fallback'' supernova explosions 
\citep{umeda03,umeda05,tominaga13},
or whose supernovae had relativistic jets \citep{tominaga07}.
These zero-metallicity supernovae are predicted 
to have seeded pristine gas clouds
with carbon, oxygen, and other metals, 
enabling low-mass star formation to occur.

We now pose a related question:\
\textit{What heavy element ($Z >$~32) abundance patterns
are found in stars in the CEMP-no class?}
We address this question using seven metal-poor stars 
identified in the abundance survey of \citet{roederer14}.
These stars are carbon-enhanced and barium-poor
(Ba, $Z =$~56).
We supplement these stars with nine 
nitrogen-enhanced metal-poor stars that are barium-poor
(defined here as NEMP-no, cf.\ \citealt{johnson07}) 
from the same survey.
Our goal is not to resolve
whether the CEMP-no or NEMP-no stars
are the immediate descendants of 
zero-metallicity stars.
Instead, we aim to characterize the 
heavy element abundance signatures in these stars
to motivate studies of neutron-capture nucleosynthesis
in stars with zero or extremely low levels of metals.
Unfortunately, existing nucleosynthesis calculations for such stars
rarely extend to nuclei heavier than the iron group.
For example, the reaction network of
\citet{tominaga13} is truncated at
bromine (Br, $Z =$~35), which is
several mass units short of the lightest 
neutron-capture element commonly studied in
metal-poor stars, strontium (Sr, $Z =$~38).

Throughout this work we
adopt the standard definitions of elemental abundances and ratios.
For element X, the logarithmic abundance is defined
as the number of atoms of X per 10$^{12}$ hydrogen atoms,
$\log\epsilon$(X)~$\equiv \log_{10}(N_{\rm X}/N_{\rm H}) +$~12.0.
For elements X and Y, the logarithmic abundance ratio relative to the
solar ratio, denoted
[X/Y], is defined as $\log_{10} (N_{\rm X}/N_{\rm Y}) -
\log_{10} (N_{\rm X}/N_{\rm Y})_{\odot}$.
Abundances or ratios denoted with the ionization state
indicate the total elemental abundance as derived from transitions of
that particular state 
after ionization corrections have been applied.
When reporting relative abundance ratios for elements X and Y,
these ratios 
compare the total abundances of X 
and Y derived from like ionization states;
i.e., neutrals with neutrals and ions with ions.

\section{Sample Selection}
\label{sample}

\begin{figure*}
\begin{center}
\includegraphics[angle=90,width=4.5in]{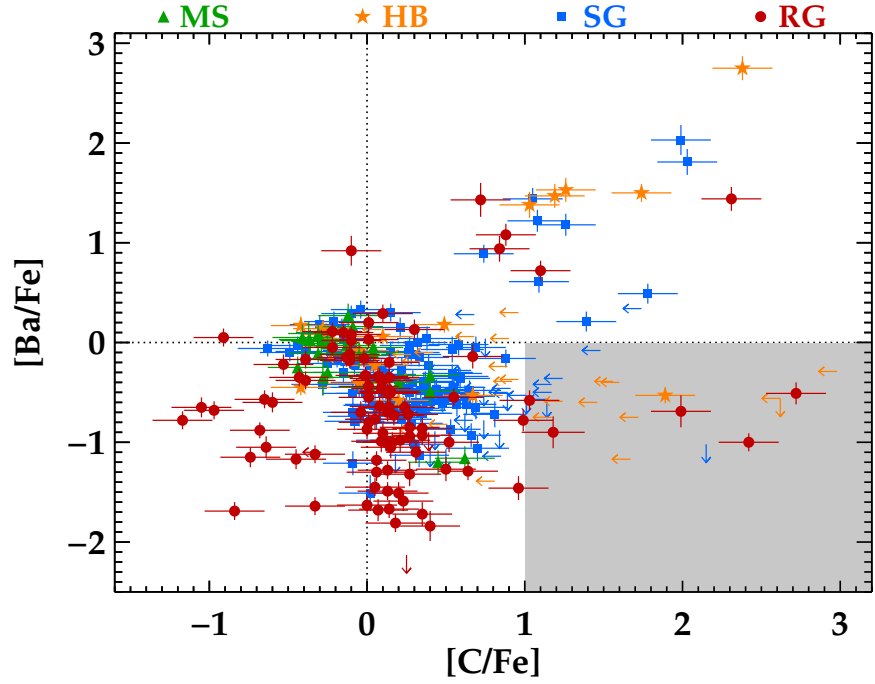}
\end{center}
\caption{
\label{cbafig}
[Ba/Fe] ratios as a function of [C/Fe] ratios
for the full sample of stars from \citet{roederer14}.
Green triangles represent unevolved dwarf stars on
the main sequence (MS),
orange stars represent stars on the horizontal branch (HB),
blue squares represent stars beyond the main sequence turn off
or on the subgiant branch (SG), and 
red circles represent stars on the red giant branch (RG).
The dotted lines represent the solar ratios.
The shaded region marks one criterion for inclusion in our sample,
[C/Fe]~$> +$1.0 and [Ba/Fe]~$<$~0.0.
}
\end{figure*}

\begin{figure*}
\begin{center}
\includegraphics[angle=90,width=4.5in]{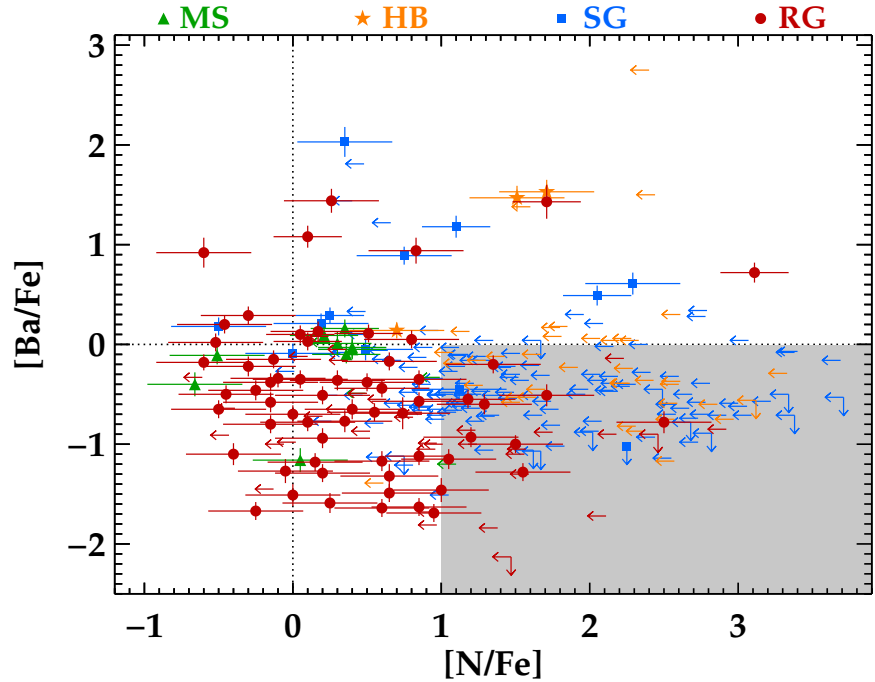}
\end{center}
\caption{
\label{nbafig}
[Ba/Fe] ratios as a function of [N/Fe] ratios
for the full sample of stars from \citet{roederer14}.
Symbols are the same as in Figure~\ref{cbafig}.
The shaded region marks one criterion for inclusion in our sample,
[N/Fe]~$> +$1.0 and [Ba/Fe]~$<$~0.0.
}
\end{figure*}

We draw our sample from the catalog of 313 metal-poor stars
observed and analyzed by \citet{roederer14}.
Adopting the classification scheme defined by \citet{beers05},
we identify CEMP or NEMP stars
with no enhancement of neutron-capture elements by 
requiring that a star have [Ba/Fe]~$<$~0.0 and either
[C/Fe]~$> +$1.0 or 
[N/Fe]~$> +$1.0.
Figures~\ref{cbafig} and \ref{nbafig} illustrate
our selection criteria.
There are 16~stars with either carbon or nitrogen detection, or both,
that lie in the shaded regions.
Barium is detected in all but one of these stars (\cse), and the
upper limit on [Ba/Fe] places this star unequivocally in the shaded region.
One star is on the horizontal branch (\csk),
two are subgiants (\cse\ and \cso), 
and the remaining 13 are red giants.
All are field stars and not associated with any known clusters or streams.
Our study marks the first detailed abundance study based on
high-resolution spectroscopic data for three of these stars
(\csb, \csk, and \cso).
Many of the remaining 13~stars have been analyzed repeatedly 
over the last 20~years by
\citet{primas94},
\citet{thorburn94},
\citet{mcwilliam95b},
\citet{norris97,norris01,norris02,norris13},
\citet{bonifacio98},
\citet{giridhar01},
\citet{preston01},
\citet{aoki02a,aoki02b,aoki02c,aoki04},
\citet{carretta02},
\citet{depagne02},
\citet{cayrel04},
\citet{spite05},
\citet{sivarani06},
\citet{francois07},
\citet{cohen08,cohen13},
\citet{lai08},
\citet{ito09,ito13},
\citet{hollek11},
\citet{ruchti11a}, and
\citet{yong13a}.

\begin{figure}
\includegraphics[angle=270,width=3.35in]{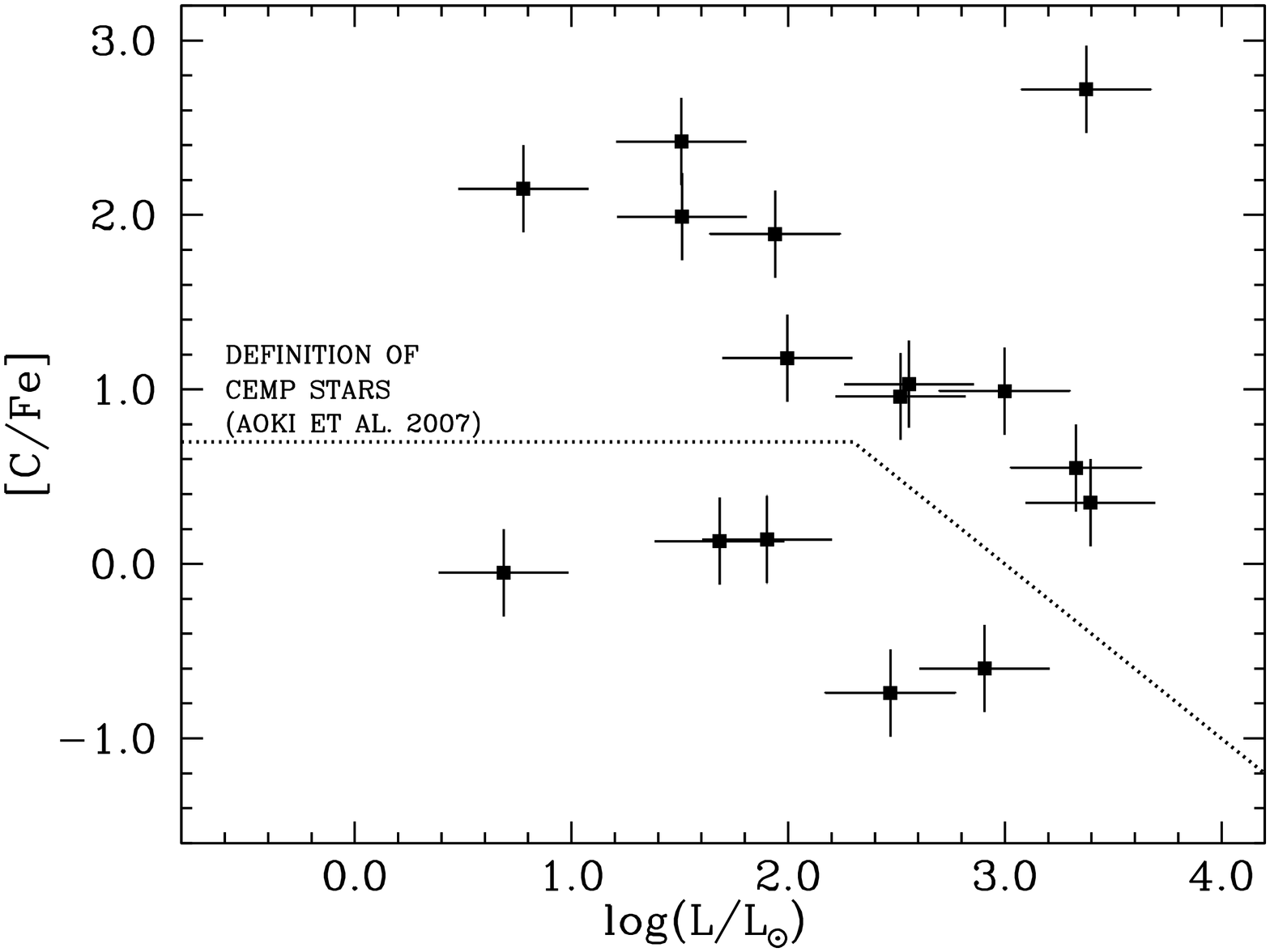}
\caption{
\label{crichdeffig}
[C/Fe] ratios as a function of luminosity.
The squares illustrate the 16~stars in our sample.
The dotted line represents the lower bound of
the class of CEMP stars as defined by \citet{aoki07}.
The five stars below this line are considered NEMP stars
since they all show [N/Fe]~$> +$1.0.
}
\end{figure}

Figure~\ref{crichdeffig} compares the [C/Fe] ratios in our
sample with the CEMP definition given by \citet{aoki07}.
Here, we see that 11~stars in our sample would 
be considered CEMP by their definition.
Five stars have low [C/Fe] ratios but
show [N/Fe]~$> +$1.0:\
\csi, \csb, \csl, \csm, and \cso.
The NEMP stars may or may not be related to the CEMP stars.
Including NEMP stars in our sample
allows for the possibility that these stars were, at some time
in the past, CEMP stars.  
In this scenario, these stars would have undergone internal
mixing, dredging CN-cycled nitrogen-rich and carbon-poor material
to the surface.
In Section~\ref{histograms}, we
consider whether these NEMP stars 
constitute a sample distinct from the CEMP
stars with regard to their [Fe/H], [Sr/Fe], or [Ba/Fe] ratios.

\section{Summary of Observations and Analysis Techniques 
Used by Roederer et al.\ (2014)}
\label{r13summary}

This section summarizes the observations and analysis techniques
used by \citet{roederer14} to measure radial velocities,
derive stellar parameters, and derive abundances.
All of the values discussed here are presented
in the tables in that work.
Several of those tables are only available online,
so we feel it is
helpful to reproduce this information here
for the 16~CEMP-no and NEMP-no stars 
for easy reference.

\subsection{Observations}
\label{obs}

Most observations were made with the Magellan Inamori Kyocera Echelle (MIKE)
spectrograph \citep{bernstein03}
on the 
6.5~m Walter Baade and Landon Clay Telescopes
at Las Campanas Observatory.
These spectra were taken with the 0\farcs7\,$\times$\,5\farcs0 slit, 
yielding
a resolving power of $R \equiv \lambda/\Delta\lambda \sim$~41,000 in the blue 
and $R \sim$~35,000 in the red as measured from isolated ThAr lines
in the comparison lamp images.
A dichroic splits the two arms at $\approx$~4950~\AA.
This setup achieves complete wavelength coverage from 
3350--9150~\AA.
Data reduction,
extraction, sky subtraction,
and wavelength calibration were performed using 
the MIKE data reduction pipeline
written by Dan Kelson
(see \citealt{kelson03}).

Observations of \mbox{BD$+$44~493} 
were made with the 
Robert G.\ Tull Coud\'{e} Spectrograph
\citep{tull95} on the 2.7~m Harlan J.\ Smith
Telescope at McDonald Observatory.
These spectra were taken with the 2\farcs4\,$\times$\,8\farcs0 slit, yielding 
a resolving power $R \sim$~33,000.
This setup delivers complete wavelength coverage from 3700--5700~\AA,
with small gaps between the echelle orders further to the red.
For our analysis we only use the spectra blueward of 8000~\AA.
Data reduction, extraction, sky subtraction,
and wavelength calibration 
were performed using the REDUCE software package
\citep{piskunov02}.
Coaddition and continuum normalization for 
both sets of spectra were performed within the 
IRAF environment.

Table~\ref{obstab} presents a record of observations.
Signal to noise (S/N) estimates, listed in Table~\ref{rvtab}, are based on
Poisson statistics of the photons collected in the continuum
at several reference wavelengths.

\begin{deluxetable*}{lcccccc}
\tablecaption{Log of Observations
\label{obstab}}
\tablewidth{0pt}
\tabletypesize{\scriptsize}
\tablehead{
\colhead{Star} &
\colhead{Telescope/} &
\colhead{Exposure} & 
\colhead{Date} &
\colhead{UT at} &
\colhead{Heliocentric} &
\colhead{Heliocentric} \\
\colhead{} &
\colhead{Instrument} &
\colhead{Length} &
\colhead{} &
\colhead{Mid-exposure} &
\colhead{Julian Date} &
\colhead{Radial Velocity} \\
\colhead{} &
\colhead{} &
\colhead{(s)} &
\colhead{} &
\colhead{} &
\colhead{} &
\colhead{(\kmsec)} }
\startdata
BD$+$44 493    & McDonald-Smith/Tull     & 7200   & 2008 Aug 14 & 10:56 & 2454692.956   & $-$149.7      \\
BD$+$44 493    & McDonald-Smith/Tull     & 3600   & 2008 Nov 04 & 09:01 & 2454774.881   & $-$150.0      \\
CS 22877--001   & Magellan-Baade/MIKE     & 1000   & 2003 Jan 14 & 06:58 & 2452653.790   & $+$169.8      \\
CS 22877--001   & Magellan-Baade/MIKE     & 1600   & 2003 Jan 15 & 07:52 & 2452654.828   & $+$167.0      \\
CS 22877--001   & Magellan-Baade/MIKE     & 1200   & 2003 Jan 20 & 08:31 & 2452659.856   & $+$166.6      \\
CS 22877--001   & Magellan-Clay/MIKE      & 1800   & 2006 Aug 02 & 23:07 & 2453950.461   & $+$166.7      \\
CS 22878--101   & Magellan-Clay/MIKE      & 2400   & 2006 Aug 05 & 23:59 & 2453953.501   & $-$128.8      \\
CS 22891--200   & Magellan-Clay/MIKE      & 4100   & 2006 Aug 06 & 01:05 & 2453953.549   & $+$137.2      \\
CS 22893--010   & Magellan-Clay/MIKE      & 3600   & 2012 Aug 27 & 02:48 & 2456166.622   & $-$51.6       \\
%CS 22943--132   & Magellan-Clay/MIKE      & 1400   & 2006 Aug 03 & 02:19 & 2453950.602   & $+$18.9       \\
CS 22943--201   & Magellan-Clay/MIKE      & 2200   & 2005 Aug 21 & 01:26 & 2453603.564   & $+$37.2       \\
CS 22943--201   & Magellan-Clay/MIKE      & 5100   & 2006 Jun 12 & 05:54 & 2453898.750   & $+$37.6       \\
CS 22943--201   & Magellan-Clay/MIKE      & 1500   & 2007 Aug 03 & 06:22 & 2454315.771   & $+$35.4       \\
CS 22943--201   & Magellan-Clay/MIKE      & 1800   & 2007 Aug 04 & 04:04 & 2454316.675   & $+$36.4       \\
CS 22948--066   & Magellan-Clay/MIKE      & 1500   & 2007 Aug 22 & 02:48 & 2454334.622   & $-$171.2      \\
CS 22948--066   & Magellan-Clay/MIKE      & 900    & 2009 May 15 & 10:32 & 2454966.940   & $-$170.8      \\
CS 22949--037   & Magellan-Clay/MIKE      & 5000   & 2006 Aug 06 & 08:32 & 2453953.860   & $-$125.4      \\
CS 22949--037   & Magellan-Clay/MIKE      & 1350   & 2009 Jul 25 & 08:04 & 2455037.840   & $-$125.3      \\
CS 22957--027   & Magellan-Clay/MIKE      & 1000   & 2008 Sep 11 & 06:13 & 2454720.765   & $-$74.8       \\
CS 22958--042   & Magellan-Baade/MIKE     & 4800   & 2003 Jan 16 & 01:30 & 2452655.561   & $+$165.2      \\
CS 22958--042   & Magellan-Baade/MIKE     & 2400   & 2003 Jan 17 & 01:10 & 2452656.547   & $+$165.7      \\
CS 22958--042   & Magellan-Baade/MIKE     & 2400   & 2003 Jan 19 & 01:14 & 2452658.550   & $+$165.6      \\
CS 22960--064   & Magellan-Clay/MIKE      & 3000   & 2006 Jul 22 & 06:58 & 2453938.795   & $-$86.4       \\
CS 29498--043   & Magellan-Clay/MIKE      & 6000   & 2013 Apr 22 & 08:54 & 2456404.870   & $-$32.3       \\
CS 29498--043   & Magellan-Clay/MIKE      & 3300   & 2013 Apr 23 & 09:20 & 2456405.888   & $-$32.3       \\
CS 29502--092   & Magellan-Clay/MIKE      & 1800   & 2009 Oct 26 & 01:32 & 2455130.567   & $-$66.6       \\
CS 30314--067   & Magellan-Clay/MIKE      & 1800   & 2013 Apr 21 & 09:34 & 2456403.898   & $+$145.6      \\
CS 30492--001   & Magellan-Clay/MIKE      & 1650   & 2008 Sep 10 & 01:18 & 2454719.558   & $-$116.2      \\
HE 1012$-$1540  & Magellan-Clay/MIKE      & 1800   & 2013 Apr 21 & 23:40 & 2456404.489   & $+$225.4      \\
HE 1012$-$1540  & Magellan-Clay/MIKE      & 5400   & 2013 Apr 23 & 01:45 & 2456405.576   & $+$225.8      \\
\enddata                                                                                                        
\end{deluxetable*} 
  % 2 col

\begin{deluxetable*}{lcccccc}
\tablecaption{Observational Stellar Data
\label{rvtab}}
\tablewidth{0pt}
\tabletypesize{\scriptsize}
\tablehead{
\colhead{Star} &
\colhead{Total exp.} &
\colhead{No.} & 
\colhead{S/N} & 
\colhead{S/N} & 
\colhead{S/N} & 
\colhead{S/N} \\  
\colhead{} &
\colhead{time (s)} &
\colhead{obs.} &
\colhead{3950\AA} &
\colhead{4550\AA} &
\colhead{5200\AA} &
\colhead{6750\AA} }
\startdata
BD$+$44 493                      & 10800 & 2   & 85    & 170   & 220   & 285    \\ 
CS 22877--001                    & 5600  & 4   & 150   & 230   & 265   & 405    \\ 
CS 22878--101                    & 2400  & 1   & 60    & 95    & 75    & 115    \\ 
CS 22891--200                    & 4100  & 1   & 65    & 105   & 85    & 130    \\ 
CS 22893--010                    & 3600  & 1   & 45    & 75    & 65    & 115    \\ 
%CS 22943--132                   & 1400  & 1   & 75    & 105   & 75    & 95     \\ 
CS 22943--201                    & 10600 & 4   & 85    & 120   & 85    & 120    \\ 
CS 22948--066\tablenotemark{a}   & 2400  & 2   & 60    & 95    & 80    & 130    \\ 
CS 22949--037\tablenotemark{b}   & 6350  & 2   & 65    & 100   & 85    & 135    \\ 
CS 22957--027\tablenotemark{c}   & 1000  & 1   & 40    & 55    & 60    & 95     \\ 
CS 22958--042                    & 9600  & 3   & 60    & 90    & 110   & 155    \\ 
CS 22960--064                    & 3000  & 1   & 75    & 110   & 80    & 110    \\ 
CS 29498--043                    & 9300  & 2   & 70    & 140   & 135   & 280    \\ 
CS 29502--092                    & 1800  & 1   & 50    & 85    & 90    & 165    \\ 
CS 30314--067                    & 1800  & 1   & 85    & 165   & 165   & 350    \\ 
CS 30492--001                    & 1650  & 1   & 55    & 75    & 70    & 115    \\ 
HE 1012$-$1540                   & 7200  & 2   & 60    & 95    & 85    & 150    \\ 
\enddata 
\tablenotetext{a}{CS~30343--064}
\tablenotetext{b}{HE~2323$-$0256}
\tablenotetext{c}{HE~2356$-$0410}
\end{deluxetable*}

  % 2 col

\subsection{Radial Velocity Measurements}
\label{rv}

\citet{roederer14} measured radial velocities
by cross-correlating the spectral
order containing the Mg~\textsc{i}~b lines
against metal-poor template standards.
Heliocentric corrections 
were computed using the IRAF \textit{rvcorrect} task.
Table~\ref{obstab} lists the
heliocentric velocity measurements for each observation.
Typical uncertainties are
$\approx$~0.6--0.8~\kmsec\
per observation.

\bd, \csa, \csk, \csl, \csc, \csf, \csn, and \csg\
are consistent with no velocity variations
at the $\approx$~2~\kmsec\ level.
This conclusion is based on
comparisons with velocity measurements by other investigators
\citep{
primas94,
mcwilliam95a,
aoki02a,aoki02b,
norris01,
cohen02,cohen08,cohen13,
depagne02,
carney03,
lai04,
bonifacio09,
hollek11,
ito13}
or repeat observations separated by 
more than 2~months.

\csh\ exhibits radial velocity variations at a 
full amplitude of $\approx$~4~\kmsec\ 
\citep{giridhar01,aoki02a,tsangarides04}.
Our own measurements span a range of 3.2~\kmsec\ 
over 3.5~years.

Radial velocity measurements of \csi\ have been reported
by several authors, including
\citet{mcwilliam95b}, \citet{cohen02}, \citet{lai04}, and \citet{bonifacio09}.
These measurements span a range of 3~\kmsec\ and may be 
expected to have internal precisions of better than 1~\kmsec,
though it is difficult to assess the systematic uncertainty
from one author to another.
\csi\ may exhibit low-amplitude radial velocity variations.

Our measured velocity of \csb\ is 24~\kmsec\ different 
than that measured from a medium-resolution spectrum
obtained by \citet{lai04}.
They report a measurement 
uncertainty of 4.2~\kmsec.
It seems probable that \csb\ exhibits radial velocity variations,
but no other radial velocity information is available
for this star.

Variations in the velocity of \csd\ have been
confirmed by \citet{preston01}, who supplemented 
their own observations with measurements by
\citet{norris97} and \citet{bonifacio98}.
Subsequent observations through May, 2012, by G.\ Preston 
(to be published elsewhere)
reveal that the original orbital period reported
by \citeauthor{preston01} is an alias;
the current best-fit orbit has a period of $\approx$~1078~days
and a systemic velocity of $-$67.7~\kmsec.

\citet{sivarani06} reported radial velocity variations 
in \cse\ among their observations that spanned less than 1~hour.
Our radial velocity measurements, spanning 3~days,
are consistent with a single value and
fall within the range of their observations.

\citet{cohen08} report a 4~$\sigma$ difference in their two
radial velocity measurements of \he,
although these measurements taken 3~years apart
are different by only 1.5~\kmsec.
One subsequent observation of this star 10~years later
by \citet{cohen13} expands the range to 1.9~\kmsec.
Our radial velocity measurements are separated by only 1~day,
but they agree with each other and fall within the
range reported by \citeauthor{cohen08} %.

We are aware of only our single epoch radial velocities
measured from high-resolution spectroscopic observations
for \csm\ and \cso.

In summary, 8 of the 16 stars in our sample show no 
evidence for radial velocity variations.
A binary orbital solution has been determined previously for one star.
Another shows tentative evidence for large-amplitude velocity variations.
Four stars show evidence of low-amplitude velocity variations.
Two stars have only been observed at high spectral resolution 
at a single epoch.
Followup velocity observations of
\csh, \csi, \csb, \cse, \csm, \cso, and \he\
would be of interest.

\subsection{Equivalent Width Measurements}
\label{ew}

\citet{roederer14} measured
equivalent widths using a
semi-automatic routine that fits Voigt absorption line profiles
to continuum-normalized spectra.
Comparison with equivalent widths measured by
\citet{johnson02},
\citet{cayrel04},
\citet{honda04a}, and
\citet{lai08}
indicates that 
the standard deviation of the residuals
is 3.5~m\AA\ for 3087~lines with
equivalent width $<$~100~m\AA.

\subsection{Model Atmospheres}
\label{modelatm}

\citet{roederer14} used model atmospheres interpolated from the
grid of one-dimensional MARCS models
\citep{gustafsson08}
and performed the analysis using the
latest version of the 
spectral line analysis code MOOG 
(\citealt{sneden73}; see also \citealt{sobeck11}).
The stars in our sample show strong 
absorption from molecular bands, so it is advisable to
avoid deriving the model atmosphere parameters
from color-temperature relations based
on broadband photometry.
Instead, effective temperatures (\teff) were derived by requiring that
abundances derived from Fe~\textsc{i} lines showed no trend with
the E.P.\ of the lower level of the transition.
Microturbulent velocities (\vt) were derived by requiring that 
abundances derived from Fe~\textsc{i} lines showed no trend with
line strength.
Surface gravities (\logg, in cgs units) were calculated from the 
relationship between \teff\ and \logg\ given by theoretical
isochrones in the Y$^{2}$ grid \citep{demarque04};
an age of 12~$\pm$~1.5~Gyr was assumed for all stars. 
The iron abundance derived 
from Fe~\textsc{ii} lines was taken to represent
the overall metallicity, [M/H].
This method was used for the 15~stars not on the horizontal branch.
For the one star on the horizontal branch, 
\logg\ was derived by requiring that the iron abundance derived
from neutral iron lines matches that derived from ionized iron lines.
The derived model parameters and their statistical (internal)
uncertainties are presented in Table~\ref{atmtab}.

\begin{deluxetable*}{lccccc}
\tablecaption{Magnitudes and Atmospheric Parameters
\label{atmtab}}
\tablewidth{0pt}
\tabletypesize{\scriptsize}
\tablehead{
\colhead{Star} &
\colhead{$V$} &
\colhead{\teff} &
\colhead{\logg} &
\colhead{\vt} &
\colhead{[M/H]\tablenotemark{a}} \\
\colhead{} &
\colhead{} &
\colhead{(K)} &
\colhead{} &
\colhead{(\kmsec)} &
\colhead{} }
\startdata                                                                                                           
 BD$+$44 493     &  9.11 & 5040 (36) & 2.10 (0.14) & 1.35 (0.06) & $-$4.26 (0.03) \\
 CS~22877--001   & 12.16 & 4790 (34) & 1.45 (0.14) & 1.55 (0.06) & $-$3.24 (0.06) \\
 CS~22878--101   & 13.73 & 4650 (35) & 1.05 (0.14) & 1.90 (0.06) & $-$3.30 (0.06) \\
 CS~22891--200   & 13.93 & 4490 (33) & 0.50 (0.12) & 1.70 (0.06) & $-$3.88 (0.08) \\
 CS~22893--010   & 14.74 & 5150 (44) & 2.45 (0.20) & 1.35 (0.06) & $-$2.93 (0.07) \\
 CS~22943--201   & 15.98 & 5970 (52) & 2.45 (0.39) & 1.60 (0.06) & $-$2.69 (0.06) \\
 CS~22948--066   & 13.47 & 4830 (34) & 1.55 (0.15) & 2.00 (0.06) & $-$3.18 (0.07) \\
 CS~22949--037   & 14.36 & 4630 (34) & 0.95 (0.13) & 1.70 (0.06) & $-$4.21 (0.05) \\
 CS~22957--027   & 13.60 & 5220 (39) & 2.65 (0.23) & 1.45 (0.06) & $-$3.00 (0.07) \\
 CS~22958--042   & 14.52 & 5760 (57) & 3.55 (0.18) & 0.95 (0.08) & $-$2.99 (0.15) \\
 CS~22960--064   & 13.94 & 5060 (36) & 2.20 (0.14) & 1.40 (0.06) & $-$2.77 (0.07) \\
 CS~29498--043   & 13.72 & 4440 (20) & 0.50 (0.13) & 1.75 (0.06) & $-$3.85 (0.08) \\
 CS~29502--092   & 11.87 & 4820 (34) & 1.50 (0.14) & 1.50 (0.06) & $-$3.20 (0.07) \\
 CS~30314--067   & 11.85 & 4320 (12) & 0.50 (0.10) & 1.85 (0.06) & $-$3.01 (0.06) \\
 CS~30492--001   & 14.20 & 5790 (50) & 3.65 (0.15) & 0.85 (0.07) & $-$2.35 (0.07) \\
 HE~1012$-$1540  & 14.04 & 5230 (32) & 2.65 (0.20) & 1.70 (0.06) & $-$3.76 (0.14) \\
\enddata
\tablenotetext{a}{
[M/H] is adopted to equal [Fe/H] as derived from Fe~\textsc{ii} lines.
}
\end{deluxetable*}
  % 2 col

To evaluate the reliability of the
model atmosphere parameters derived for the full sample, 
\citet{roederer14}
compared these values with parameters derived 
by a variety of different methods for stars in 
common with previous studies.
For red giants, subgiants, and stars on the horizontal branch,
these comparisons for the full sample yielded standard deviations of 
151, 211, and 156~K in \teff,
0.40, 0.34, and 0.42 in \logg,
0.41, 0.33, and 0.26~\kmsec\ in \vt, and
0.24, 0.22, and 0.16~dex in [Fe~\textsc{ii}/H].
We adopt these as the systematic uncertainties
in the model atmosphere parameters.

\begin{deluxetable}{lcccc}
\tablecaption{Comparison of Derived Model Parameters
with Previous Work
\label{comparetab}}
%\rotate
\tablewidth{0pt}
%\tabletypesize{\tiny}
\tabletypesize{\scriptsize}
\tablehead{
\colhead{Star} &
\colhead{\teff} &
\colhead{\logg} &
\colhead{[Fe/H]\tablenotemark{a}} &
\colhead{Reference} }
\startdata
 BD$+$44 493     & 5040 & 2.10 & $-$4.26 & This study \\
                 & 5510 & 3.70 & $-$3.68 & \citet{ito09} \\
                 & 5430 & 3.40 & $-$3.82 & \citet{ito13} \\
 CS~22877--001   & 4790 & 1.45 & $-$3.24 & This study \\
                 & 5000 & 1.50 & $-$2.88 & \citet{giridhar01} \\
                 & 5100 & 2.20 & $-$2.71 & \citet{aoki02a} \\
 CS~22878--101   & 4650 & 1.50 & $-$3.30 & This study \\
                 & 4790 & 1.15 & $-$3.13 & \citet{mcwilliam95b} \\
                 & 4775 & 1.30 & $-$3.13 & \citet{carretta02} \\
                 & 4800 & 1.30 & $-$3.21 & \citet{cayrel04} \\
                 & 4789 & 1.72 & $-$3.00 & \citet{lai04} \\
                 & 4730 & 1.30 & $-$3.27 & \citet{cohen13} \\
 CS~22891--200   & 4490 & 0.50 & $-$3.88 & This study \\
                 & 4700 & 0.45 & $-$3.48 & \citet{mcwilliam95b} \\
                 & 4500 & 1.00 & $-$3.92 & \citet{hollek11} \\
 CS~22893--010   & 5150 & 2.45 & $-$2.93 & This study \\
                 & 5528 & 3.44 & $-$2.50 & \citet{lai04} \\
 CS~22948--066   & 4830 & 1.55 & $-$3.18 & This study \\
                 & 5170 & 1.80 & $-$3.16 & \citet{primas94} \\
                 & 5020 & 1.45 & $-$3.04 & \citet{mcwilliam95b} \\
                 & 5100 & 1.80 & $-$3.14 & \citet{cayrel04} \\
 CS~22949--037   & 4630 & 0.95 & $-$4.21 & This study \\
                 & 4810 & 2.10 & $-$3.99 & \citet{mcwilliam95b} \\
                 & 4900 & 1.70 & $-$3.79 & \citet{norris01} \\
                 & 4900 & 1.50 & $-$3.94 & \citet{depagne02} \\
                 & 4915 & 1.70 & $-$3.73 & \citet{cohen08} \\
                 & 4915 & 1.70 & $-$3.93 & \citet{cohen13} \\
 CS~22957--027   & 5220 & 2.65 & $-$3.00 & This study \\
                 & 4850 & 1.90 & $-$3.38 & \citet{norris97} \\
                 & 4839 & 2.25 & $-$3.43 & \citet{bonifacio98} \\
                 & 5050 & 2.00 & $-$2.96 & \citet{preston01} \\
                 & 5100 & 1.90 & $-$3.11 & \citet{aoki02b} \\
                 & 5205 & 2.50 & $-$3.13 & \citet{cohen06} \\
%                & 5205 & 2.50 & $-$3.06 & \citet{cohen13} \\ % (Fe)_sun -0.07
 CS~22958--042   & 5760 & 3.55 & $-$2.99 & This study \\
                 & 6217 & 3.50 & $-$3.34 & \citet{thorburn94} \\
                 & 6250 & 3.50 & $-$2.93 & \citet{sivarani06} \\
 CS~29498--043   & 4440 & 0.50 & $-$3.85 & This study \\
                 & 4400 & 0.60 & $-$3.75 & \citet{aoki02b} \\
                 & 4600 & 1.20 & $-$3.53 & \citet{aoki04} \\
                 & 4639 & 1.00 & $-$3.49 & \citet{yong13a} \\
 CS~29502--092   & 4820 & 1.50 & $-$3.20 & This study \\
                 & 5000 & 2.10 & $-$2.76 & \citet{aoki02a} \\
                 & 5114 & 2.51 & $-$3.00 & \citet{lai04} \\
                 & 4890 & 1.72 & $-$3.20 & \citet{lai08} \\
                 & 5123 & 2.20 & $-$2.83 & \citet{ruchti11a} \\
 CS~30314--067   & 4320 & 0.50 & $-$3.01 & This study \\
                 & 4400 & 0.70 & $-$2.85 & \citet{aoki02a} \\
 HE~1012$-$1540  & 5230 & 2.65 & $-$3.76 & This study \\
                 & 5620 & 3.40 & $-$3.71 & \citet{cohen08} \\
                 & 5745 & 3.45 & $-$3.47 & \citet{yong13a} \\
\enddata
\tablenotetext{a}{
As derived from Fe~\textsc{ii} lines, if specified
}
\end{deluxetable}
  % 1 col

Table~\ref{comparetab} compares the derived \teff\ and metallicity
with those found by previous investigators
for the CEMP-no and NEMP-no stars in our sample.
Most of these previous studies calculated \teff\ 
using color-\teff\ relations, frequently leading to
warmer \teff\ and higher metallicity for the red giants.
Even so, there is a fair amount of scatter in 
the derived values, and Table~\ref{comparetab} shows
that our values are reasonable given the different methods employed.
\citet{roederer14} found that the derived metallicities were,
on average, lower than those found by previous studies
by 0.25, 0.04, and 0.12~dex for 
red giants (108~stars),
subgiants (40~stars), and 
stars on the horizontal branch (28~stars), 
respectively.
For the 13~stars listed in Table~\ref{comparetab}, 
our metallicities are lower than those found by previous studies
by 0.17~dex ($\sigma =$~0.25).

\subsection{Abundance Analysis}
\label{abundances}

Table~8 of \citet{roederer14} lists the 
line wavelength, species identification,
excitation potential (E.P.)\ of the lower level,
and \loggf\ value
for each transition examined.
Spectrum synthesis matching was performed for lines broadened by
hyperfine splitting (hfs) or in cases where
a significant isotope shift (IS) may be present.
Damping constants were adopted from \citet{barklem00}
and \citet{barklem05b} when available,
otherwise the standard \citet{unsold55} approximation was used.
Linelists were generated using the \citet{kurucz95} lists
and updated using more recent experimental data when available.
For unblended lines, \citeauthor{roederer14}\ used MOOG to compute
theoretical equivalent widths, which were then forced to match
measured equivalent widths by adjusting the abundance.
When a line was not detected, 
\citeauthor{roederer14}\ derived 3~$\sigma$ upper limits on the abundance.
Table~11 of \citeauthor{roederer14}\ lists the abundances
derived from each line in each star.
Relative abundances were computed 
with respect to the solar ratios given by 
\citet{asplund09} and are listed in Table~13 of 
\citeauthor{roederer14} %.

Carbon abundances were derived from the CH 
$A^2\Delta - X^2\Pi$ G band.
The C$_{2}$ $d^3\Pi - a^3\Pi$ (0, 0) Swan bands were
detected in three stars,
\csd, \cse, and \csf.
The carbon abundances derived from the CH and C$_{2}$
bands only agree within a factor of $\approx$~3
with a fair amount of scatter,
and we will continue to investigate these discrepancies elsewhere.
Nitrogen abundances were derived from the 
NH $A^3\Pi - X^3\Sigma$ band, if detected.
The CN $B^2\Sigma - X^2\Sigma$ band was also detected in seven stars
in our sample.
Nitrogen abundances derived from these bands
show a consistent offset of 0.32~$\pm$~0.10~dex
with the nitrogen abundance derived from CN being higher.
Since NH was detected more frequently than CN,
we adopt the nitrogen abundances derived from NH
unless the S/N at the NH band was too low to enable a measurement
or upper limit.

\citet{roederer14} leveraged their large dataset
to identify lines yielding derived abundances
systematically lower or higher than 
other lines of the same species.
This effort minimized
systematic effects resulting from using
different lines as abundance indicators.
A list of these corrections is given in Table~16 of 
\citeauthor{roederer14} %.
Independently, that study also adopted corrections
to account for departures from local thermodynamic equilibrium (LTE) 
in the line formation regions for 
Li~\textsc{i} \citep{lind09},
O~\textsc{i} \citep{fabbian09}, 
Na~\textsc{i} \citep{lind11}, and 
K~\textsc{i} \citep{takeda02}.
These corrections are listed in Table~15 of \citeauthor{roederer14} %.

Weighted mean abundances and uncertainties were computed using the
formalism presented in \citet{mcwilliam95b},
as discussed in detail in \citet{roederer14}.
These abundances are reported for each CEMP-no and NEMP-no star in 
Tables~\ref{abundtab1}--\ref{abundtab17}.
Several sets of uncertainties are listed in each table.
The statistical uncertainty, $\sigma_{\rm statistical}$, 
accounts for uncertainties in the equivalent widths, \loggf\ values,
non-LTE corrections, and line-by-line offset
corrections.
The total uncertainty, $\sigma_{\rm total}$,
accounts for the statistical uncertainty and 
uncertainties in the model atmosphere parameters.
The other two uncertainties listed in 
Tables~\ref{abundtab1}--\ref{abundtab17}
are approximations to the abundance ratio uncertainties
given by equations~A19 and A20 of \citeauthor{mcwilliam95b} %.
The quantity $\sigma_{\rm neutrals}$ for element A 
should be added in quadrature with $\sigma_{\rm statistical}$ for
element B when computing the ratio [A/B] when B is 
derived from neutral lines.
Similarly, $\sigma_{\rm ions}$ for element A 
should be added in quadrature with $\sigma_{\rm statistical}$ for
element B when element B is derived from ionized lines.

\begin{deluxetable}{lc}
\tablecaption{$^{12}$C/$^{13}$C Ratios
\label{cisotab}}
%\rotate
\tablewidth{0pt}
\tabletypesize{\scriptsize}
%\tabletypesize{\tiny}
\tablehead{
\colhead{Star} &
\colhead{$^{12}$C/$^{13}$C} }
\startdata      
BD$+$44 493     & $>$15 \\
CS 22877--001   & 35~$\pm$~15 \\
CS 22878--101   & \nodata \\
CS 22891--200   & $>$6 \\
CS 22893--010   & $>$5 \\
%CS 22943--132   & $>$4 \\
CS 22943--201   & $>$12 \\
CS 22948--066   & \nodata \\
CS 22949--037   & $>$4 \\
CS 22957--027   & 6~$\pm$~2 \\
CS 22958--042   & 7~$\pm$~2 \\
CS 22960--064   & 15~$\pm$~5 \\
CS 29498--043   & 8~$\pm$~3 \\
CS 29502--092   & 12~$\pm$~6 \\
CS 30314--067   & 5~$\pm$~1 \\
CS 30492--001   & \nodata \\
HE 1012$-$1540  & $>$30 \\
\enddata
\end{deluxetable}
  % 1 col

We examine the $^{12}$C/$^{13}$C isotope ratio in each star in our sample
using seven isolated CH features between 4209 and 4222~\AA.
(\citealt{roederer14} did not measure the $^{12}$C/$^{13}$C ratios
for any stars in their sample.)
Six stars yielded only lower limits on 
$^{12}$C/$^{13}$C, and in three stars
the CH features were too weak to estimate this ratio reliably.  
$^{13}$CH features are identified
in the other seven stars.
These results are listed in Table~\ref{cisotab}.

\section{Results}
\label{results}

\subsection{Iron, Strontium, and Barium}
\label{histograms}

\begin{figure}
\begin{center}
\includegraphics[angle=0,width=2.8in]{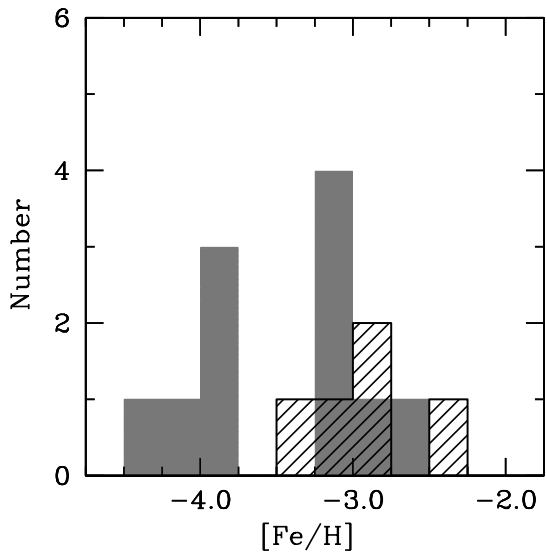} \\
\end{center}
\caption{
\label{metallicity}
The metallicity distribution of our sample.
The gray shaded histogram represents the 11 carbon-enhanced stars
and the hatched histogram represents the 5~stars 
that are nitrogen-enhanced but not carbon-enhanced.
}
\end{figure}

Figure~\ref{metallicity} shows the metallicity distribution
of our sample, which spans a range
of $-$4.3~$<$~[Fe/H]~$< -$2.3;
the median [Fe/H] is $-$3.2.
(Recall that our metallicities are $\approx$~0.17~dex
lower than those found by previous studies
that derived \teff\ by different methods;
see Section~\ref{modelatm}.)
The metallicities of our CEMP-no and NEMP-no stars are
low enough that relatively small numbers of
supernovae can be expected to have pre-enriched the gas from
which they formed.
Yet the metallicities are high enough that 
elements heavier than the 
iron group can still be detected.

\begin{figure*}
\begin{center}
\includegraphics[angle=0,width=2.8in]{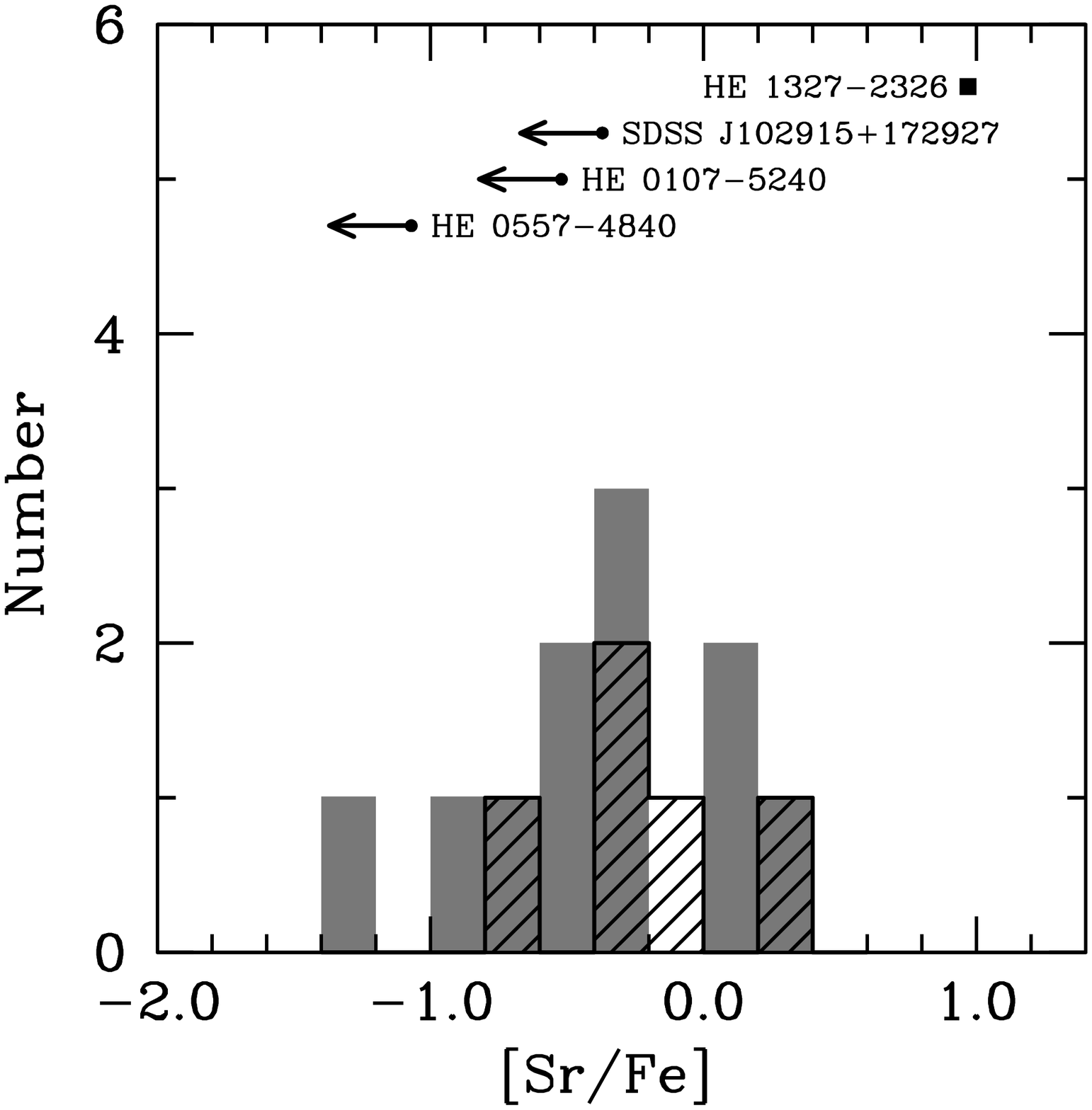} 
\hspace*{0.2in}
\includegraphics[angle=0,width=2.8in]{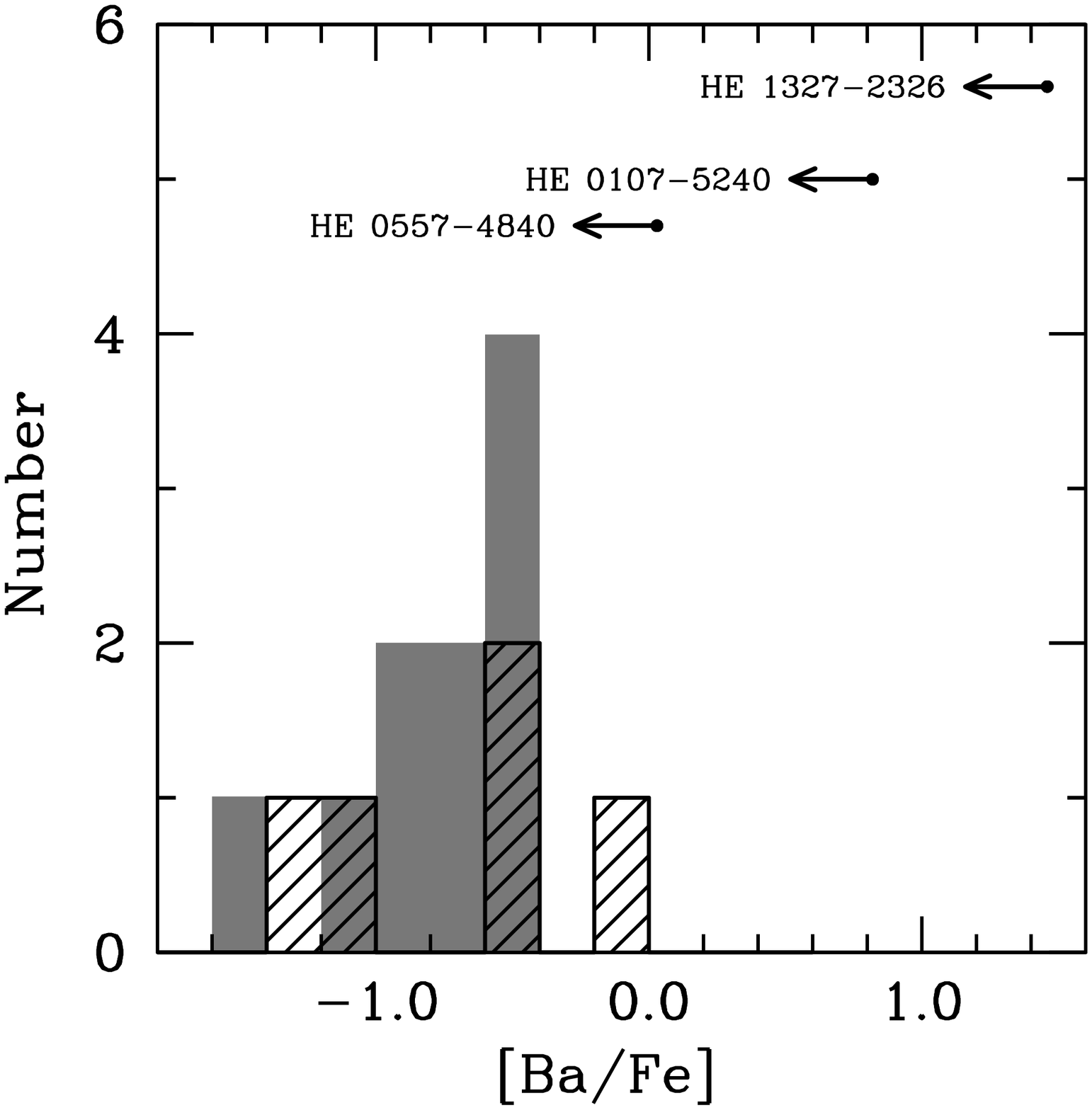} \\
\end{center}
\caption{
\label{srhist}
The [Sr/Fe] and [Ba/Fe] distributions of our sample.
The gray shaded histogram represents the 11 carbon-enhanced stars
and the hatched histogram represents the 5~stars 
that are nitrogen-enhanced but not carbon-enhanced.
Abundances or upper limits for stars with
[Fe/H]~$< -$4.5 are also shown:\
\mbox{HE~0557$-$4840} \citep{norris07},
\mbox{HE~0107$-$5240} \citep{christlieb04}, 
\mbox{SDSS~J102915$+$172927} \citep{caffau12}, and
\mbox{HE~1327$-$2326} \citep{frebel05,aoki06}.
No barium abundance or upper limit
has been reported for \mbox{SDSS~J102915$+$172927}.
}
\end{figure*}

Our CEMP-no sample
is shown by the shaded histogram in Figure~\ref{metallicity}.
Our NEMP-no sample, comprised of the 5~stars with [N/Fe]~$> +$1.0
and low [C/Fe]
is shown by the hatched black histogram.
A two-sided Kolmogorov-Smirnov (K-S) test rejects the 
null hypothesis that the metallicity distributions of
the CEMP-no and NEMP-no samples are drawn from the same distribution
at the 77\% confidence level.
Figure~\ref{srhist} shows similar histograms for the
[Sr/Fe] and [Ba/Fe] ratios.
For [Sr/Fe], [Ba/Fe], and [Sr/Ba] (not shown in
Figure~\ref{srhist}), the two-sided K-S test rejects the
null hypothesis that the CEMP-no and NEMP-no samples
are drawn from the same distribution
at the 8\%, 46\%, and 14\% confidence levels.   
% 29% --> 8% because the star with Ba u.l.
% had been excluded in the older sm calculations by accident.
These tests indicate that the CEMP-no and NEMP-no samples do not exhibit
significantly different distributions of
[Fe/H], [Sr/Fe], [Ba/Fe], and [Sr/Ba].

We also examine whether the 
[Sr/Fe], [Ba/Fe], or [Sr/Ba] distributions 
in either the CEMP-no, NEMP-no, or combined sample
are different than ``normal''
stars in the \citet{roederer14} comparison sample.
We exclude from the \citeauthor{roederer14}\ sample
all stars that are carbon- or \spro\ rich,
are included in the CEMP-no or NEMP-no samples, or
lack detection of both Sr~\textsc{ii} and Ba~\textsc{ii} lines.
This leaves 266~stars for comparison.
When comparing [Sr/Fe] in the CEMP-no and normal samples, for example,
we select 11~stars at random from the normal sample.
We then perform a K-S test on the [Sr/Fe] distributions in 
the CEMP-no and this subset of 11~normal stars
to calculate the probability of rejecting the null hypothesis
that these samples are drawn from the same distribution.
We conduct 1000 such trials, finding that the null hypothesis
can be rejected only at the 76\% confidence level on average.
Similar tests conducted for [Sr/Fe] in the NEMP-no and combined
samples can reject the null hypothesis at only the 
41\% and 80\% confidence levels.
For [Ba/Fe] in the CEMP-no, NEMP-no, and combined samples,
this test can only reject the null hypothesis at the
88\%, 42\%, and 87\% confidence levels.
For [Sr/Ba] in the CEMP-no, NEMP-no, and combined samples,
this test can only reject the null hypothesis at the
32\%, 52\%, and 43\% confidence levels.

In summary, 
the CEMP-no, NEMP-no, and normal star samples are
not distinct from one another from the
perspective of the distributions of [Sr/Fe], [Ba/Fe], or [Sr/Ba] ratios.
We will thus proceed to analyze both the CEMP-no and NEMP-no
samples identically.
Discussion of the heavy elements in the stars without
carbon- and nitrogen enhancements is deferred for future work.

\subsection{Detection of Strontium and Barium}
\label{detection}

\begin{figure}
\begin{center}
\includegraphics[angle=0,width=3.35in]{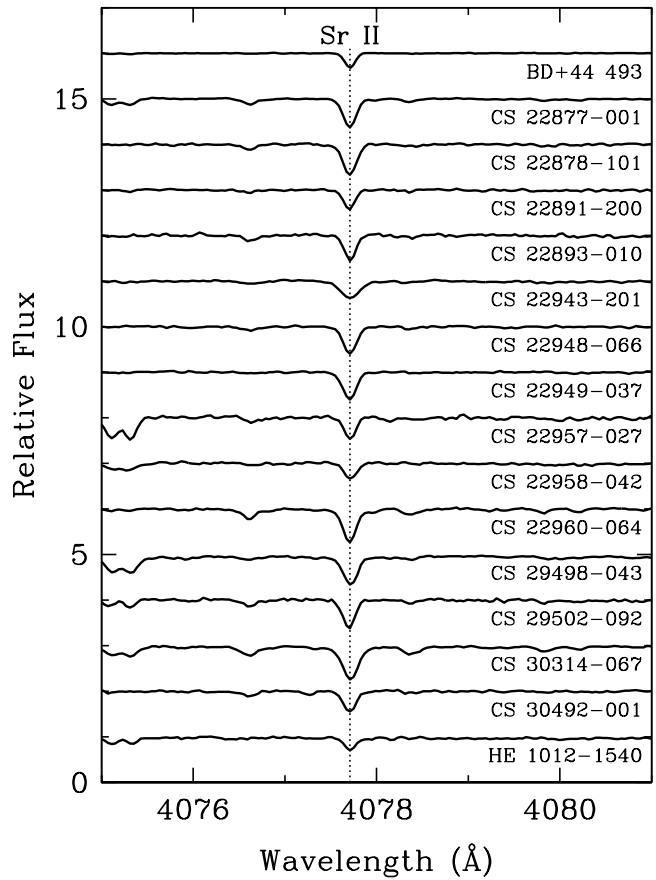} \\
\end{center}
\caption{
\label{srfig}
Spectra of the Sr~\textsc{ii} 4077~\AA\ line
for all 16~stars in our sample.
The spectra have been offset vertically by intervals of 1.0.
}
\end{figure}

\begin{figure}
\begin{center}
\includegraphics[angle=0,width=3.35in]{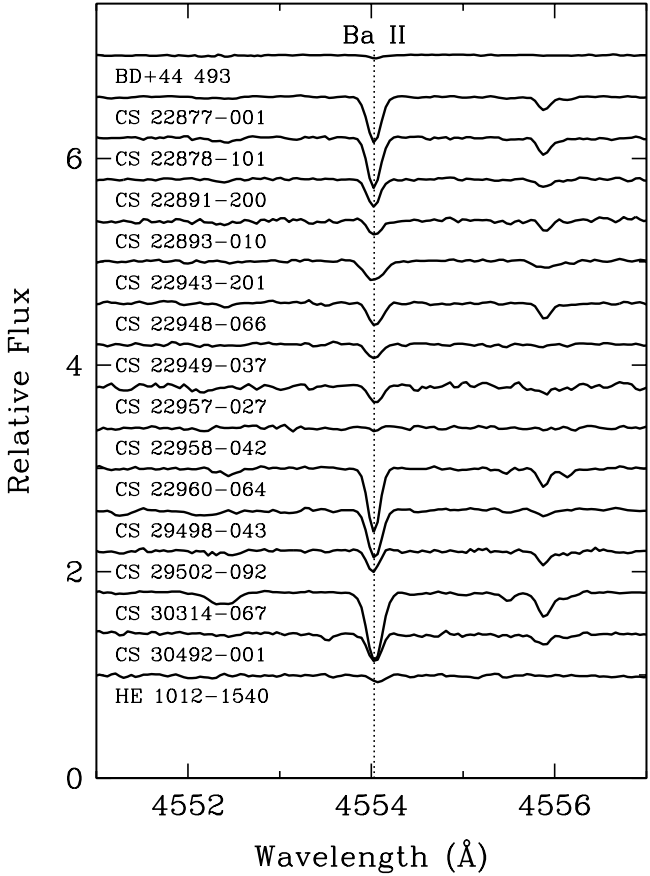} \\
\end{center}
\caption{
\label{bafig}
Spectra of the Ba~\textsc{ii} 4554~\AA\ line
for all 16~stars in our sample.
The spectra have been offset vertically by intervals of 0.4.
}
\end{figure}

\begin{figure}
\begin{center}
\includegraphics[angle=0,width=3.35in]{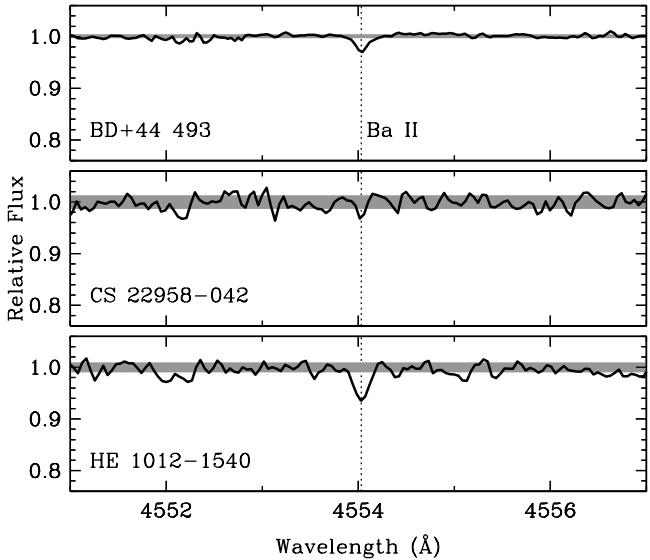} \\
\end{center}
\caption{
\label{ba2fig}
Spectra of the Ba~\textsc{ii} 4554~\AA\ line
for three stars with weak Ba~\textsc{ii} lines.
The gray shaded bands illustrate the noise level in each spectrum.
Although the Ba~\textsc{ii} line appears weak in 
\mbox{BD$+$44~493}, \mbox{CS~22958--042}, and \mbox{HE~1012$-$1540},
it is detected at the 7~$\sigma$ level in each of
\mbox{BD$+$44~493} and \mbox{HE~1012$-$1540}.
This line is only detected at the 1.3~$\sigma$ level in 
\mbox{CS~22958--042}, which is not a significant detection.
}
\end{figure}

We detect strontium in all 16~stars.
Figure~\ref{srfig} illustrates the spectral region
around the Sr~\textsc{ii} 4077~\AA\ line.
We detect barium in 15~stars,
as shown in Figure~\ref{bafig}.
At the scale shown in Figure~\ref{bafig},
the Ba~\textsc{ii} 4554~\AA\ line 
is difficult to see in 
\bd, \cse, and \he, so Figure~\ref{ba2fig} 
illustrates these spectra on a magnified scale.
The Ba~\textsc{ii} 4554~\AA\ line is detected
at the 7~$\sigma$ level in each of \bd\ and \he.
\citet{ito13} and \citet{cohen08} also 
detected this line in their spectra of these stars.
Meanwhile, the Ba~\textsc{ii} 4554~\AA\ line 
is only detected at the 1.3~$\sigma$ level in \cse,
which is not significant.
\cse\ is a warm subgiant (\teff~$=$~5760~K), and 
the 3~$\sigma$ upper limit derived from this line
constrains [Ba/Fe]~$< -$1.02.
The [Ba/Fe] ratio is even lower
in three stars in our sample, 
so it is possible that barium is present 
but not detected in \cse.
The strontium and barium detections form
our first main observational result:\
\textit{an element at or beyond the first neutron-capture peak
(i.e., strontium) is detected in all stars,
and an element at or beyond the second neutron-capture peak
(i.e., barium) is detected in nearly all stars.}

In contrast, other studies have shown that
it is observationally challenging to
detect strontium or barium in the four most iron-poor stars known,
those with [Fe/H]~$< -$4.5.
Three of these stars are carbon-enhanced, and two of these three
are nitrogen-enhanced.
The fourth star,
\object[SDSS J102915.14+172927.9]{SDSS~J102915$+$172927},
is not carbon- or nitrogen-enhanced, and no interesting
upper limits have been placed on its oxygen abundance
\citep{caffau12}.
Barium has not been detected in any of these stars,
and none of the upper limits constrains [Ba/Fe] to be sub-solar.
By construction all stars in our sample show sub-solar [Ba/Fe].
Among these four stars,
strontium has only been detected in 
\object[HE 1327-2326]{HE~1327$-$2326},
the most iron-poor star known,
where [Sr/Fe]~$= +$0.97 \citep{frebel05,aoki06}.
[Sr/Fe] is constrained to be sub-solar in the other three stars, 
and the upper limits are within the range of
[Sr/Fe] ratios found for our sample.
These upper limits are marked in Figure~\ref{srhist}.
As noted previously by \citeauthor{aoki06},
the [Sr/Fe] ratios span a range greater than 2~dex
in the four stars with [Fe/H]~$< -$4.5.
This is comparable
to the range of [Sr/Fe] ratios
found in our sample.
This forms our second main observational result:\
\textit{from the perspective of heavy elements ($Z \geq$~38),
our sample could represent 
higher-metallicity analogs of 
the carbon- and nitrogen-enhanced stars with [Fe/H]~$< -$4.5.}

\subsection{Light Elements}
\label{lightelements}

The abundance patterns of the light elements
(lithium through the iron group)
in our sample have been 
discussed in many of the studies referenced in
Section~\ref{sample}.
We build on those excellent studies 
by leveraging the large comparison sample of \citet{roederer14}
to place the abundance patterns of the CEMP-no and NEMP-no stars
in the context of other stars at similar metallicity
and evolutionary state.
This enables us to identify element ratios that
are outliers relative to the majority of 
carbon-normal metal-poor stars.
The advantage of this differential 
approach is that uncertainties 
related to the analysis techniques largely cancel out.

We identify all stars in the 
\citet{roederer14} sample that have similar \teff\ and [Fe/H]
to each CEMP-no or NEMP-no star in our sample.
In most cases, we select comparison stars that have \teff\ within
$\pm$~200~K and [Fe/H] within $\pm$~0.3~dex. 
This typically yields 10--20~stars for comparison
(minimum, 3~stars; maximum, 29~stars).
In a few cases, we broaden the range of \teff\ or [Fe/H]
to include sufficient numbers of comparison stars.
We exclude other CEMP-no or NEMP-no stars in our sample
from the comparison samples, and we also exclude stars
that exhibit high levels of \spro\ enrichment.
This signature indicates 
pollution by a companion that passed through the 
TP-AGB phase of evolution,
so the present-day abundances of these stars
are not representative of the 
interstellar medium from which they formed.

\begin{figure}
\begin{center}
\includegraphics[angle=0,width=3.35in]{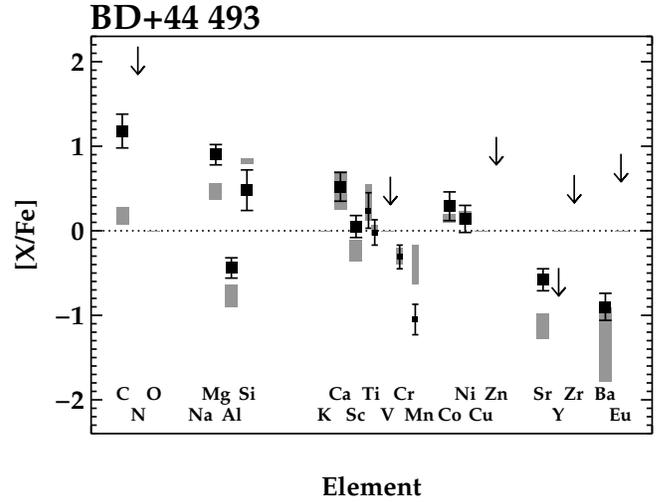} \\
\vspace*{0.3in}
\includegraphics[angle=0,width=3.35in]{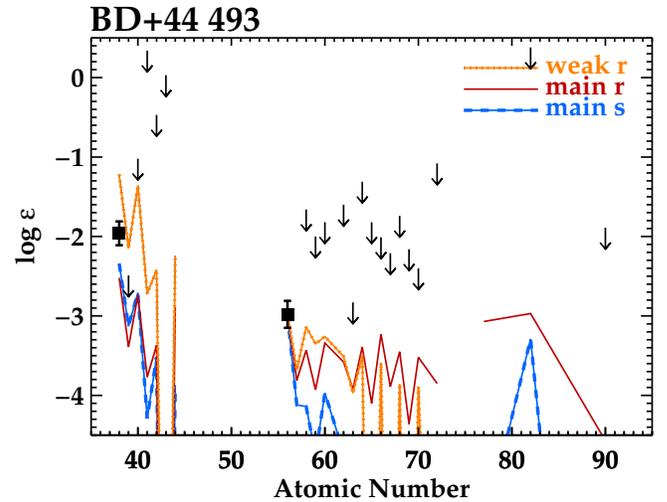}
\end{center}
\caption{
\label{bdp440493fig}
\scriptsize
TOP:\ 
Comparison of abundances in 
\mbox{BD$+$44~493}
(filled black squares signifying detections or arrows 
signifying upper limits)
with the average abundances of 
three other stars with 
\teff\ within $\pm$~250~K and
[Fe/H] within $\pm$~0.5~dex of 
\mbox{BD$+$44~493}.
This comparison sample, shown by the shaded gray boxes,
represents the mean $\pm$~1~$\sigma$ standard deviations.
The comparison sample is only 
shown if it is derived from three or more stars.
Smaller symbols are shown for
titanium, vanadium, chromium, and manganese
to accommodate ratios from both the neutral and ionized 
states, which may differ.
The dotted line marks the solar ratios.
BOTTOM:\
The heavy element distribution in 
\mbox{BD$+$44~493}.
Filled squares mark detections, and arrows mark 3~$\sigma$ upper limits
derived from non-detections.
The studded orange line marks the scaled heavy element
distribution found in the metal-poor giant \mbox{HD~122563}
\citep{honda06,roederer12b},
frequently referred to as the distribution
produced by the weak component of the \rpro.
The solid red line marks the scaled heavy element
distribution found in the metal-poor giant \mbox{CS~22892--052}
\citep{sneden03,sneden09,roederer09b},
frequently associated with the distribution
produced by the main component of the \rpro.
The long-dashed blue line marks the scaled heavy element
distribution predicted by the main and strong components
of the \spro\ \citep{sneden08,bisterzo11}.
Each of the three curves has been renormalized
to the barium abundance in
\mbox{BD$+$44~493}.
}
\end{figure}

\begin{figure}
\begin{center}
\includegraphics[angle=0,width=3.35in]{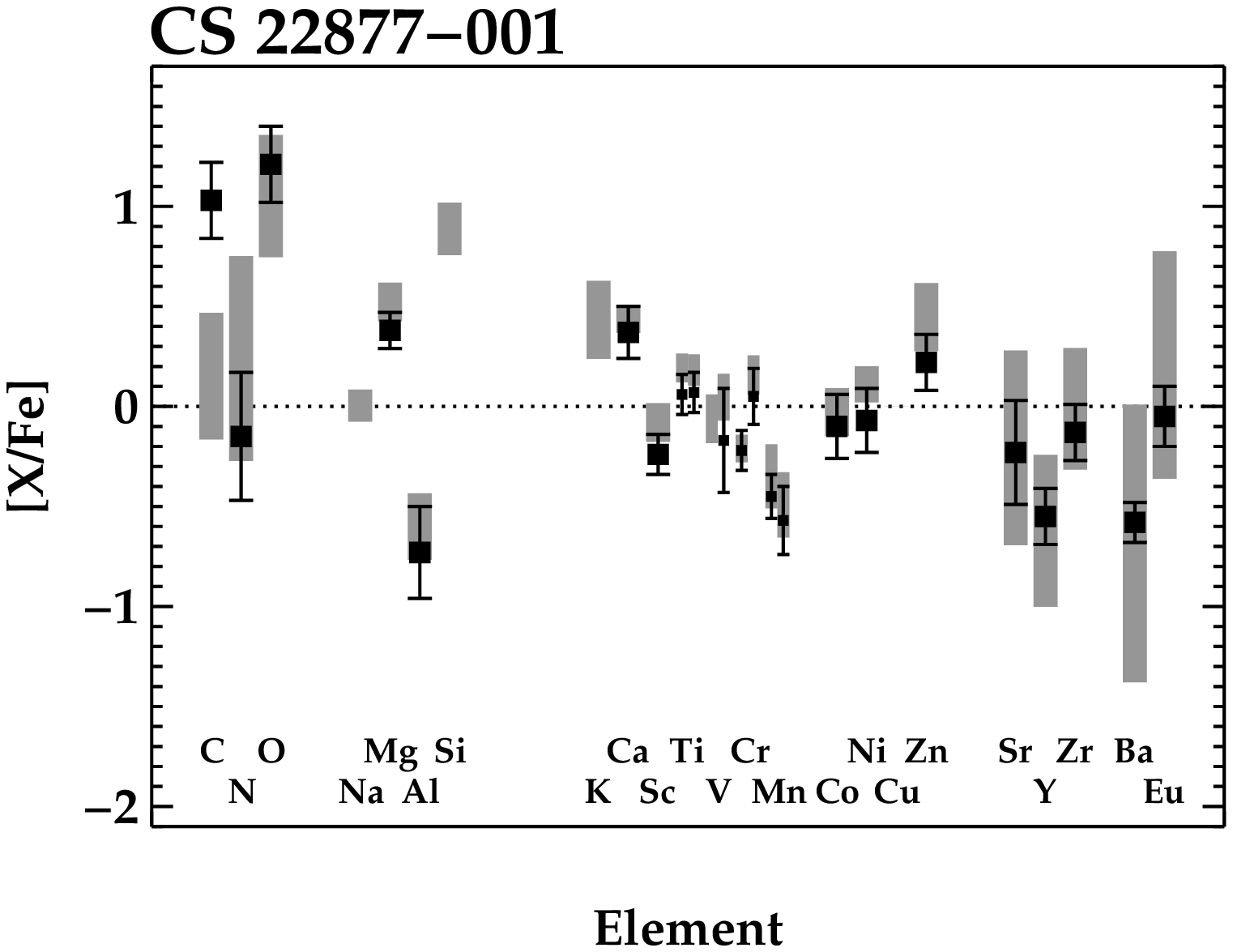} \\
\vspace*{0.3in}
\includegraphics[angle=0,width=3.35in]{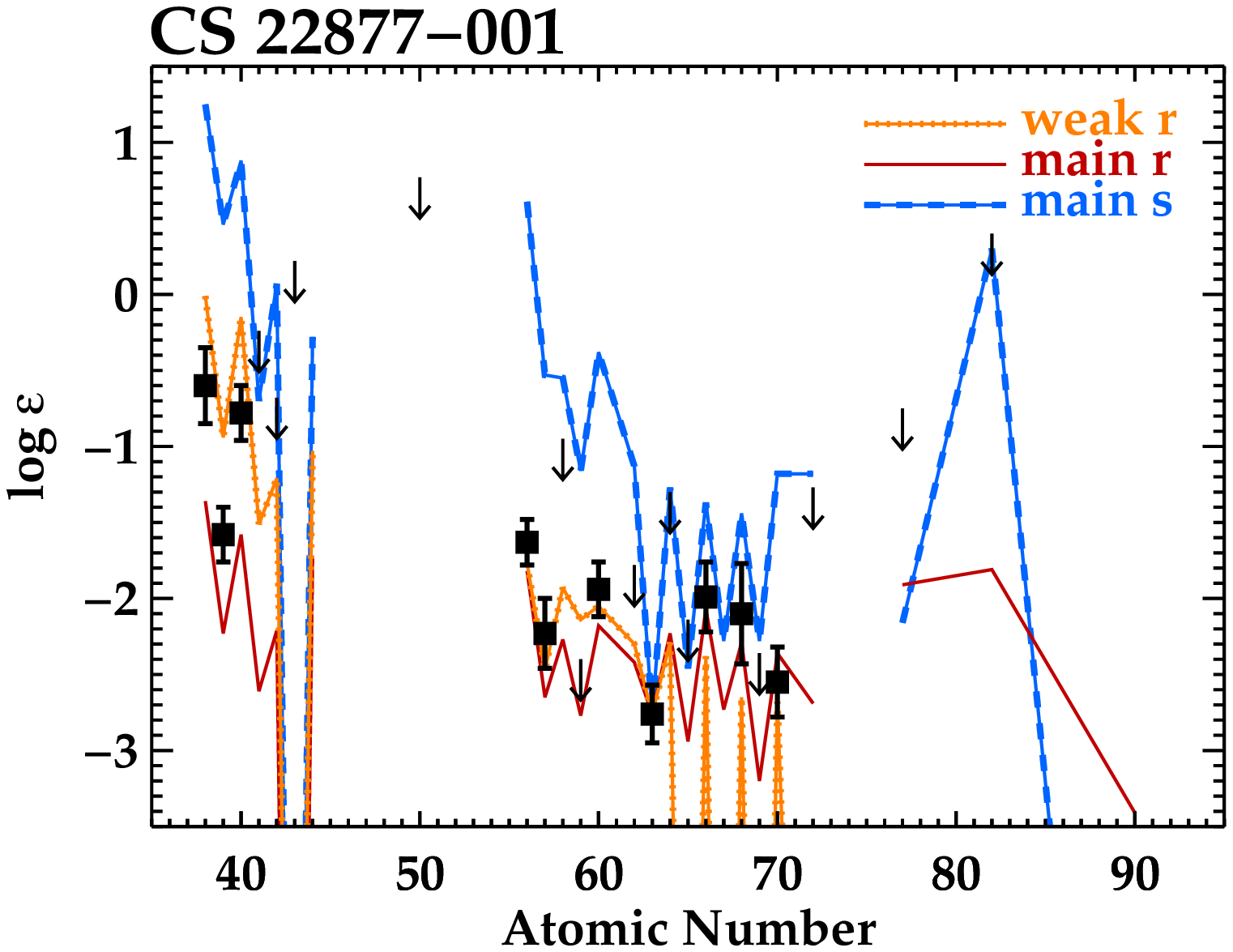}
\end{center}
\caption{
\label{cs22877m001fig}
\scriptsize
TOP:\ 
Comparison of abundances in 
\mbox{CS~22877--001}
with the average abundances of 
24 other stars with 
\teff\ within $\pm$~200~K and
[Fe/H] within $\pm$~0.3~dex of 
\mbox{CS~22877--001}.
BOTTOM:\
The heavy element distribution in 
\mbox{CS~22877--001}.
Each of the three curves has been renormalized
to the europium abundance in
\mbox{CS~22877--001}.
Symbols in both panels
are the same as in Figure~\ref{bdp440493fig}.
}
\end{figure}

\begin{figure}
\begin{center}
\includegraphics[angle=0,width=3.35in]{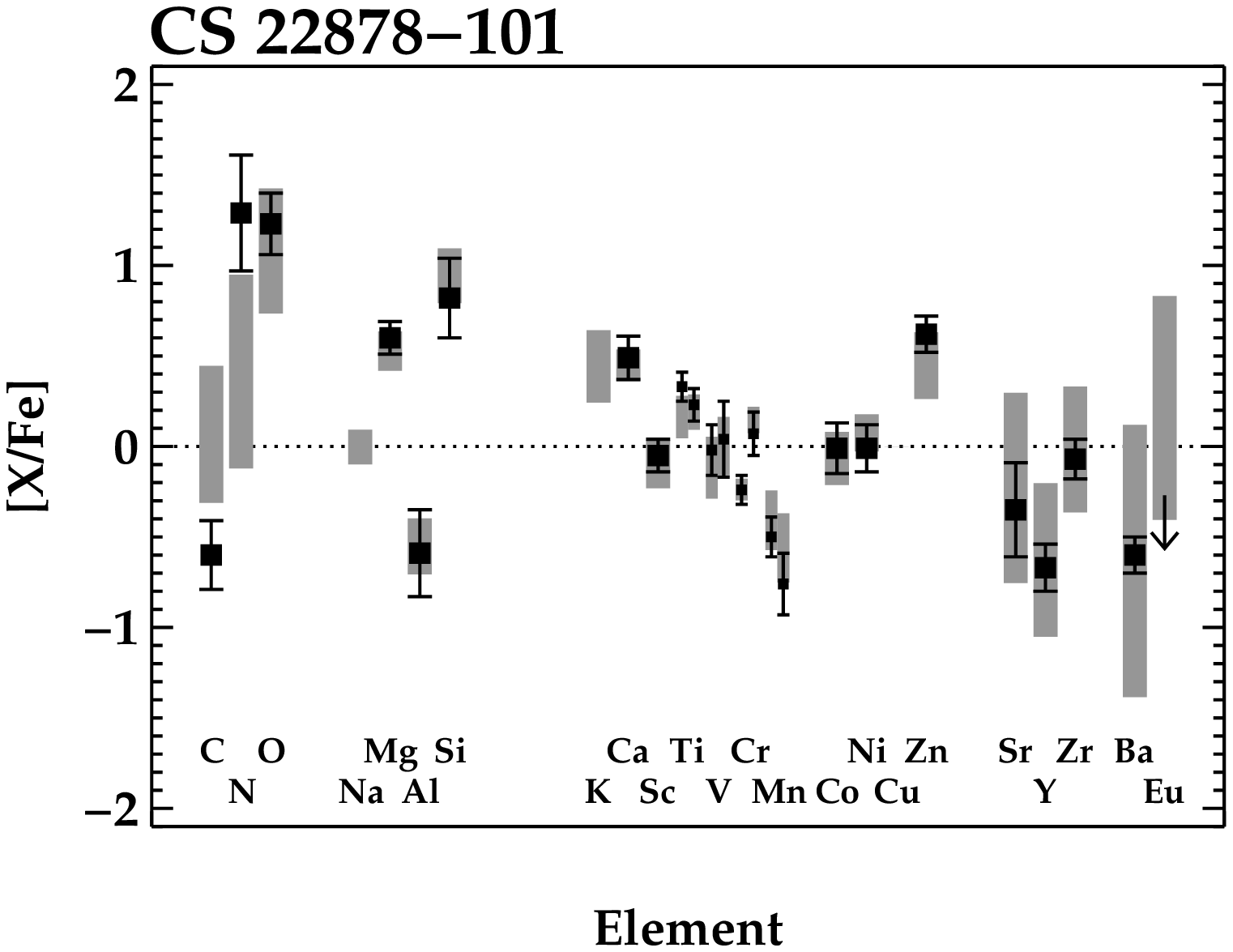} \\
\vspace*{0.3in}
\includegraphics[angle=0,width=3.35in]{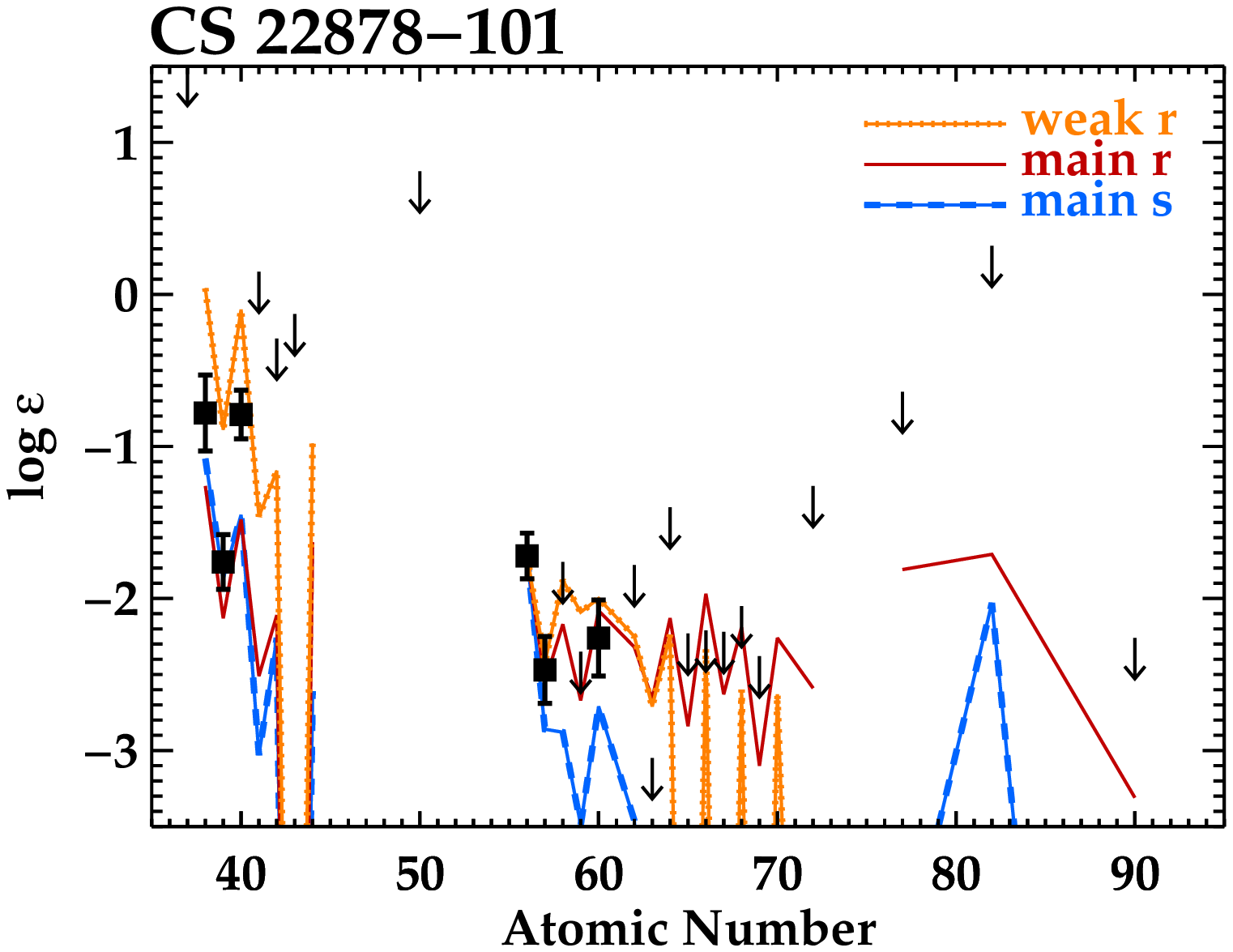}
\end{center}
\caption{
\label{cs22878m101fig}
\scriptsize
TOP:\ 
Comparison of abundances in 
\mbox{CS~22878--101}
with the average abundances of 
18 other stars with 
\teff\ within $\pm$~200~K and
[Fe/H] within $\pm$~0.3~dex of 
\mbox{CS~22878--101}.
BOTTOM:\
The heavy element distribution in 
\mbox{CS~22878--101}.
Each of the three curves has been renormalized
to the barium abundance in
\mbox{CS~22878--101}.
Symbols in both panels
are the same as in Figure~\ref{bdp440493fig}.
}
\end{figure}

\begin{figure}
\begin{center}
\includegraphics[angle=0,width=3.35in]{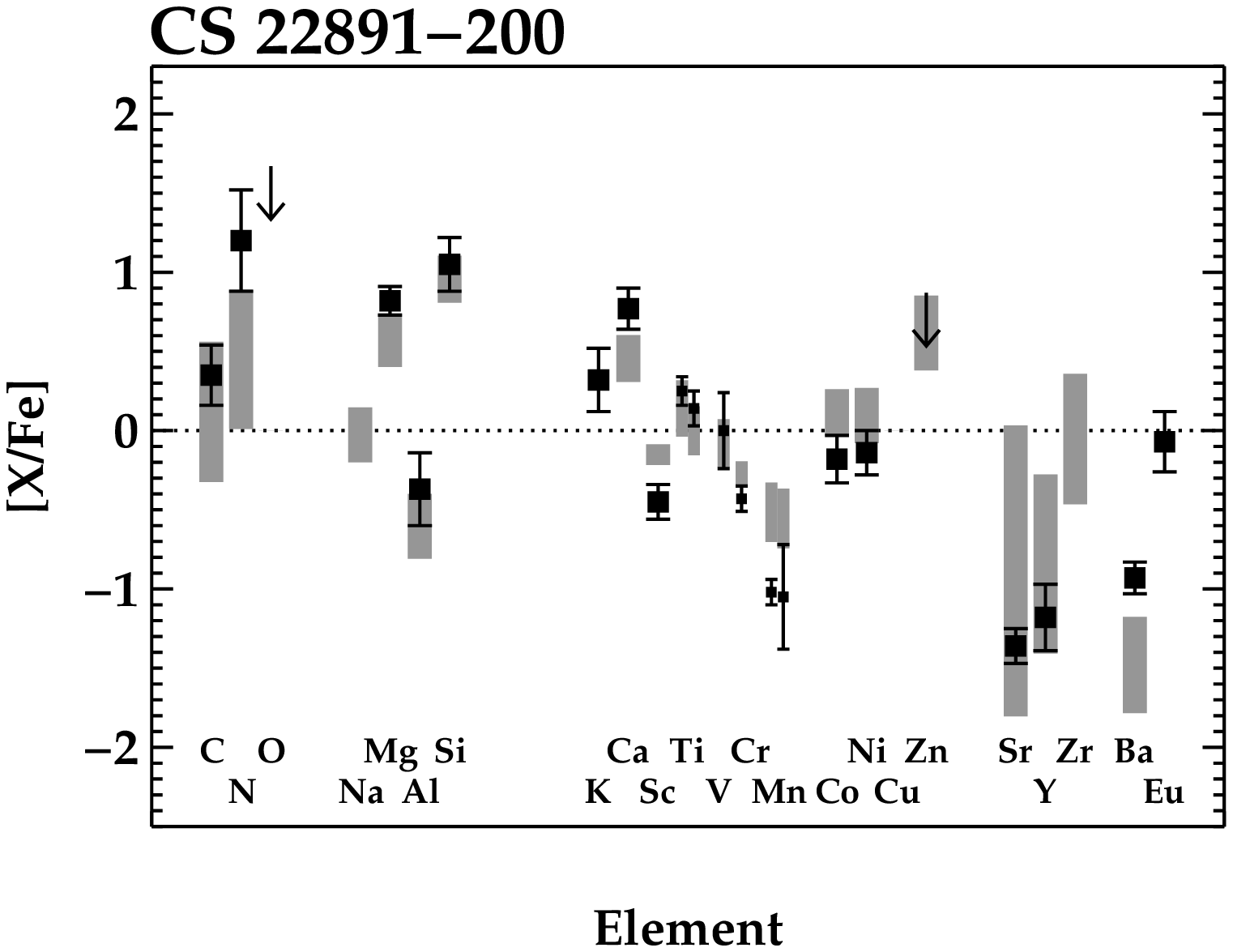} \\
\vspace*{0.3in}
\includegraphics[angle=0,width=3.35in]{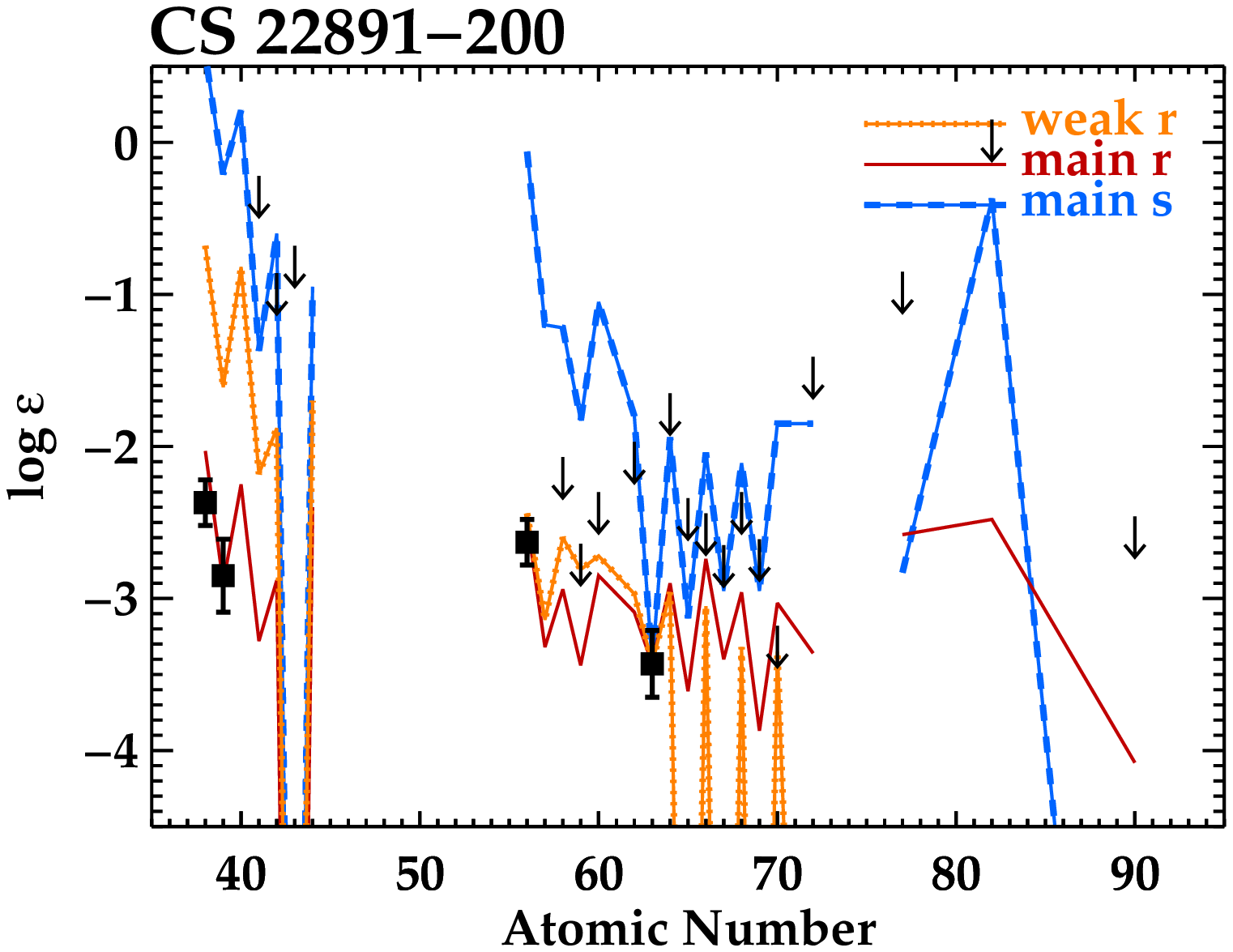}
\end{center}
\caption{
\label{cs22891m200fig}
\scriptsize
TOP:\ 
Comparison of abundances in 
\mbox{CS~22891--200}
with the average abundances of 
eight other stars with 
\teff\ within $\pm$~200~K and
[Fe/H] within $\pm$~0.3~dex of 
\mbox{CS~22891--200}.
BOTTOM:\
The heavy element distribution in 
\mbox{CS~22891--200}.
Each of the three curves has been renormalized
to the europium abundance in
\mbox{CS~22891--200}.
Symbols in both panels
are the same as in Figure~\ref{bdp440493fig}.
}
\end{figure}

\begin{figure}
\begin{center}
\includegraphics[angle=0,width=3.35in]{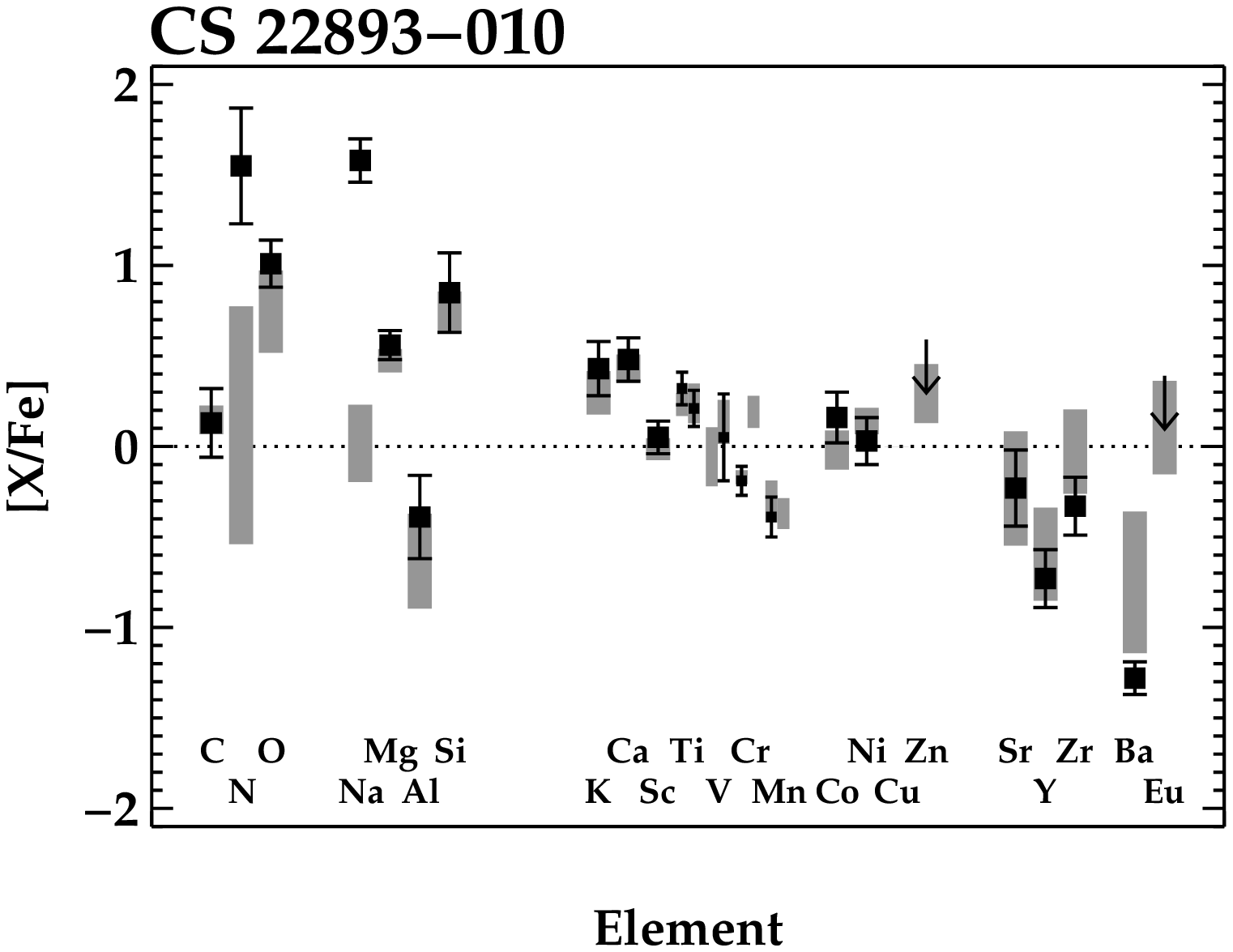} \\
\vspace*{0.3in}
\includegraphics[angle=0,width=3.35in]{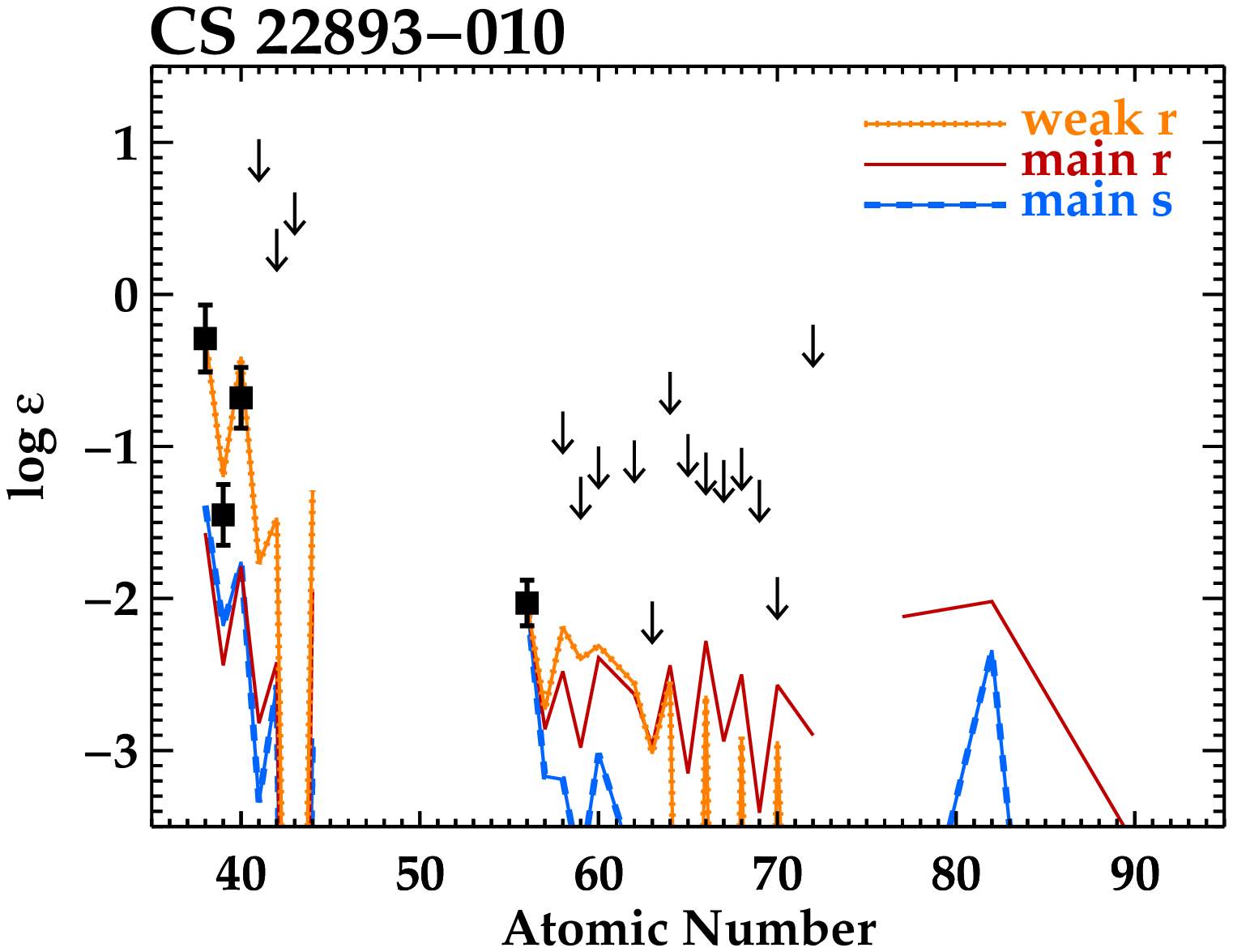}
\end{center}
\caption{
\label{cs22893m010fig}
\scriptsize
TOP:\ 
Comparison of abundances in 
\mbox{CS~22893--010}
with the average abundances of 
12 other stars with 
\teff\ within $\pm$~200~K and
[Fe/H] within $\pm$~0.3~dex of 
\mbox{CS~22893--010}.
BOTTOM:\
The heavy element distribution in 
\mbox{CS~22893--010}.
Each of the three curves has been renormalized
to the barium abundance in
\mbox{CS~22893--010}.
Symbols in both panels
are the same as in Figure~\ref{bdp440493fig}.
}
\end{figure}

\begin{figure}
\begin{center}
\includegraphics[angle=0,width=3.35in]{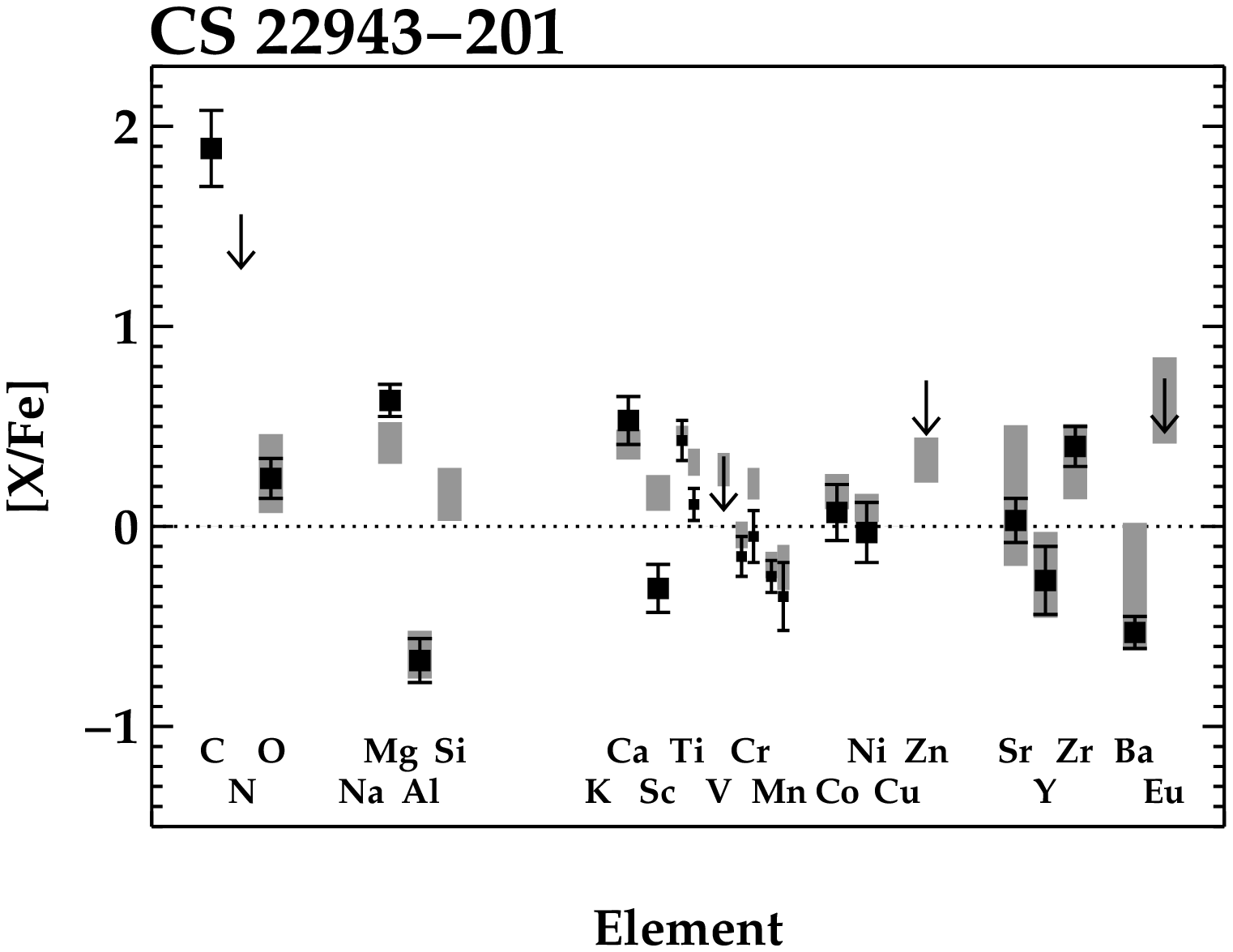} \\
\vspace*{0.3in}
\includegraphics[angle=0,width=3.35in]{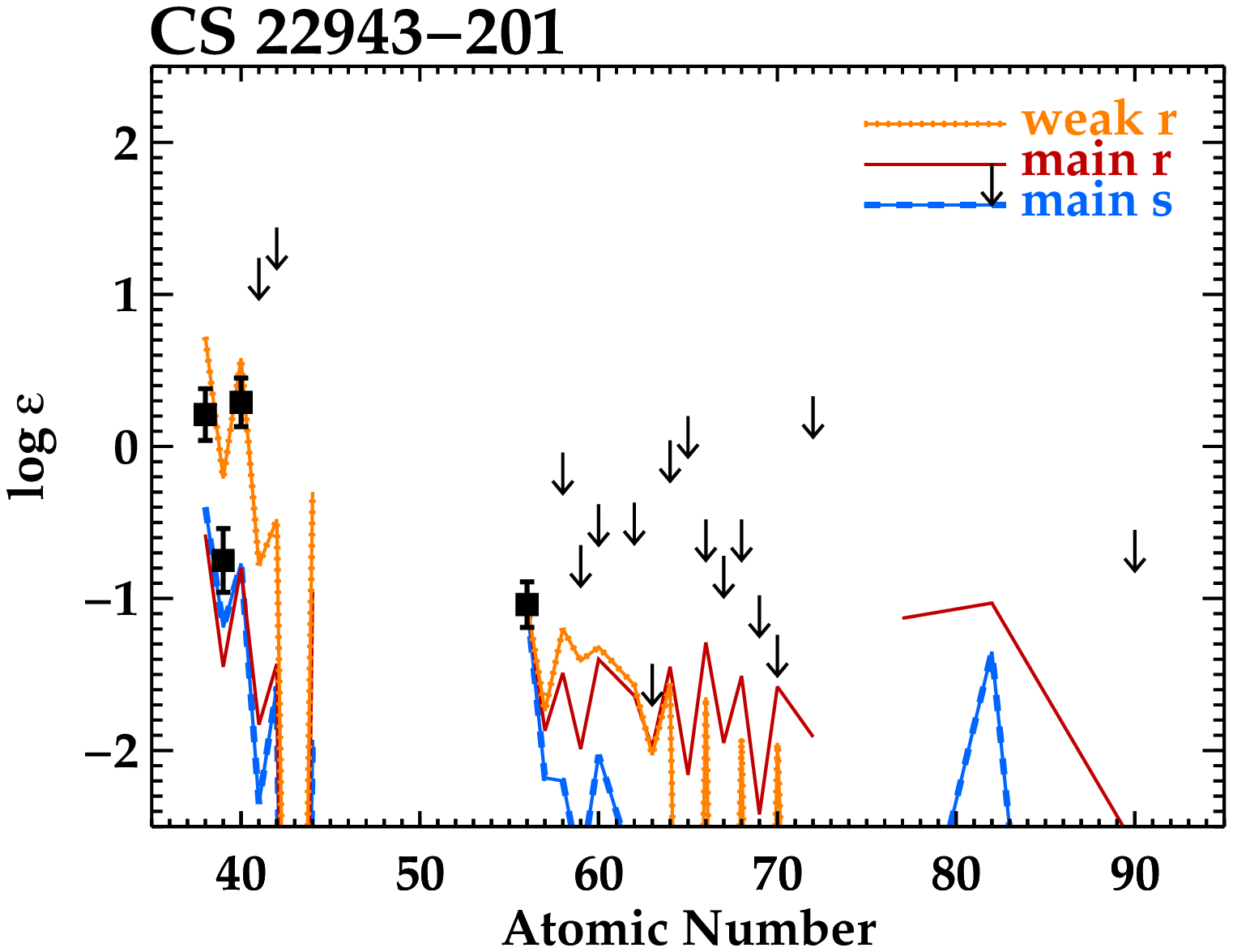}
\end{center}
\caption{
\label{cs22943m201fig}
\scriptsize
TOP:\ 
Comparison of abundances in 
\mbox{CS~22943--201}
with the average abundances of 
11 other stars with 
\teff\ within $\pm$~200~K and
[Fe/H] within $\pm$~0.3~dex of 
\mbox{CS~22943--201}.
BOTTOM:\
The heavy element distribution in 
\mbox{CS~22943--201}.
Each of the three curves has been renormalized
to the barium abundance in
\mbox{CS~22943--201}.
Symbols in both panels
are the same as in Figure~\ref{bdp440493fig}.
}
\end{figure}

\begin{figure}
\begin{center}
\includegraphics[angle=0,width=3.35in]{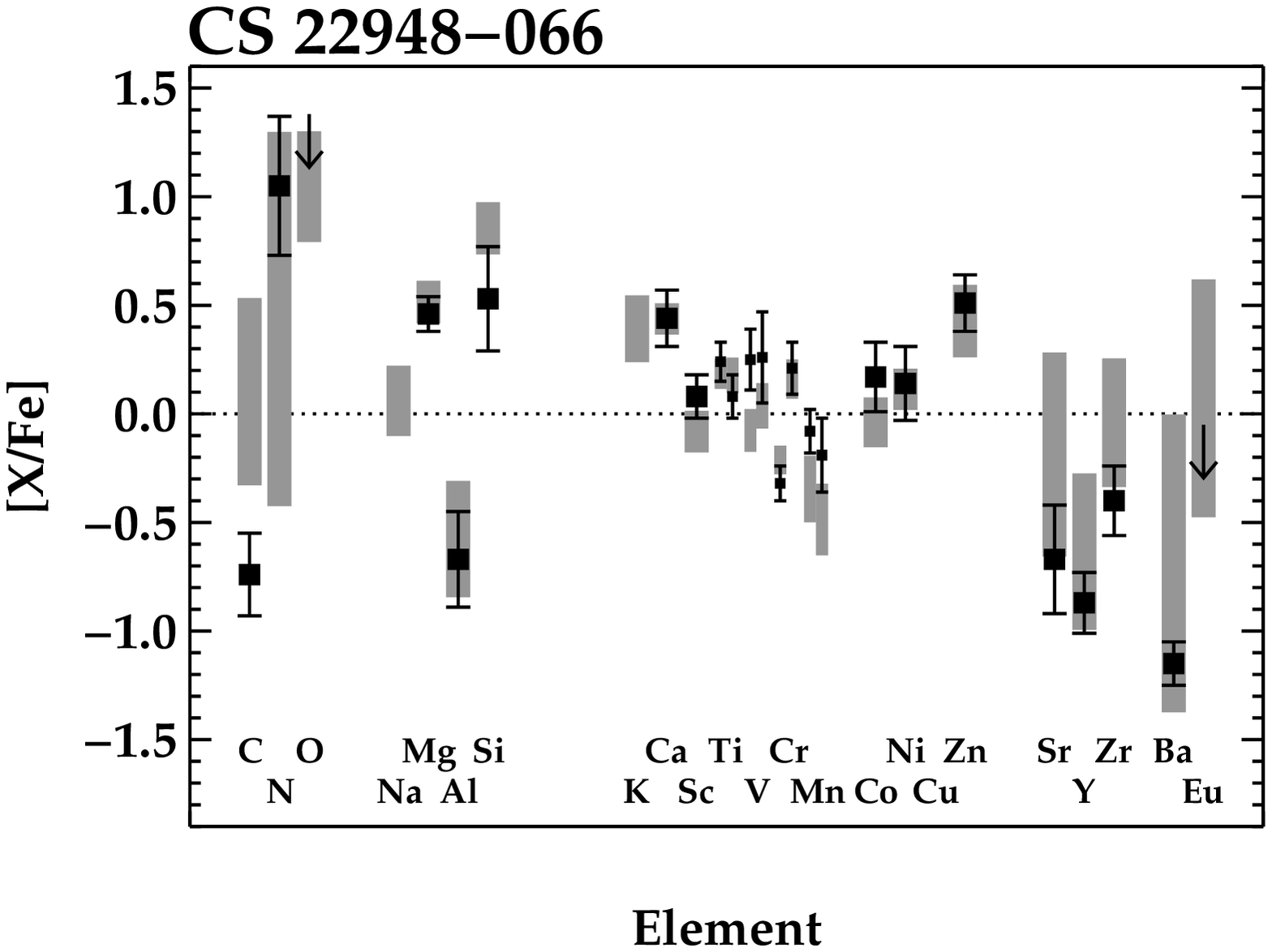} \\
\vspace*{0.3in}
\includegraphics[angle=0,width=3.35in]{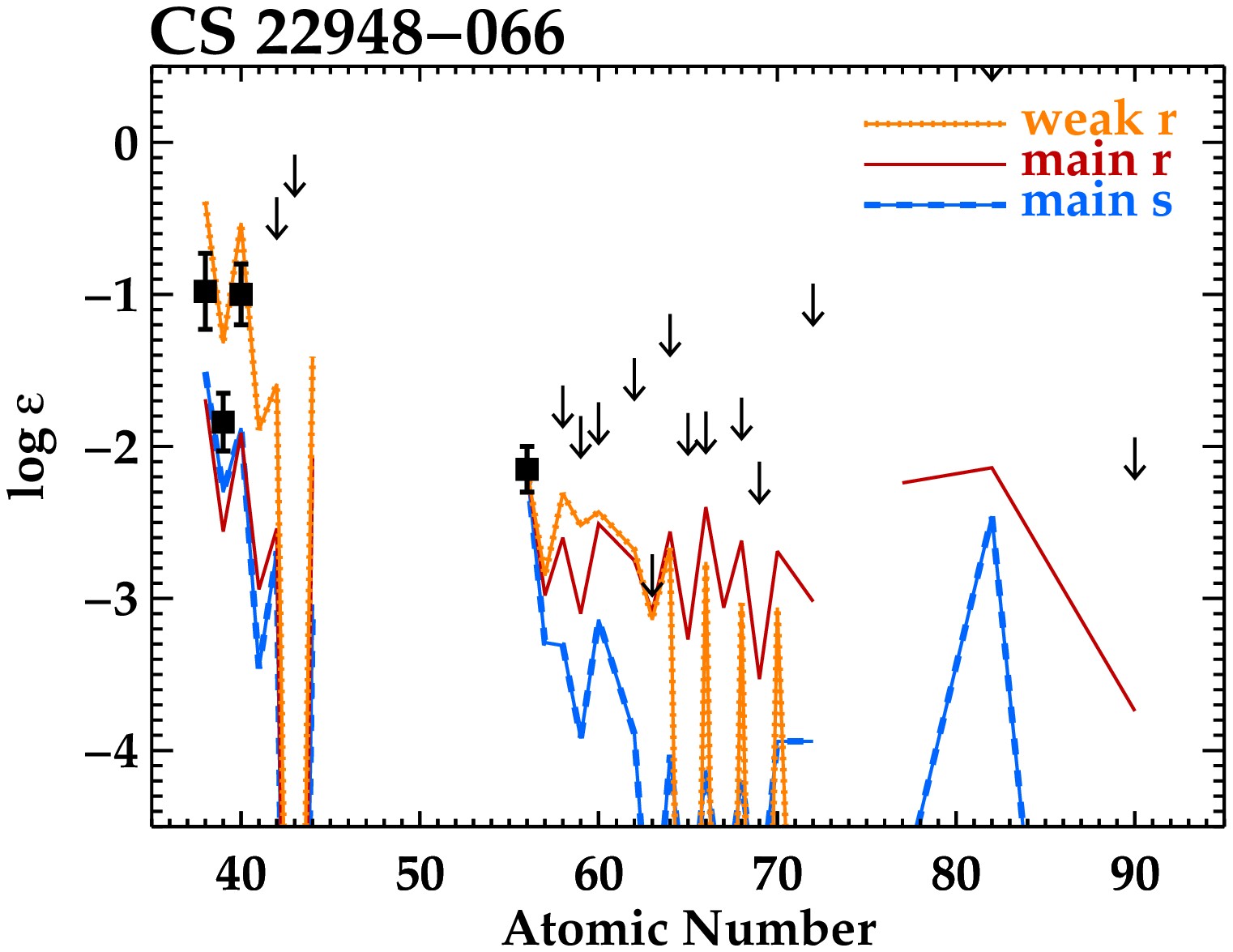}
\end{center}
\caption{
\label{cs22948m066fig}
\scriptsize
TOP:\ 
Comparison of abundances in 
\mbox{CS~22948--066}
with the average abundances of 
29 other stars with 
\teff\ within $\pm$~200~K and
[Fe/H] within $\pm$~0.3~dex of 
\mbox{CS~22948--066}.
BOTTOM:\
The heavy element distribution in 
\mbox{CS~22948--066}.
Each of the three curves has been renormalized
to the barium abundance in
\mbox{CS~22948--066}.
Symbols in both panels
are the same as in Figure~\ref{bdp440493fig}.
}
\end{figure}

\begin{figure}
\begin{center}
\includegraphics[angle=0,width=3.35in]{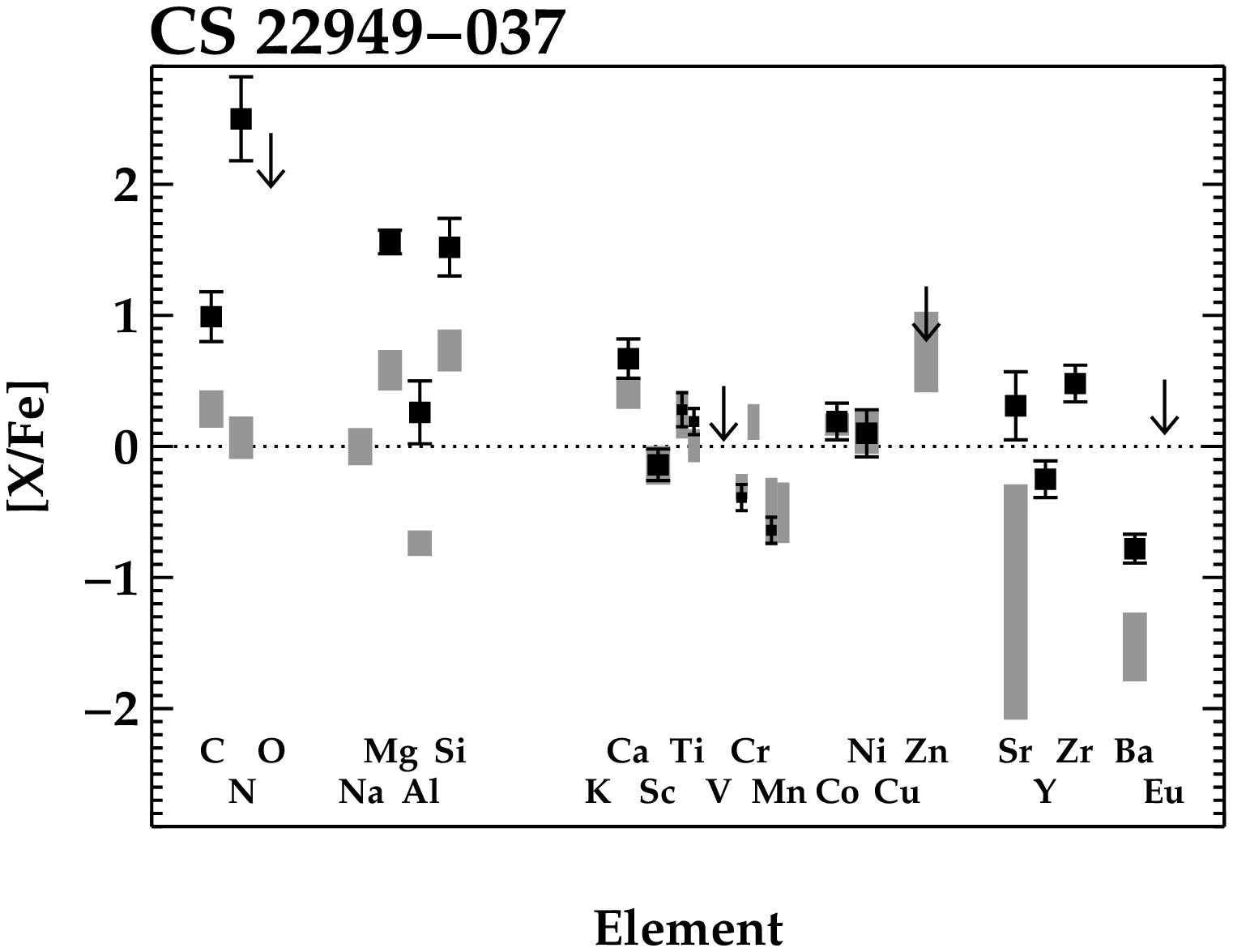} \\
\vspace*{0.3in}
\includegraphics[angle=0,width=3.35in]{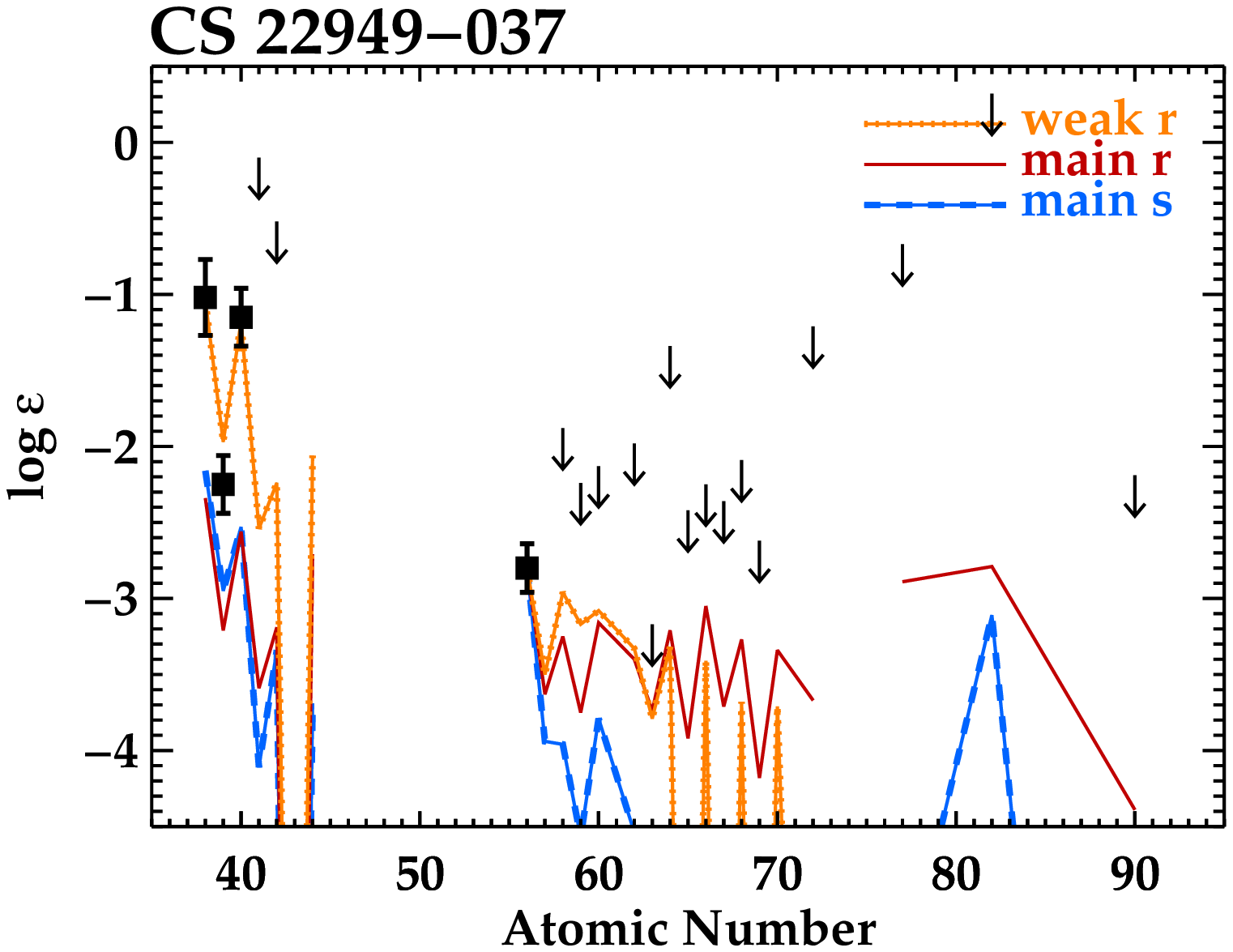}
\end{center}
\caption{
\label{cs22949m037fig}
\scriptsize
TOP:\ 
Comparison of abundances in 
\mbox{CS~22949--037}
with the average abundances of 
eight other stars with 
\teff\ within $\pm$~200~K and
[Fe/H] within $\pm$~0.5~dex of 
\mbox{CS~22949--037}.
BOTTOM:\
The heavy element distribution in 
\mbox{CS~22949--037}.
Each of the three curves has been renormalized
to the barium abundance in
\mbox{CS~22949--037}.
Symbols in both panels
are the same as in Figure~\ref{bdp440493fig}.
}
\end{figure}

\begin{figure}
\begin{center}
\includegraphics[angle=0,width=3.35in]{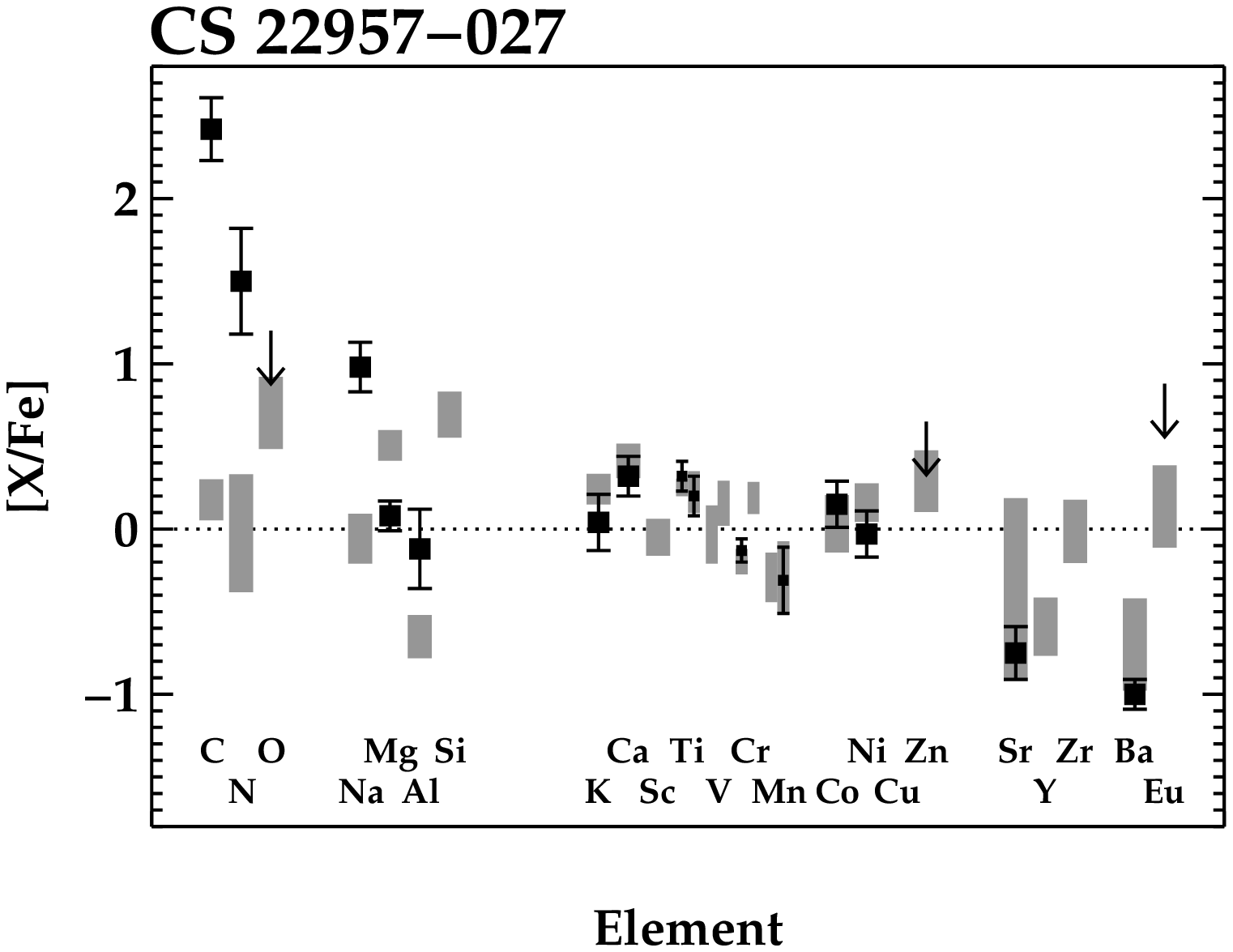} \\
\vspace*{0.3in}
\includegraphics[angle=0,width=3.35in]{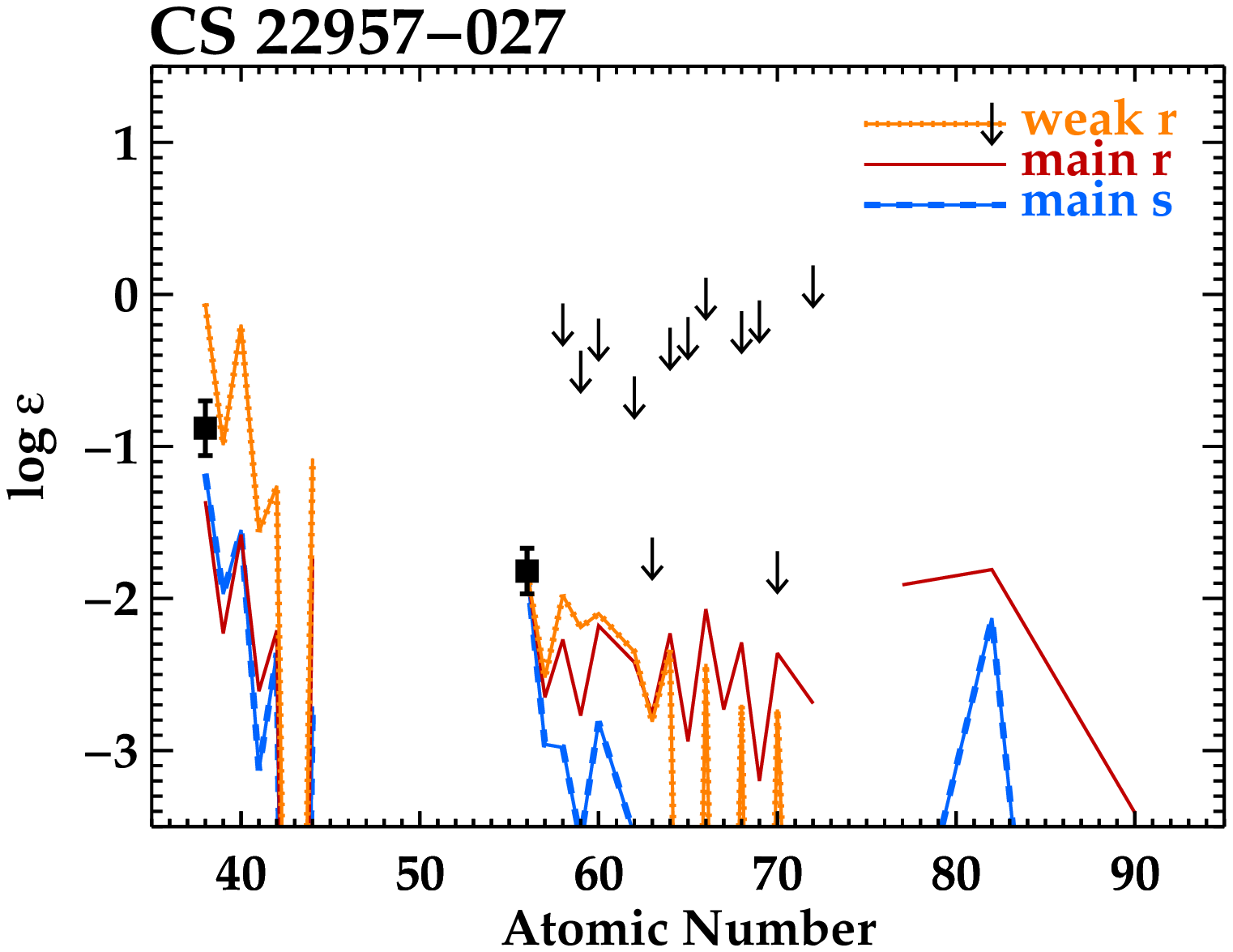}
\end{center}
\caption{
\label{cs22957m027fig}
\scriptsize
TOP:\ 
Comparison of abundances in 
\mbox{CS~22957--027}
with the average abundances of 
11 other stars with 
\teff\ within $\pm$~200~K and
[Fe/H] within $\pm$~0.4~dex of 
\mbox{CS~22957--027}.
BOTTOM:\
The heavy element distribution in 
\mbox{CS~22957--027}.
Each of the three curves has been renormalized
to the barium abundance in
\mbox{CS~22957--027}.
Symbols in both panels
are the same as in Figure~\ref{bdp440493fig}.
}
\end{figure}

\begin{figure}
\begin{center}
\includegraphics[angle=0,width=3.35in]{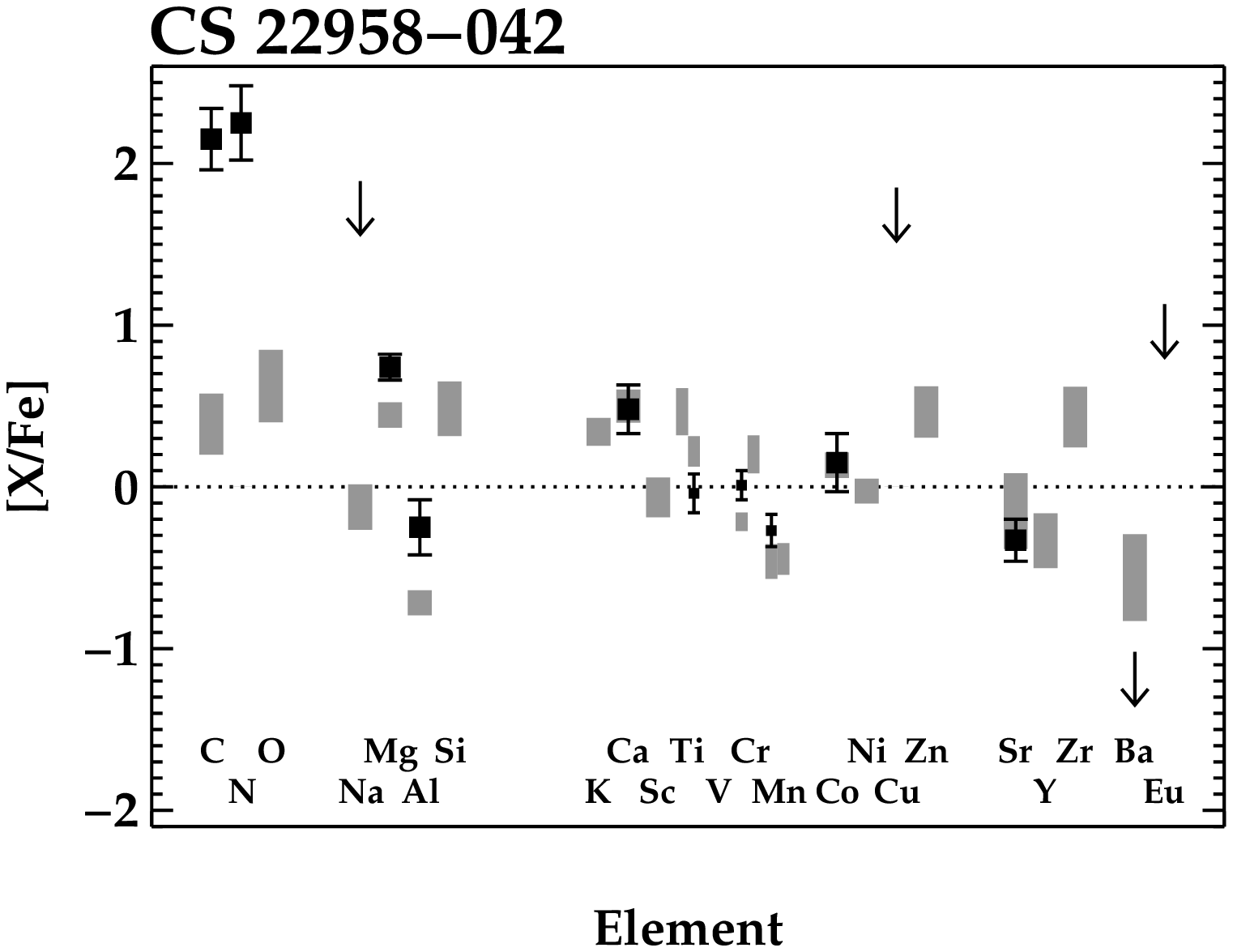} \\
\vspace*{0.3in}
\includegraphics[angle=0,width=3.35in]{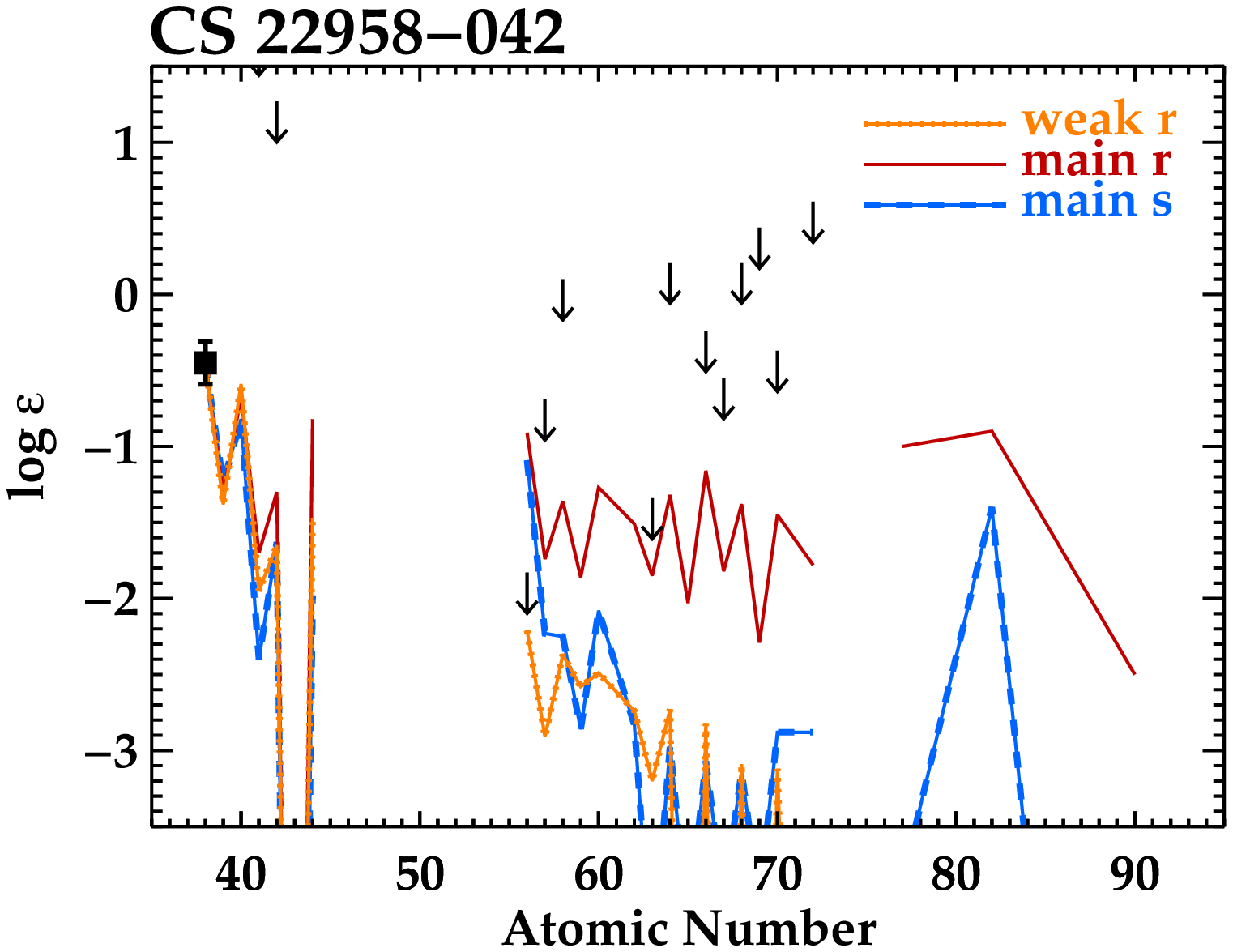}
\end{center}
\caption{
\label{cs22958m042fig}
\scriptsize
TOP:\ 
Comparison of abundances in 
\mbox{CS~22958--042}
with the average abundances of 
28 other stars with 
\teff\ within $\pm$~200~K and
[Fe/H] within $\pm$~0.3~dex of 
\mbox{CS~22958--042}.
BOTTOM:\
The heavy element distribution in 
\mbox{CS~22958--042}.
Each of the three curves has been renormalized
to the strontium abundance in
\mbox{CS~22958--042}.
Symbols in both panels
are the same as in Figure~\ref{bdp440493fig}.
}
\end{figure}

\begin{figure}
\begin{center}
\includegraphics[angle=0,width=3.35in]{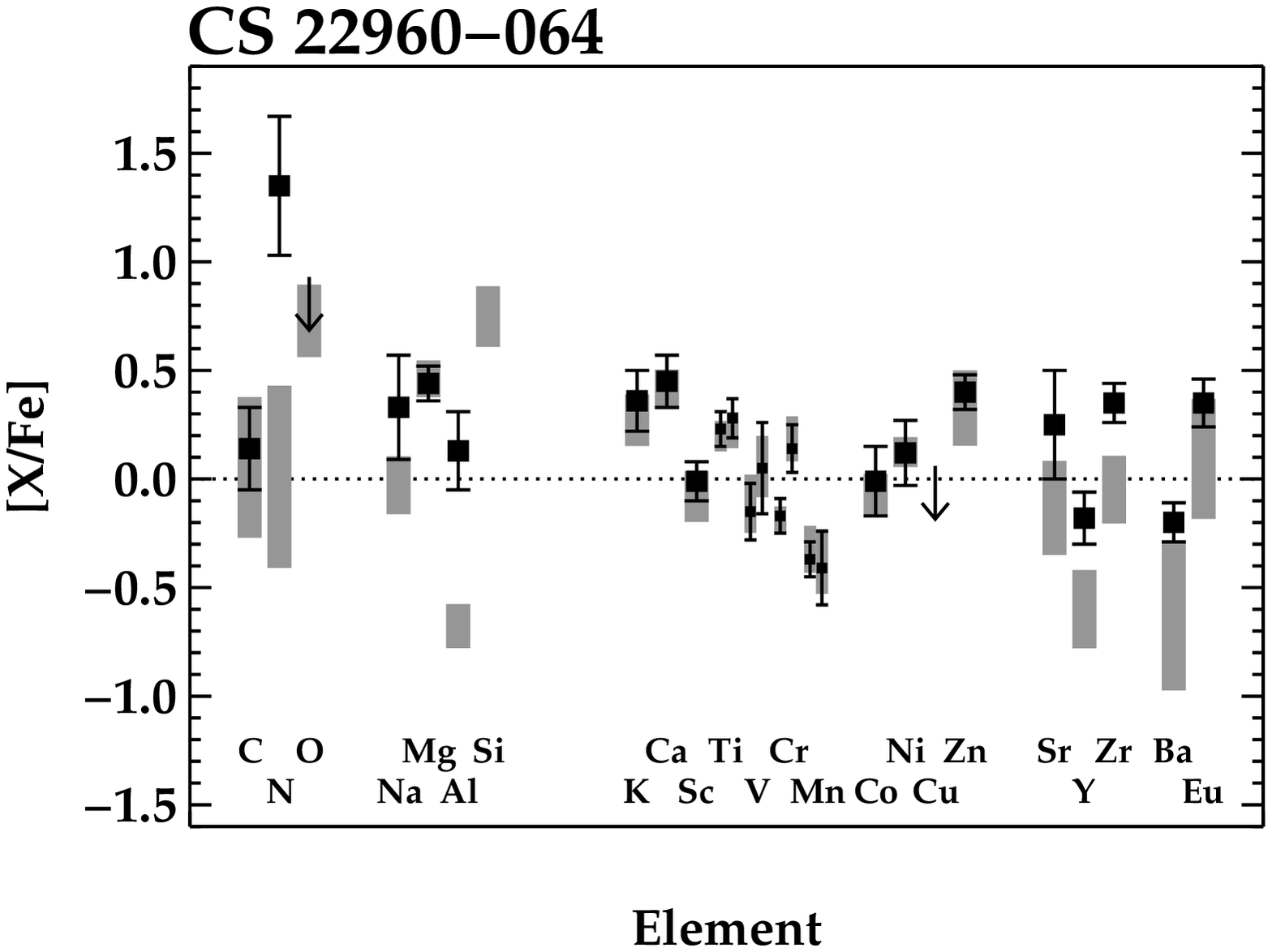} \\
\vspace*{0.3in}
\includegraphics[angle=0,width=3.35in]{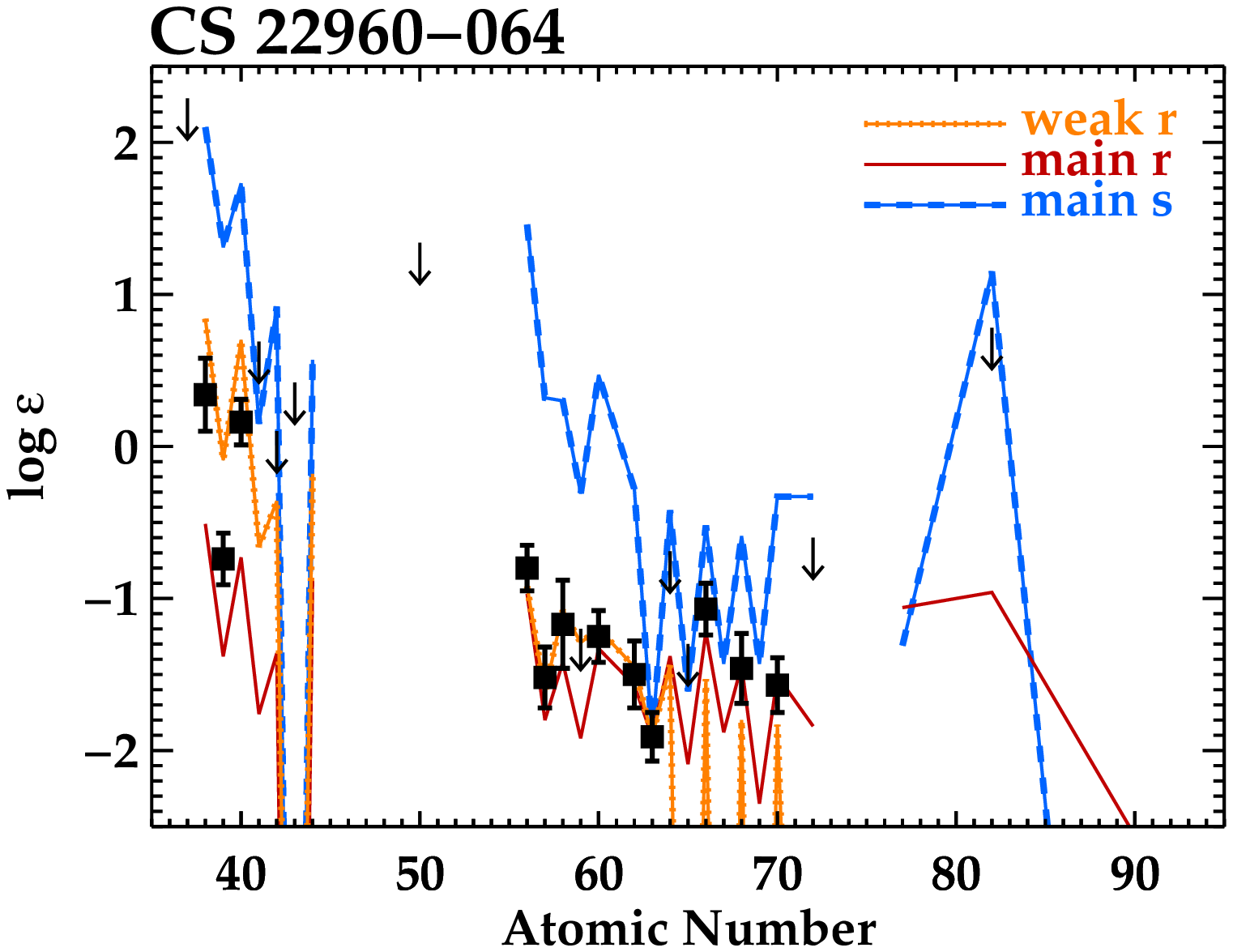}
\end{center}
\caption{
\label{cs22960m064fig}
\scriptsize
TOP:\ 
Comparison of abundances in 
\mbox{CS~22960--064}
with the average abundances of 
18 other stars with 
\teff\ within $\pm$~200~K and
[Fe/H] within $\pm$~0.3~dex of 
\mbox{CS~22960--064}.
BOTTOM:\
The heavy element distribution in 
\mbox{CS~22960--064}.
Each of the three curves has been renormalized
to the europium abundance in
\mbox{CS~22960--064}.
Symbols in both panels
are the same as in Figure~\ref{bdp440493fig}.
}
\end{figure}

\begin{figure}
\begin{center}
\includegraphics[angle=0,width=3.35in]{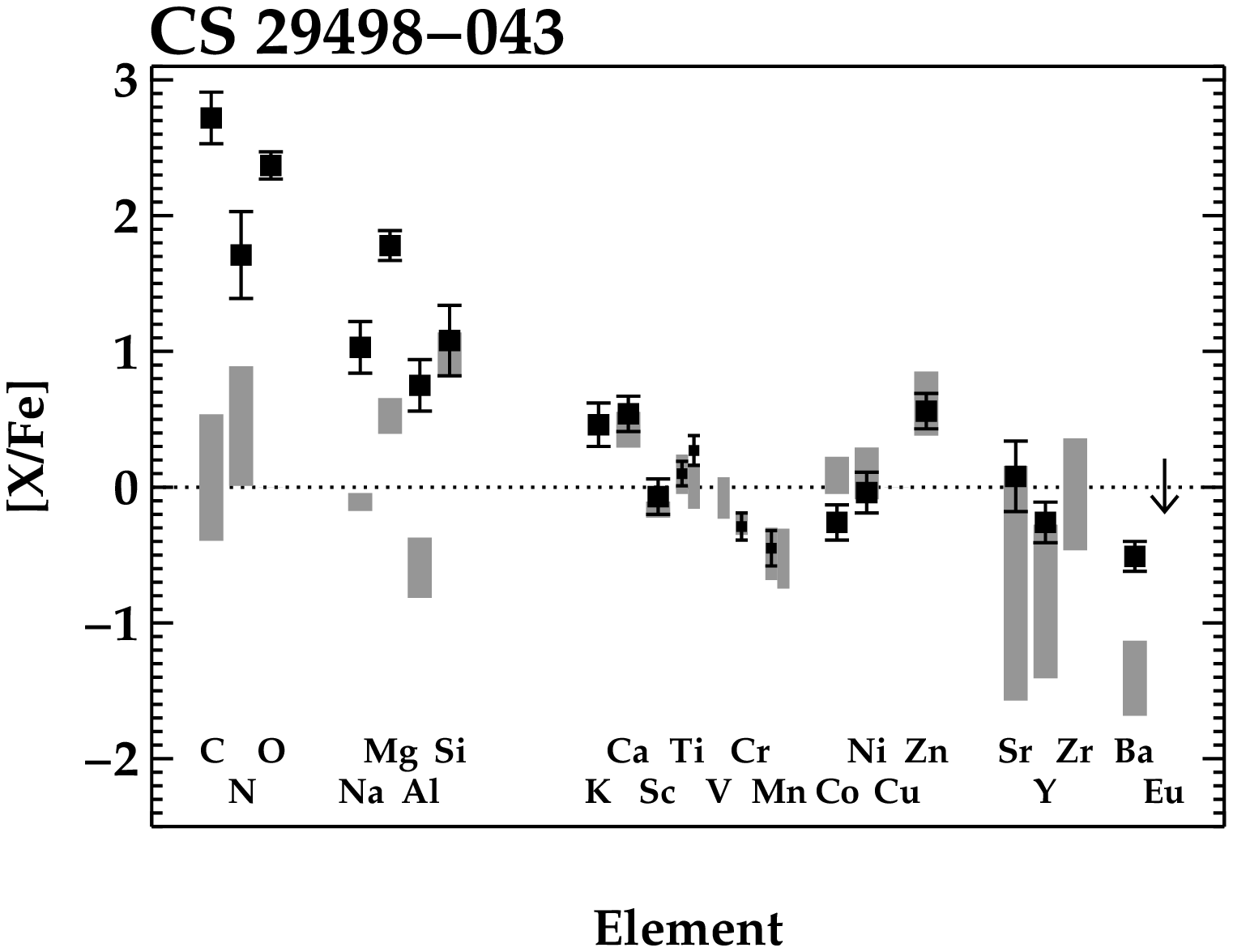} \\
\vspace*{0.3in}
\includegraphics[angle=0,width=3.35in]{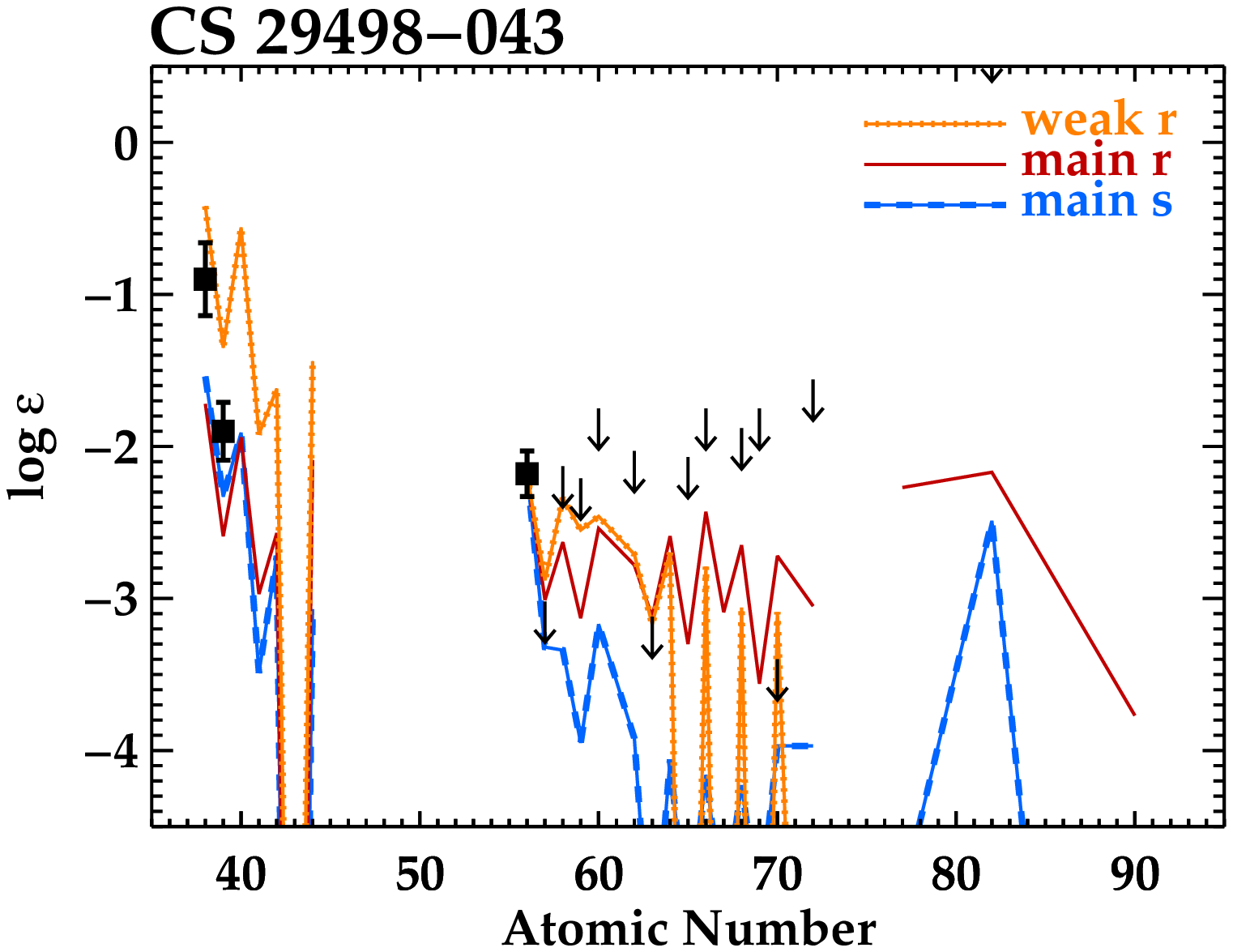}
\end{center}
\caption{
\label{cs29498m043fig}
\scriptsize
TOP:\ 
Comparison of abundances in 
\mbox{CS~29498--043}
with the average abundances of 
six other stars with 
\teff\ within $\pm$~200~K and
[Fe/H] within $\pm$~0.3~dex of 
\mbox{CS~29498--043}.
BOTTOM:\
The heavy element distribution in 
\mbox{CS~29498--043}.
Each of the three curves has been renormalized
to the barium abundance in
\mbox{CS~29498--043}.
Symbols in both panels
are the same as in Figure~\ref{bdp440493fig}.
}
\end{figure}

\begin{figure}
\begin{center}
\includegraphics[angle=0,width=3.35in]{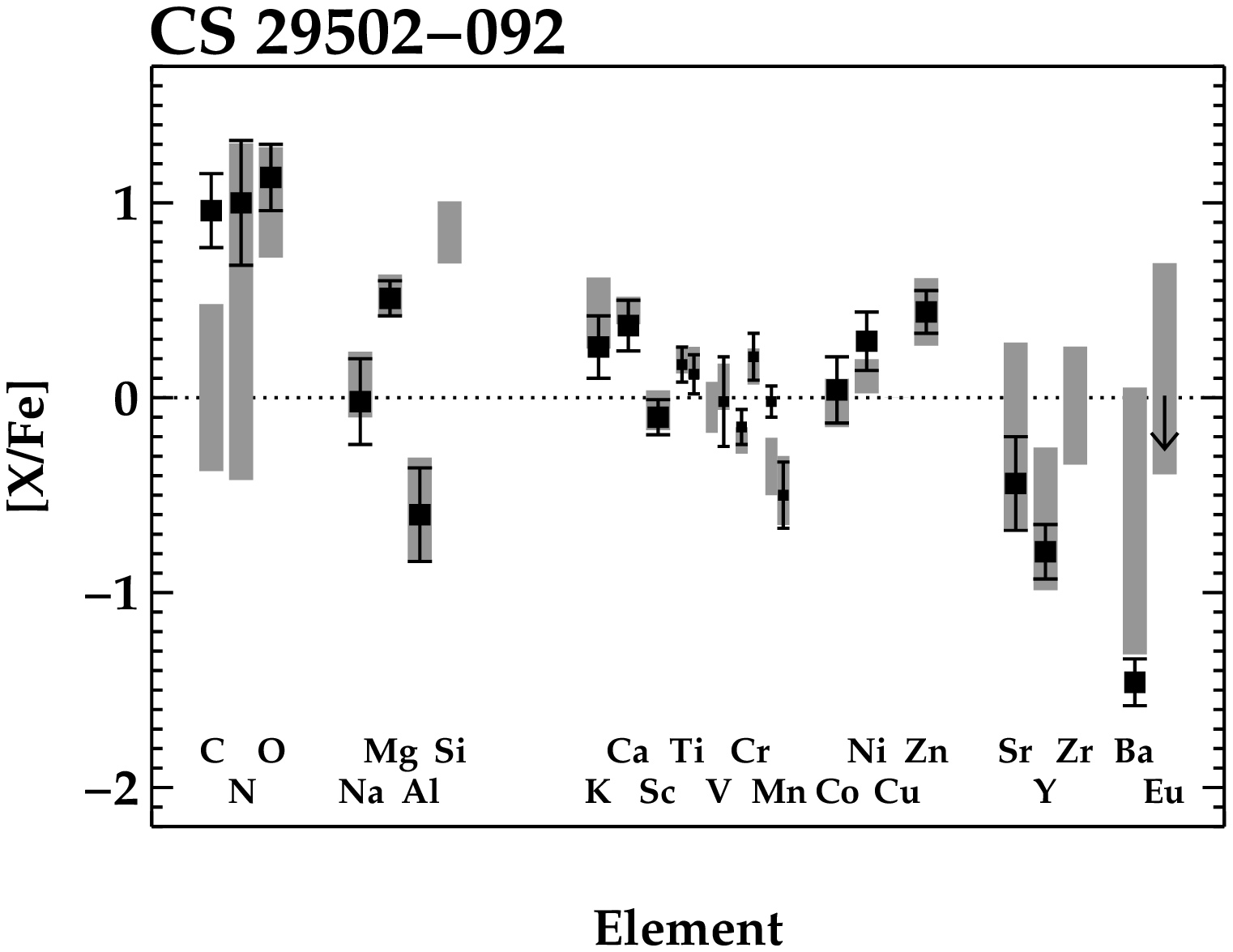} \\
\vspace*{0.3in}
\includegraphics[angle=0,width=3.35in]{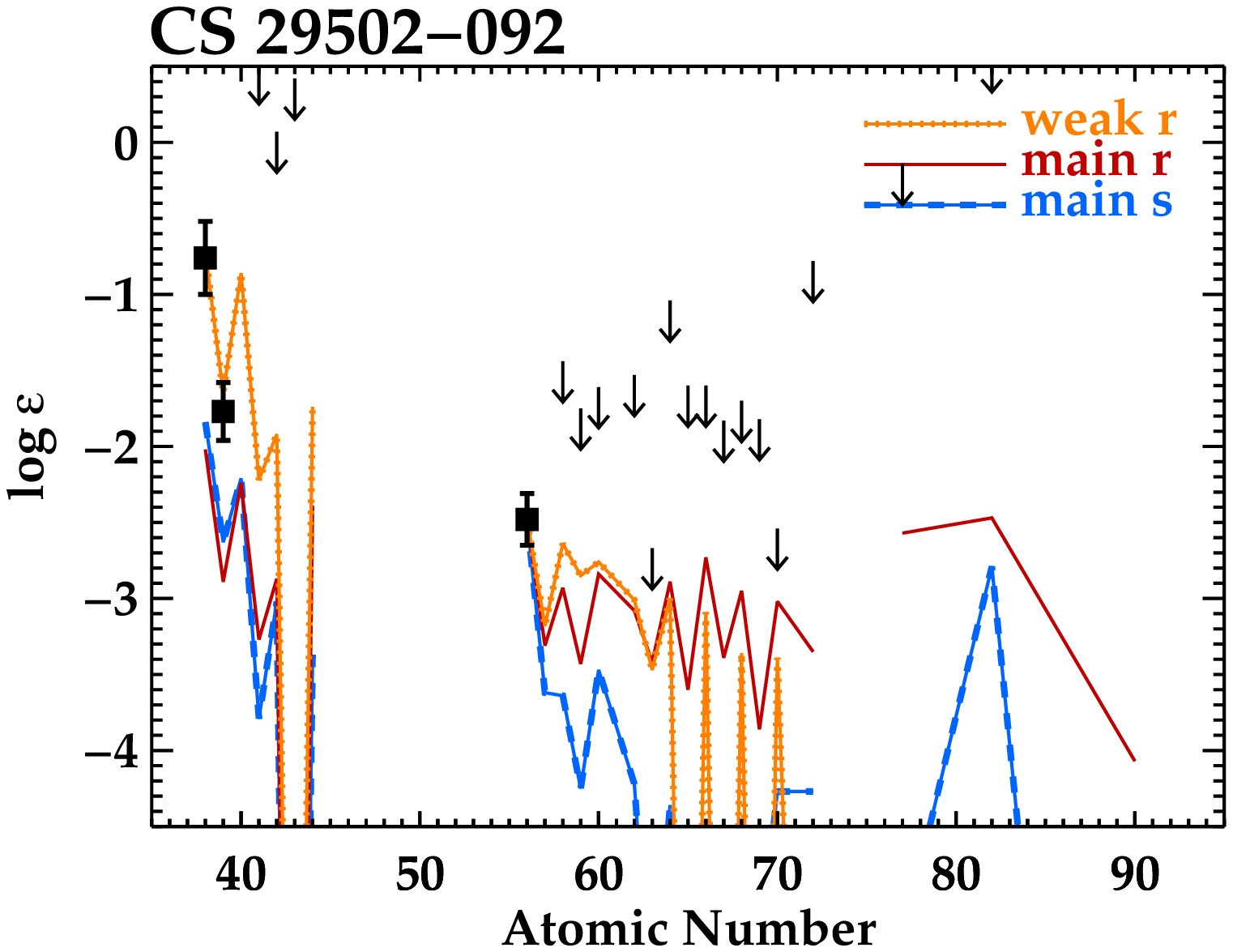}
\end{center}
\caption{
\label{cs29502m092fig}
\scriptsize
TOP:\ 
Comparison of abundances in 
\mbox{CS~29502--092}
with the average abundances of 
29 other stars with 
\teff\ within $\pm$~200~K and
[Fe/H] within $\pm$~0.3~dex of 
\mbox{CS~29502--092}.
BOTTOM:\
The heavy element distribution in 
\mbox{CS~29502--092}.
Each of the three curves has been renormalized
to the barium abundance in
\mbox{CS~29502--092}.
Symbols in both panels
are the same as in Figure~\ref{bdp440493fig}.
}
\end{figure}

\begin{figure}
\begin{center}
\includegraphics[angle=0,width=3.35in]{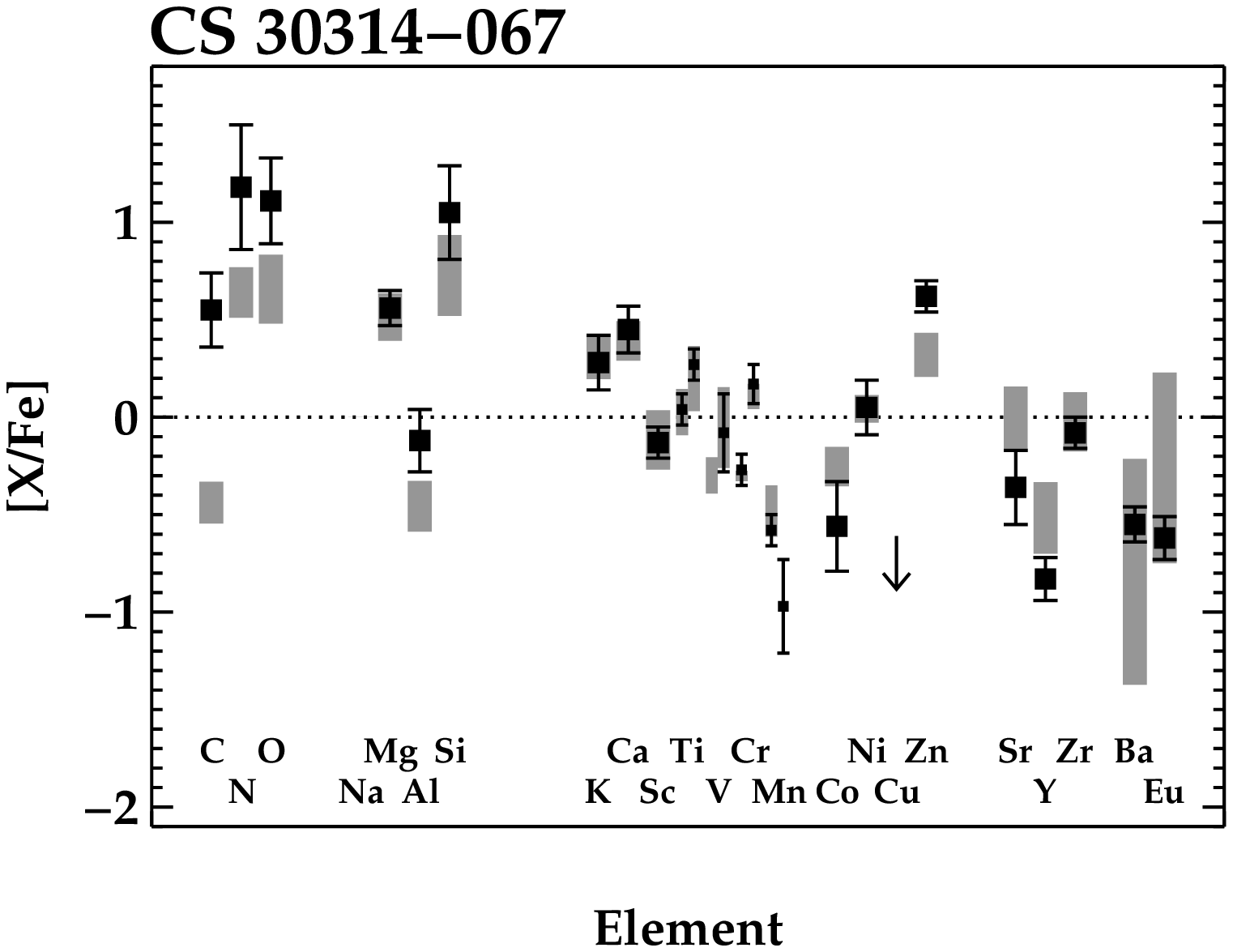} \\
\vspace*{0.3in}
\includegraphics[angle=0,width=3.35in]{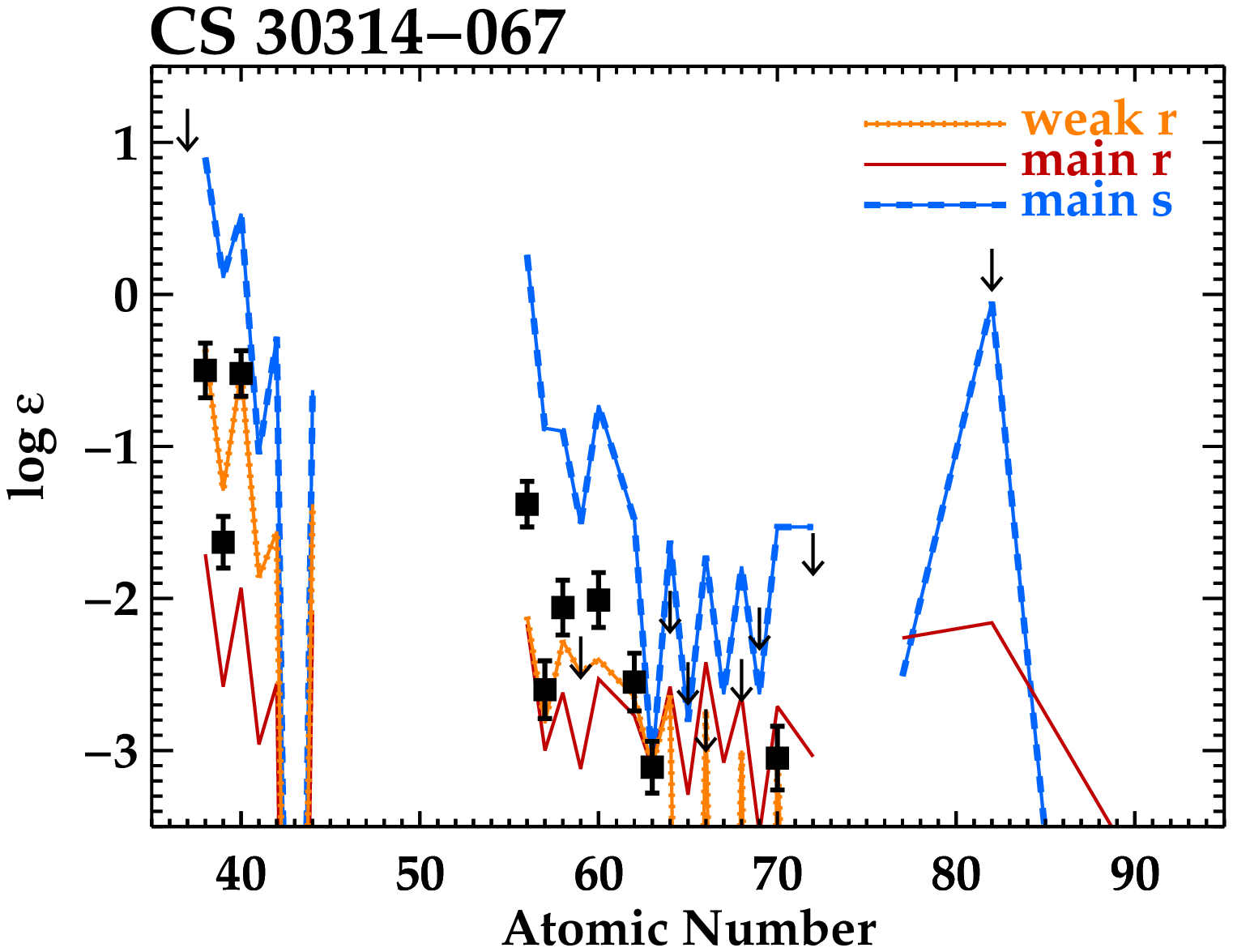}
\end{center}
\caption{
\label{cs30314m067fig}
\scriptsize
TOP:\ 
Comparison of abundances in 
\mbox{CS~30314--067}
with the average abundances of 
six other stars with 
\teff\ within $\pm$~200~K and
[Fe/H] within $\pm$~0.3~dex of 
\mbox{CS~30314--067}.
BOTTOM:\
The heavy element distribution in 
\mbox{CS~30314--067}.
Each of the three curves has been renormalized
to the europium abundance in
\mbox{CS~30314--067}.
Symbols in both panels
are the same as in Figure~\ref{bdp440493fig}.
}
\end{figure}

\begin{figure}
\begin{center}
\includegraphics[angle=0,width=3.35in]{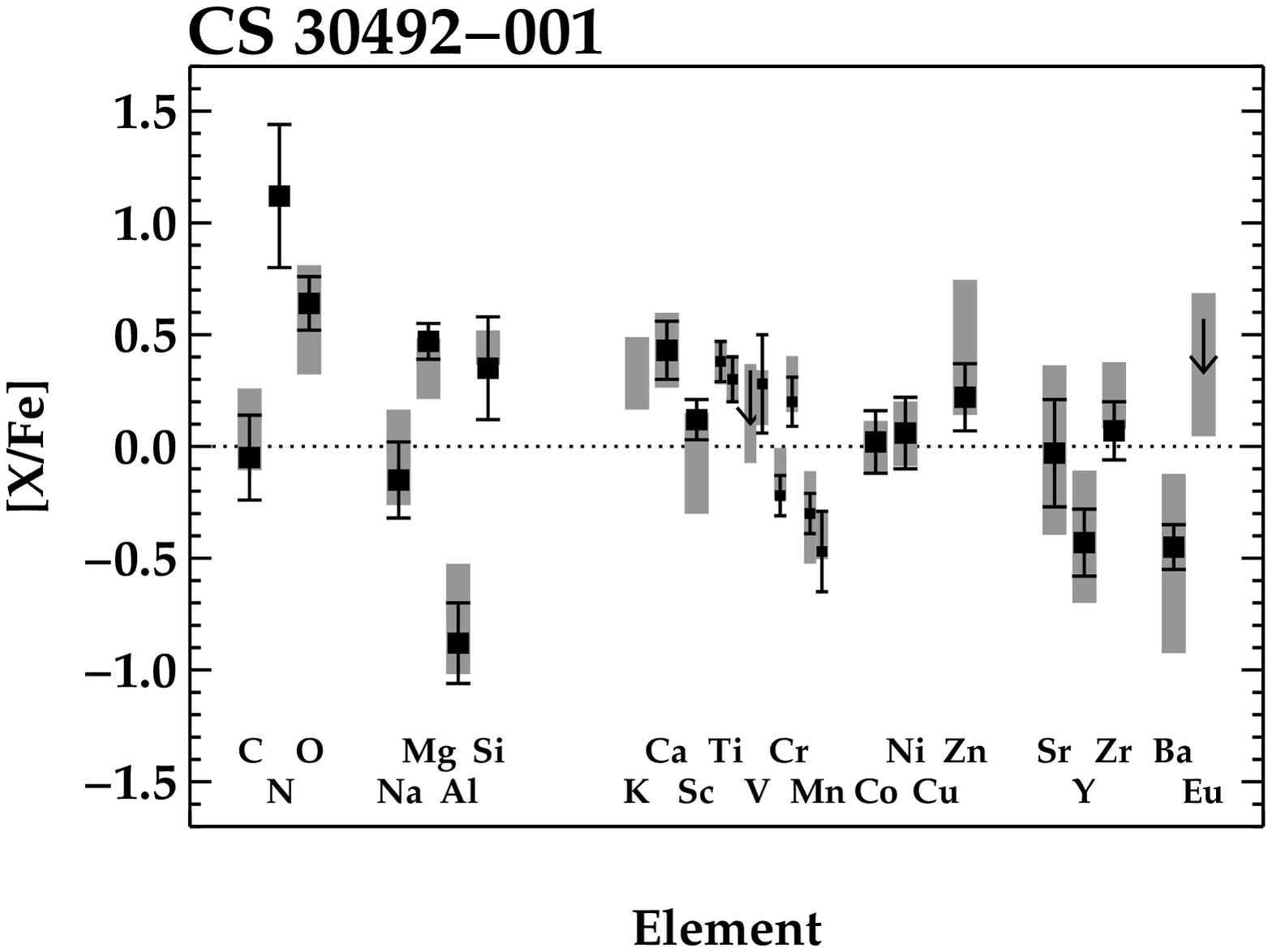} \\
\vspace*{0.3in}
\includegraphics[angle=0,width=3.35in]{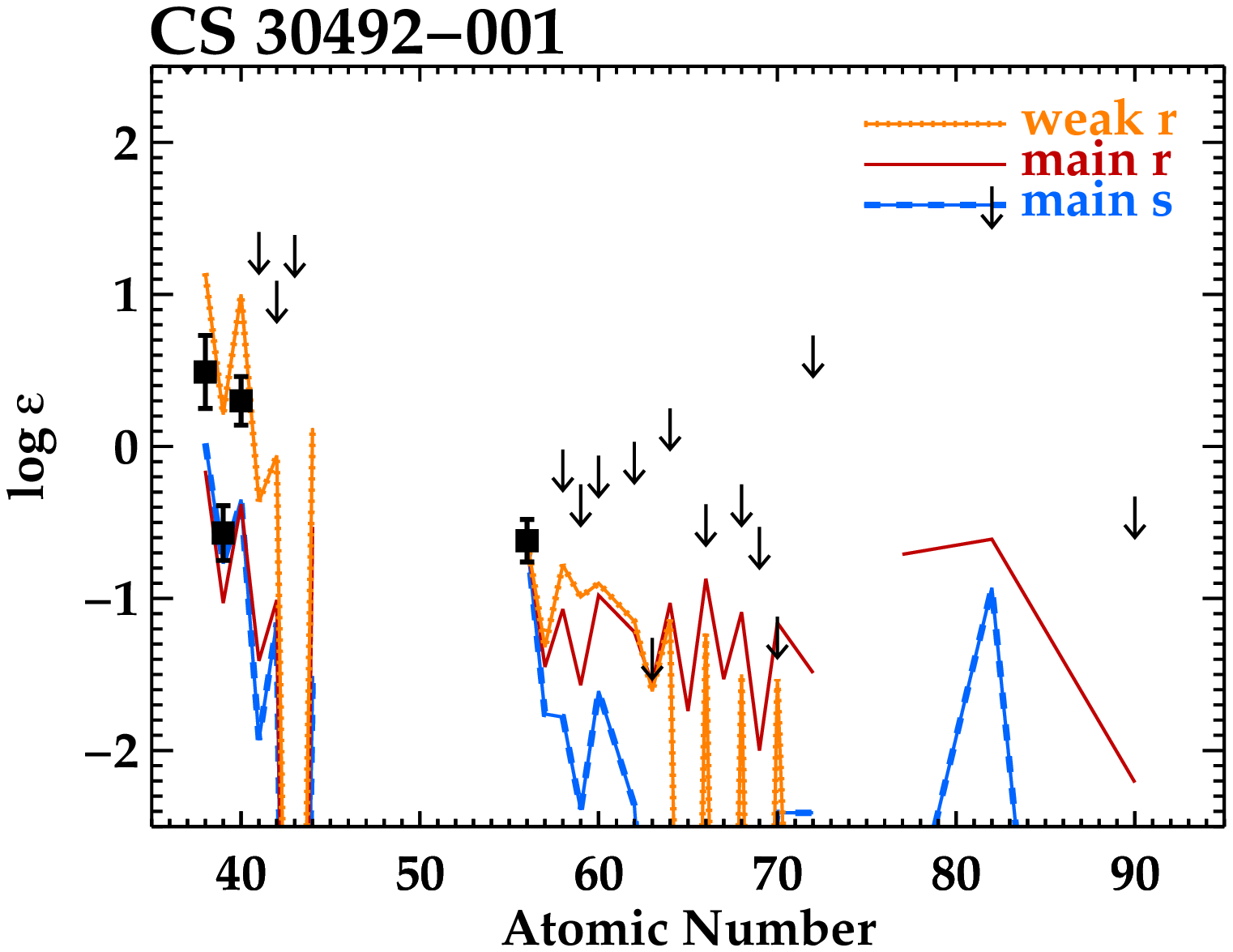}
\end{center}
\caption{
\label{cs30492m001fig}
\scriptsize
TOP:\ 
Comparison of abundances in 
\mbox{CS~30492--001}
with the average abundances of 
19 other stars with 
\teff\ within $\pm$~200~K and
[Fe/H] within $\pm$~0.3~dex of 
\mbox{CS~30492--001}.
BOTTOM:\
The heavy element distribution in 
\mbox{CS~30492--001}.
Each of the three curves has been renormalized
to the barium abundance in
\mbox{CS~30492--001}.
Symbols in both panels
are the same as in Figure~\ref{bdp440493fig}.
}
\end{figure}

\begin{figure}
\begin{center}
\includegraphics[angle=0,width=3.35in]{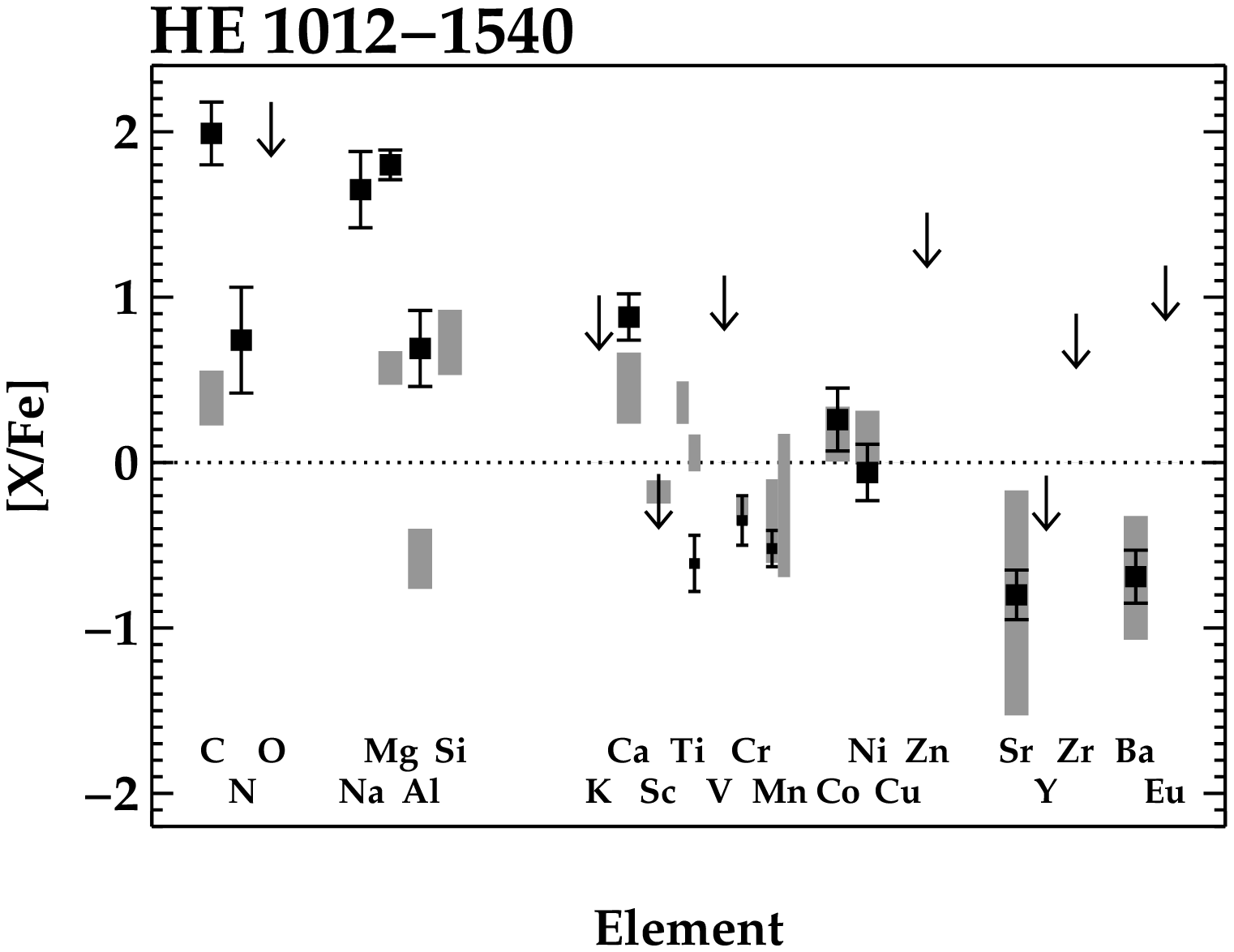} \\
\vspace*{0.3in}
\includegraphics[angle=0,width=3.35in]{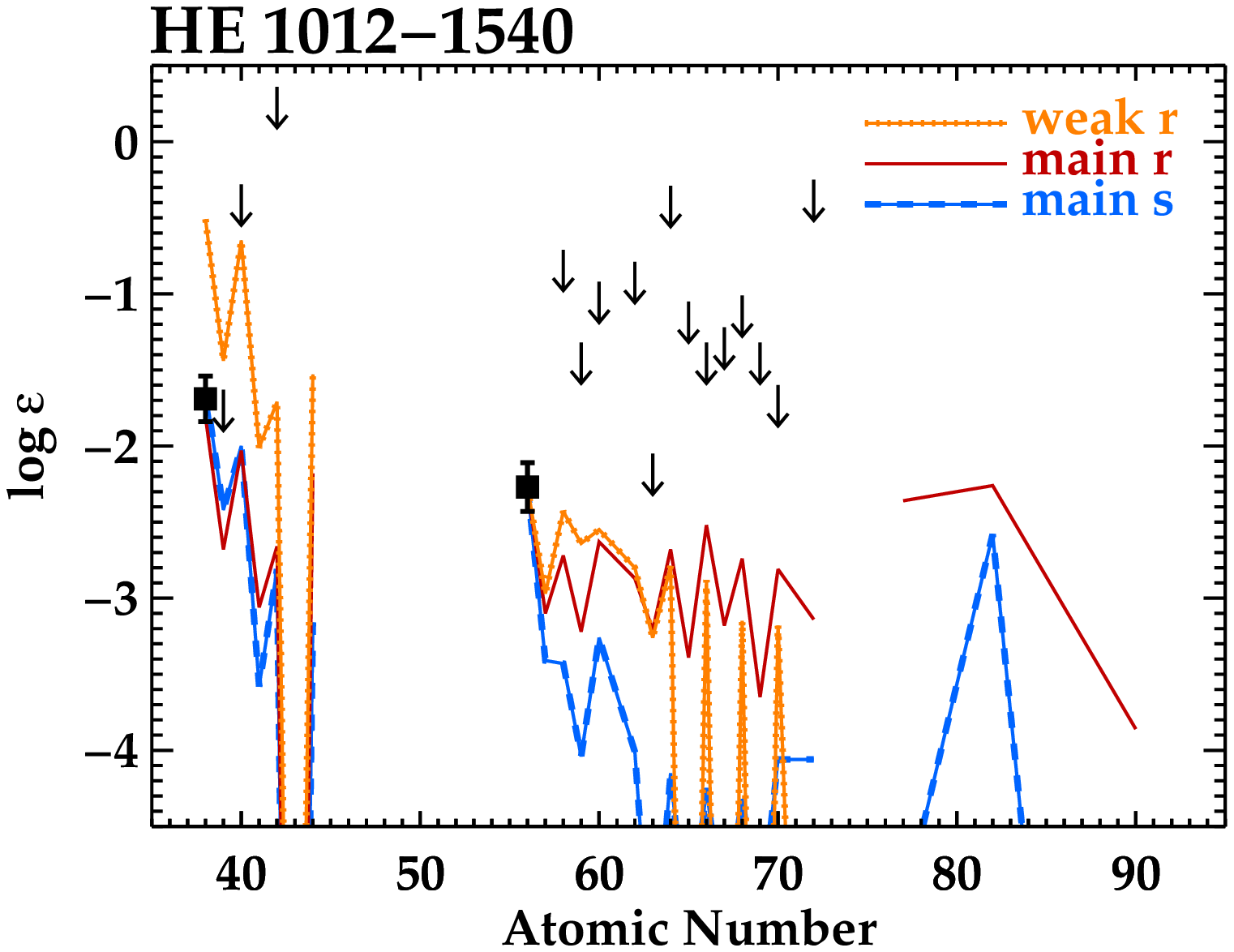}
\end{center}
\caption{
\label{he1012m1540fig}
\scriptsize
TOP:\ 
Comparison of abundances in 
\mbox{HE~1012$-$1540}
with the average abundances of 
five other stars with 
\teff\ within $\pm$~200~K and
[Fe/H] within $\pm$~0.5~dex of 
\mbox{HE~1012$-$1540}.
BOTTOM:\
The heavy element distribution in 
\mbox{HE~1012$-$1540}.
Each of the three curves has been renormalized
to the barium abundance in
\mbox{HE~1012$-$1540}.
Symbols in both panels
are the same as in Figure~\ref{bdp440493fig}.
}
\end{figure}

The top panels of 
Figures~\ref{bdp440493fig} through \ref{he1012m1540fig}
illustrate this comparison for all 16~stars in our sample.
The number of comparison stars and the range in \teff\ and [Fe/H]
considered are given in each figure caption.
For each light element X in each CEMP-no or NEMP-no star,
the [X/Fe] ratio is compared to the mean and standard deviation
of the [X/Fe] ratios for stars in the comparison sample.
Multiple ionization states of the same element 
are indicated separately.
Only detections are considered in the comparison sample,
thus the means may be overestimated 
for a few elements (e.g, [N/Fe]).

These comparisons reveal that 
the [X/Fe] ratios for all elements from potassium
through zinc (19~$\leq Z \leq$~30)
in each CEMP-no or NEMP-no star 
do not differ by more than $\approx$~2~$\sigma$ 
from the comparison samples.
Most are alike at the 1~$\sigma$ level.
The few $\approx$~2~$\sigma$ differences do not
show any consistent patterns from one
star to another, and within each star they do not
occur for elements adjacent in atomic number.
Thus we conclude that the elements
from potassium to zinc are produced
in similar proportions
by the progenitors responsible for enriching the
CEMP-no or NEMP-no stars and the comparison samples of halo stars.
The [X/Fe] ratios for the iron-group elements in the CEMP-no 
and NEMP-no stars
and comparison halo stars
generally agree within factors of $\approx$~2--3
with the iron group elements in
\object[HE 0107-5240]{HE~0107$-$5240},
\object[HE 1327-2326]{HE~1327$-$2326}, and
\object[HE 0557-4840]{HE~0557$-$4840}
\citep{christlieb04,aoki06,norris07}.

For the lighter elements oxygen, 
sodium, magnesium, aluminum, and silicon,
at least one of these elements is $>$~2~$\sigma$
higher in eight of the CEMP-no or NEMP-no 
stars than in the comparison samples.
Three stars show more that one of these elements
high by $>$~2~$\sigma$, and three of these elements
are high by $>$~2~$\sigma$~ in two of these three stars.
Enhanced levels of oxygen through silicon
sometimes accompany the carbon and nitrogen enhancement.
Previous studies have identified 
the similarity of potassium through zinc and the
enhancement of oxygen through silicon in the CEMP-no
and NEMP-no
stars relative to carbon-normal stars.
It is reassuring that these effects are
confirmed by our analysis.

\begin{figure}
\includegraphics[angle=90,width=3.35in]{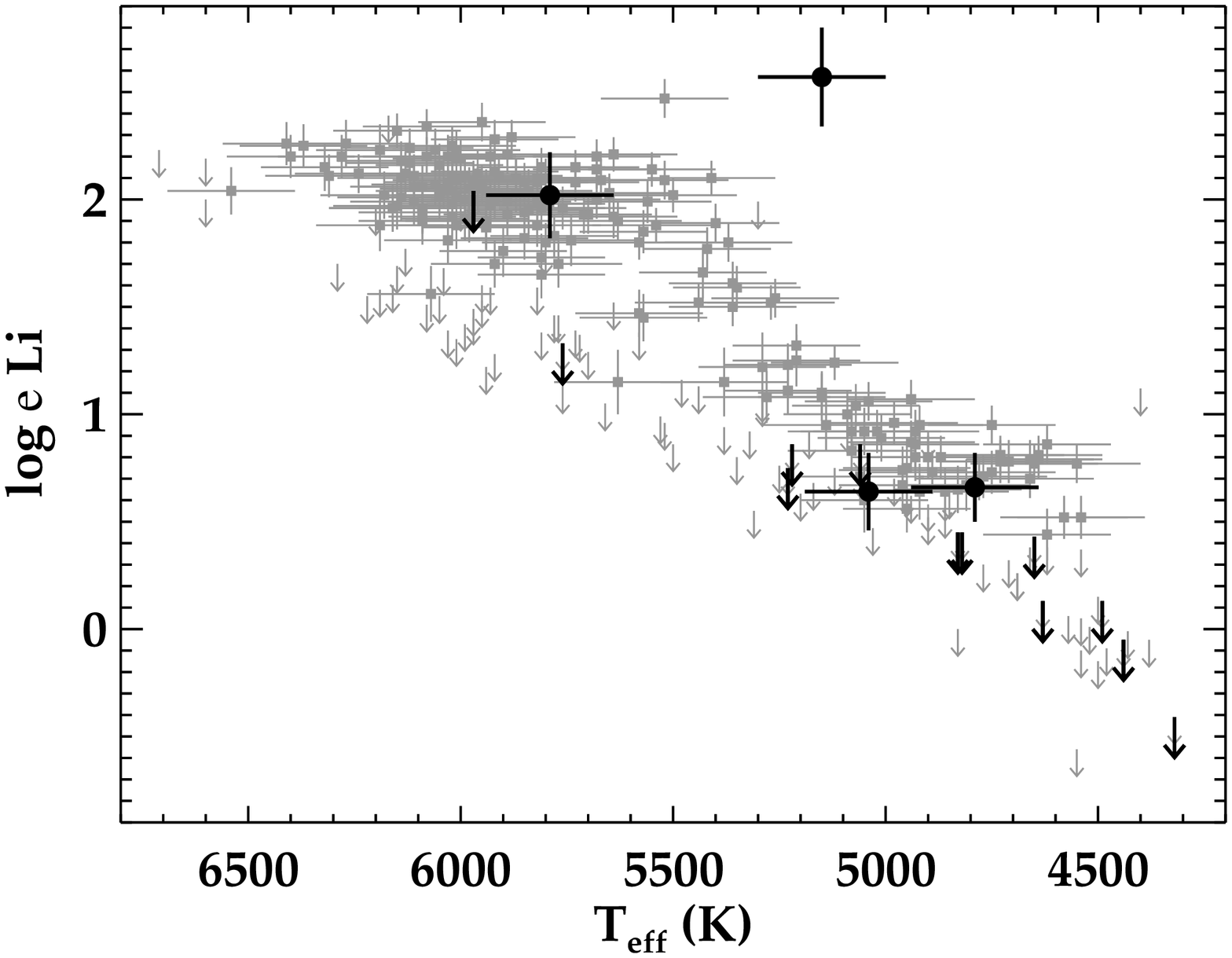}
\caption{
\label{lifig}
Log~$\epsilon$ lithium abundances as a function of \teff.
Large black symbols mark the 16~stars in our sample.
Small gray symbols mark stars from the main sequence turnoff 
to the tip of the red giant branch from the 
full sample of \citet{roederer14}.
Filled symbols represent detections, and 
arrows indicate upper limits.
}
\end{figure}

Figure~\ref{lifig} illustrates the lithium abundances of the CEMP-no 
and NEMP-no stars
as a function of \teff.
Lithium abundances for
the full sample of halo stars analyzed by \citet{roederer14} are
shown for comparison.
Lithium is detected in 4 of the 16~stars.
In 15 of the 16~stars, these abundances or upper limits
fall within the range of lithium abundances or upper limits
of the halo star sample.
In one~star, \csb\ (\teff~$=$~5150~K), 
lithium is significantly enhanced,
\logeps~(Li)~$= +$2.57~$\pm$~0.23.
For stars in a similar state of evolution on the red giant branch,
the comparison sample shows
\logeps~(Li)~$\approx +$1.0.
Nitrogen and sodium are significantly enhanced in \csb,
[N/Fe]~$= +$1.55~$\pm$~0.36 and
[Na/Fe]~$= +$1.58~$\pm$~0.17.
No other element ratios in \csb\ are different from the comparison star
sample, as shown in Figure~\ref{cs22893m010fig}.
Such enhanced [N/Fe] or [Na/Fe] ratios are not a universal
feature of lithium-enhanced field giants
(e.g., \citealt{lambert84}, \citealt{ruchti11b}).
A few such stars are found among the metal-poor red giants,
but the [N/Fe] or [Na/Fe] enhancements are probably 
attributable to enrichment from an evolved companion
(e.g., \citealt{roederer08}, \citeauthor{ruchti11b}).
This does not appear to be the case with \csb.

Previous studies of lithium in (first ascent) red giant stars
have typically shown that no more than $\approx$~1\%
of these stars presently exhibit
lithium enhancement relative to their peers
\citep{brown89,pilachowski00,ruchti11b,kirby12,lebzelter12,martell13},
although the mechanism that produces the lithium enhancement is
not known with certainty.
There are 98 red giant stars in the 
\citeauthor{roederer14}\ sample,
and our identification of one lithium-enhanced red giant
is consistent with this frequency.
If the lithium enhancement is not related to the 
nitrogen and sodium enhancement in \csb,
we may conclude
that nitrogen- (and carbon-)enhanced
stars are also capable of
going through a lithium-enhanced phase.

\subsection{Elements Beyond the Iron Group}
\label{heavyelements}

The top panels of 
Figures~\ref{bdp440493fig} through \ref{he1012m1540fig}
illustrate the [Sr/Fe], [Y/Fe],
[Zr/Fe], [Ba/Fe], and [Eu/Fe] ratios
for all 16~stars in our sample with the comparison stars
from the large sample of \citet{roederer14}.
The CEMP-no and NEMP-no stars do not appear unusual 
with respect to the carbon- and nitrogen-normal 
stars in this regard.

The bottom panels of 
Figures~\ref{bdp440493fig} through \ref{he1012m1540fig}
illustrate the heavy element abundance patterns
in these 16~stars.
Three standard abundance patterns are shown for comparison.
One pattern traces the heavy element abundances
in the metal-poor halo star \object[HD 122563]{HD~122563},
which has normal abundances of the lighter 
neutron-capture elements and a deficiency
of the heaviest neutron-capture elements.
The pattern found in this star may be
considered representative of the 
weak component of the rapid neutron-capture process (the \rpro).
Another pattern traces the heavy element abundances
in the \rpro\ rich metal-poor halo star 
\object[BPS CS 22892-052]{CS~22892--052},
which is a well-known representative of 
a group of stars enriched by the
main component of the \rpro.
The third pattern traces one outcome of \spro\ nucleosynthesis.
This is derived from models of \spro\ nucleosynthesis
in TP-AGB stars that are fit to the isotopic solar system 
\spro\ abundance pattern.
In each figure, these patterns are rescaled to 
europium (if detected), or barium (if detected), or strontium.

We emphasize that these three curves are not intended to be
rigid representations of distinct nucleosynthetic processes.
Stars passing through the TP-AGB phase of evolution will produce different
\spro\ abundance patterns that depend on the stellar
mass, metallicity, mass loss rate, availability of 
neutrons from the $^{13}$C($\alpha$,$n$)$^{16}$O reaction,
and so on.
The abundance patterns produced by the
weak and main components of the \rpro\ may
represent two extreme outcomes
that result from variations in the physical conditions
at the time of nucleosynthesis.
While the general trends of these curves may 
help identify the nucleosynthetic processes responsible for
creating the heavy elements
in the CEMP-no and NEMP-no stars,
we refrain from drawing conclusions from more detailed comparisons.

In five stars, sufficient numbers of key
elements in the rare earth element domain\footnote{
The rare earth domain formally spans
lanthanum through lutetium (57~$\leq Z \leq$~71) and includes
scandium ($Z =$~21) and yttrium ($Z =$~39).
For our purposes scandium and yttrium are irrelevant,
but we extend the lanthanide range to include 
barium ($Z =$~56) and hafnium ($Z =$~72).
Our working definition thus encompasses
56~$\leq Z \leq$~72.} 
are detected, enabling us to 
assign a probable nucleosynthetic origin to their heavy elements.
The pattern in 
\csa\ (Figure~\ref{cs22891m200fig})
resembles the main component of the \rpro\
as exemplified by \object[BPS CS 22892-0520]{CS~22892--052}.
\csh, \csm, and \csg\ 
(Figures~\ref{cs22877m001fig}, 
\ref{cs22960m064fig}, and \ref{cs30314m067fig})
show patterns that resemble the weak component of the \rpro\
as exemplified by \object[HD 122563]{HD~122563}.
The heavy element abundance pattern in \csi\ shows 
evidence for at least a partial \spro\ origin.
As shown in Figure~\ref{cs22878m101fig}, 
the upper limit on europium in \csi\ is strong enough
to rule out an exclusive \rpro\ origin for the heavy elements,
though a mix of $r$- and \spro\ material is possible.
This star also exhibits a low level of radial velocity variations,
suggesting that it may be in a binary system,
and the unobserved companion could have passed through
the TP-AGB phase of evolution and transferred a 
small, yet detectable, amount of \spro\ material to \csi.
% (For reference, [Sr/Fe]~$= -$0.35, 
% [Ba/Fe]~$= -$0.60, and [Eu/Ba]~$< +$0.33.)

\begin{figure}
\includegraphics[angle=0,width=3.35in]{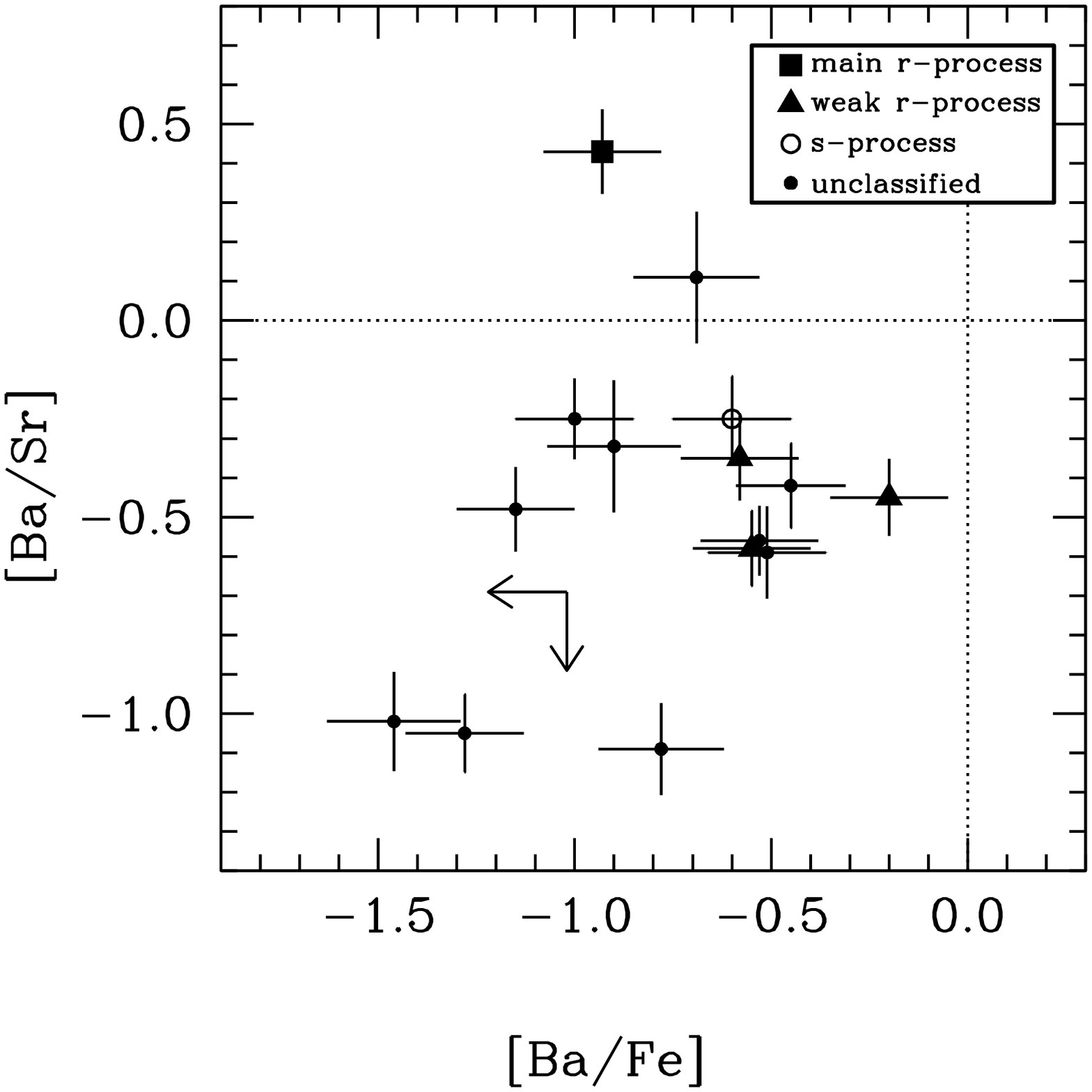}
\caption{
\label{basrfig}
[Ba/Sr] ratios as a function of [Ba/Fe] for our sample.
The large square indicates one star likely to have been enriched
with products of the main component of the \rpro\ 
(\mbox{CS~22891--200}),
the large triangles indicate stars likely to have been enriched
with products of the weak component of the \rpro\
(\mbox{CS~22877--001}, \mbox{CS~22960--064}, 
\mbox{CS~30314--067}), 
the open circle indicates one star likely to have been 
partly enriched with products of the \spro\
(\mbox{CS~22878--101}), and
the small filled circles indicate all other stars in our sample.
The dotted lines mark the solar ratios.
}
\end{figure}

Figure~\ref{basrfig} illustrates the 
[Ba/Sr] ratio as a function of [Ba/Fe] for all 16~stars in our sample.
The five stars whose neutron-capture patterns we can reliably assess
are highlighted.
The abscissa in Figure~\ref{basrfig}, 
[Ba/Fe], may be thought of as a dilution axis,
reflecting the dilution of material produced by 
neutron-capture nucleosynthesis into differing amounts of iron.
The ordinate in Figure~\ref{basrfig}, 
[Ba/Sr], may be thought of as reflecting
properties intrinsic to the neutron-capture process itself.

\csa\ is one of two stars in our sample with 
[Ba/Sr]~$>$~0, and the pattern revealed 
by its strontium, yttrium, barium, and europium abundances
is similar to that observed in the \rpro\ rich standard star
\object[BPS CS 22892-052]{CS~22892--052}
(Figure~\ref{cs22891m200fig}).
[Ba/Sr] is also super-solar in \he.
The [Ba/Sr] ratios in these two stars are similar to that
found by \citet{sneden08} for 16 metal-poor stars
with high levels of \rpro\ enrichment, 
[Ba/Sr]~$= +$0.3~$\pm$~0.2 (see Figure~7 there).
While carbon-enhanced stars with high levels of \rpro\ 
enrichment like \object[BPS CS 22892-052]{CS~22892--052}
would be excluded from our sample
by the requirement that [Ba/Fe]~$<$~0, 
we may consider \csa\ and \he\ to be
related but enriched at a much lower level.
Per hydrogen atom, europium atoms are approximately 100~times 
less common
in \csa\ than \object[BPS CS 22892-052]{CS~22892--052},
yet the ratios among the detected neutron-capture
elements in these two stars are nearly identical.
Finally, we point out that 
\object[BPS CS 22892-052]{CS~22892--052}
is also carbon- and nitrogen-enhanced, 
[C/Fe]~$= +$0.88 and 
[N/Fe]~$= +$1.01 \citep{sneden03},
though there is no consensus in the literature
regarding the origin of its carbon, nitrogen, 
and \rpro\ enhancement.

The three stars with patterns reminiscent of the weak component 
of the \rpro\ show
[Ba/Sr]~$\approx -$0.4~$\pm$~0.2.  
%  This is not by construction!
%  Need at least something else beside Sr, Ba 
%  to identify this pattern initially.               
Six of the stars whose neutron-capture patterns are 
yet unclassified also fall within this range
(\bd, \csk, \csl, \csd, \csf, \cso).
The three stars we have classified are found
on the right side of the diagram, 
where the overall neutron-capture element abundances
are highest, thus affording us the best opportunity
to detect other elements in the rare earth domain.
It is therefore not surprising that the six
unclassified stars with [Ba/Sr]~$\approx -$0.4~$\pm$~0.2
have relatively low [Ba/Fe] ratios.
The remaining four stars in Figure~\ref{basrfig} 
show the lowest levels of [Ba/Sr] in our sample, 
[Ba/Sr]~$\approx -$1.0
(\csb, \csc, \cse, and \csn; 
note that \cse\ shows an upper limit on barium 
that suggests membership in this group).
It is possible that the 13~stars with [Ba/Sr]~$<$~0
(except \csi)
exhibit the results of a range of \rpro\
nucleosynthetic conditions.
We now consider alternative explanations.

In low-metallicity stars passing through the TP-AGB phase of evolution,
\spro\ nucleosynthesis
tends to produce [Ba/Fe]~$>$~0 and [Ba/Sr]~$>$~0.
By construction,
none of the stars in our sample
fall into this regime.
Some $^{12}$C/$^{13}$C ratios of stars in our sample
are low, but they are not uniformly as low
as predicted by CN cycle equilibrium, 
$^{12}$C/$^{13}$C~$\approx$~3--4.
With the exception of \csi,
this indicates that pollution by material produced in
\spro\ nucleosynthesis and
accompanied by CN-cycled material from an AGB star
is not a likely origin of the
strontium and barium in our sample.

Lead (Pb, $Z =$~82) enjoys special status as the only 
readily observable element at the terminus of the
\spro\ nucleosynthesis path.
Enhanced lead is a clear signature of
\spro\ nucleosynthesis in low-metallicity environments,
where the high neutron-to-seed ratio drives 
the \spro\ flow to lead (e.g., \citealt{gallino98}).
We do not detect lead in any star in our sample,
and most of the upper limits are uninteresting, even in \csi.
No upper limit approaches within an order of magnitude of
the estimated minimum [Pb/Eu] ratio 
encountered in the case of \spro\ contributions
([Pb/Eu]~$\approx +$0.3; see Figure~2 of \citealt{roederer10}).
\citet{ito13} report a stronger upper limit
on lead in \bd, but even this does not thoroughly 
exclude all traces of \spro\ material 
produced in a low-metallicity environment.
Unfortunately the observational limits on lead in these 
stars allow us only to say that prodigious lead production
did not occur in the progenitors.

These considerations form our third main observational result:\
\textit{some form of \rpro\ nucleosynthesis is responsible for 
the abundance patterns in four of the stars in our sample.}
The limited neutron-capture element abundance patterns in all but one
of the remaining stars are consistent with an \rpro\ origin,
but the observations cannot offer compelling evidence
against an \spro\ origin, either.
The heavy elements in one star, \csi, suggest
that material produced by \spro\ nucleosynthesis
may be present.
This result is not surprising, and it has long been 
suspected (e.g., \citealt{truran81}) that \rpro\ nucleosynthesis
dominated the production of elements heavier than iron
in the early Galaxy.
\citeauthor{tumlinson06}'s (2006) chemical evolution model
predicts that the average field star with [Fe/H]~$= -$3 
and normal abundance ratios has $\sim$~10 
(zero-metallicity) progenitors,
ranging anywhere from 1 to $\approx$~20.
While this prediction is difficult to test observationally,
the unusual ratios of oxygen through silicon in the CEMP-no and
NEMP-no stars
hint that the number of progenitors should be on the low end of this range.
Yet even in these stars elements heavier than the iron group are found,
and in some cases they appear to have been produced by \rpro\ 
nucleosynthesis.

\section{Discussion}
\label{discussion}

Elements produced by neutron-capture reactions
are found in all stars in our sample.
These heavy elements could not have been manufactured
in situ by the low-mass stars where they are found today.
At least some of these heavy elements appear to have 
been produced via \rpro\ nucleosynthesis.
Our sample is not constructed to represent an 
unbiased sample of stars,
so distributions of, e.g., [Ba/Sr] ratios
should not be used as diagnostic tools
to attempt to discern the origin of these heavy elements.

Limited samples of stars in
a few Local Group ultra-faint dwarf galaxies suggest these systems
may have been enriched by metals from a single supernova
\citep{simon10,frebel12}.
Previous work has shown that
strontium and barium are detected in nearly all
low-metallicity stars and at least some 
stars in all galaxies examined in sufficient detail
\citep{roederer10,roederer13a}.
In the \citet{roederer14} abundance survey of low-metallicity stars,
strontium and barium are detected\footnote{
\citet{roederer14} did not detect barium
in \object[BPS CS 22968-014]{CS~22968--014},
but \citet{francois07} did.
}
in all 107~stars 
with [Fe/H]~$< -$2.0 and
\teff~$<$~5400~K 
(i.e., the stars with the lowest continuous opacity and thus
strongest lines for a given composition).

Sub-solar [Ba/Sr] ratios are not
unique to \rpro\ nucleosynthesis predictions.
The presence of one star with at least a partial contribution
from \spro\ nucleosynthesis, \csi, 
serves as a reminder that some \spro\ signatures may offer 
an alternative, if perhaps less frequent, explanation.
Models of the \spro\ operating in low-metallicity
rapidly-rotating massive stars,
affectionately known as spinstars,
also predict a range of sub-solar [Ba/Sr] ratios
\citep{frischknecht12,cescutti13}.
The enhanced production of primary
$^{22}$Ne in the fast rotating models 
provides a neutron source that enables 
\spro\ nucleosynthesis.
These models require seed nuclei
from the iron group, 
so this non-standard \spro\ nucleosynthesis 
still operates as a secondary process (\citeauthor{frischknecht12}).
Spinstars and other sources of \spro\ material at 
extremely low metallicities may be considered possible
sources of the neutron-capture elements in our sample
only if their progenitors
were not zero-metallicity stars.

Spinstars can be excluded outright
in the four stars where the rare earth elements
reveal a clear \rpro\ origin.
If, however, models of these massive stars are
shown to host some form of \rpro\ nucleosynthesis
during the subsequent supernova explosions, 
this would revive their candidacy for
having enriched the gas from which the stars in
our sample formed.

Models of pair-instability supernovae predict 
no neutron-capture element production
(e.g. \citealt{heger02}).
If so, then the results presented here and the
lack of environments devoid of neutron-capture
elements suggest that
pair-instability supernovae were not 
frequent contributors to the metals in the earliest generations of stars.
We are not the first to point out this situation
(e.g., \citealt{umeda02}), but
our efforts to improve the observational data
on neutron-capture elements in low-metallicity stars
reaffirm earlier conclusions based on other observational signatures.

\citet{heger02} emphasize 
(in the context of pair-instability supernovae)
that the simplest explanation for the
presence of neutron-capture elements 
is an additional contribution from normal supernovae. % (p.\ 11).
We encourage investigators computing supernova yields 
to extend their reaction networks to include 
nuclei produced by neutron-capture nucleosynthesis.
This, of course, is challenging because of the
number of nuclei involved and the availability of
relevant nuclear data for radioactive nuclei far from stability.
At the very least, we recommend that such analyses
report whether the physical conditions present 
are capable of supporting neutron-capture nucleosynthesis.
These comparisons are essential to
determine whether a single supernova can account for
all metals observed in each star
or whether multiple supernovae are required.

\section{Summary}
\label{summary}

We have studied the heavy element abundance patterns
found in 16~stars with sub-solar neutron-capture element abundances
and enhanced in carbon or nitrogen
(specifically, 
[Ba/Fe]~$<$~0.0 and either
[C/Fe] or [N/Fe]~$> +$1.0).
These stars span a metallicity range 
from $-$4.3~$<$~[Fe/H]~$< -$2.3 with a median 
[Fe/H] of $-$3.2.
The abundance patterns of the lighter elements
suggest that
this sample could represent the
higher-metallicity analogs of the three known
CEMP stars with [Fe/H]~$< -$4.5.
High quality optical spectra collected with the
MIKE Spectrograph on the Magellan Telescopes and the
Tull Spectrograph on the Smith Telescope 
have allowed us to detect weak absorption lines 
and derive detailed abundance patterns of elements beyond the iron group.

Strontium is detected in all 16~stars, and
barium is detected in 15 of 16~stars.
These elements lie at (beyond) the first 
and second $s$- ($r$-)
process peaks, respectively,
indicating the operation of at least one
form of neutron-capture nucleosynthesis
in the progenitors that enriched the stars in our sample.
We also detect rare earth elements in five stars, 
and we use these abundance patterns
to characterize the nature of the
neutron-capture nucleosynthesis in each of those cases.
Some form of \rpro\ nucleosynthesis is responsible for
the abundance patterns in four of them,
and some \spro\ material may be present in another.
The [Ba/Sr] ratios in the remaining 11~stars
are similar to or lower than the [Ba/Sr] ratios
in these 5~stars.

These heavy elements could not have been manufactured
in situ by the low-mass stars where they are found today.
Rapidly-rotating massive stars (spinstars)
may be able to account for the heavy elements
in the stars without clear evidence
of \rpro\ nucleosynthesis only if 
their progenitors were not
zero-metallicity stars.
The presence of neutron-capture elements
in the CEMP-no and NEMP-no stars also suggests that 
pair instability supernovae were not 
frequent contributors to the metals in the earliest generations of stars.

Observations do not indicate that all first generation stars must have
produced large amounts of carbon and oxygen in their ejecta, but perhaps only
those that did enabled low-mass stars that are poor in iron to form.
Our conclusions regarding the possibility of
neutron-capture production are only applicable, of course,
to that set of progenitors.
Low-mass stars formed via other cooling modes, 
like thermal emission by dust grains, 
provide a glimpse into alternate enrichment environments.
The neutron-capture abundance patterns 
discussed here are not
unique to the CEMP-no and NEMP-no classes of stars.
Even if the stars considered by us
were not formed from the yields of only one
supernova each,
the nearly ubiquitous presence of strontium and barium
in these and other low-metallicity stars 
is a tantalizing hint
that at least one neutron-capture process may have 
operated frequently in the earliest stellar generations.

If enhanced levels of carbon through silicon (relative to iron)
are signatures of nucleosynthesis
in prior generations of zero-metallicity stars,
and if no other prior generations of
stars contributed to the metals,
then our results indicate that zero-metallicity stars
were also responsible for production of elements beyond the iron group.
The early onset of \rpro\ nucleosynthesis has 
long been established (e.g., \citealt{truran81}).
Our results offer new evidence
in support of \citeauthor{truran81}'s assertion that
``a single prior generation of stars can have 
been responsible for the abundances observed in the
most metal-deficient stars in our galaxy'' (p.\ 393).

\acknowledgments

I.U.R.\ thanks T.\ Beers for helpful discussions 
throughout the course of this study and
M.\ Spite for reaffirming europium upper limits in her data.
We thank the referee for a thoughtful and helpful report.
This research has made use of NASA's 
Astrophysics Data System Bibliographic Services, 
the arXiv preprint server operated by Cornell University, 
the SIMBAD and VizieR databases hosted by the
Strasbourg Astronomical Data Center, and 
the Atomic Spectra Database hosted by
the National Institute of Standards and Technology. 
IRAF is distributed by the National Optical Astronomy Observatories,
which are operated by the Association of Universities for Research
in Astronomy, Inc., under cooperative agreement with the National
Science Foundation.
I.U.R.\ thanks the Carnegie Institution for Science
for support through the Barbara McClintock Fellowship.
C.S.\ is supported by the U.S.\ National Science Foundation 
(grant AST~12-11585).

{\it Facilities:} 
\facility{Magellan:Baade (MIKE)}, 
\facility{Magellan:Clay (MIKE)}, 
\facility{Smith (Tull)}

\begin{deluxetable*}{cccccccc}
\tablecaption{Mean Abundances in BD$+$44~493
\label{abundtab1}}
%\rotate
\tablewidth{0pt}
\tabletypesize{\scriptsize}
%\tabletypesize{\tiny}
\tablehead{
\colhead{Species} &
\colhead{$N_{\rm lines}$} &
\colhead{$\log\epsilon$} &
\colhead{[X/Fe]\tablenotemark{a}} &
\colhead{$\sigma_{\rm statistical}$} &
\colhead{$\sigma_{\rm total}$} &
\colhead{$\sigma_{\rm neutrals}$} &
\colhead{$\sigma_{\rm ions}$} }
\startdata      
%\hline
%\multicolumn{8}{c}{BD$+$44~493        } \\
%\hline
%species        Nline logeps     [X/Fe]  e_stat  e_total e_I     e_II
 Fe~\textsc{i}  &  61 &     3.22 &    -4.28 &  0.09 &  0.16 &  0.00 &  0.00  \\
 Fe~\textsc{ii} &   2 &     3.24 &    -4.26 &  0.12 &  0.18 &  0.00 &  0.00  \\
 Li~\textsc{i}  &   1 &     0.64 &   \nodata&  0.12 &  0.18 &  0.15 &  0.23  \\
 C~(CH)         &   1 &     5.35 &     1.18 &  0.15 &  0.25 &  0.20 &  0.22  \\
 N~(CN)         &   1 & $<$ 5.75 & $<$ 2.18 &\nodata&\nodata&\nodata&\nodata \\
 O~\textsc{i}   &   0 &  \nodata &   \nodata&\nodata&\nodata&\nodata&\nodata \\
 Na~\textsc{i}  &   0 &  \nodata &   \nodata&\nodata&\nodata&\nodata&\nodata \\
 Mg~\textsc{i}  &   7 &     4.21 &     0.90 &  0.09 &  0.16 &  0.12 &  0.22  \\
\enddata
\tablenotetext{a}{[Fe/H] is indicated for Fe~\textsc{i} and Fe~\textsc{ii}}
\end{deluxetable*}
  % 2 col 

\begin{deluxetable*}{cccccccc}
\tablecaption{Mean Abundances in CS~22877--001
\label{abundtab2}}
%\rotate
\tablewidth{0pt}
\tabletypesize{\scriptsize}
%\tabletypesize{\tiny}
\tablehead{
\colhead{Species} &
\colhead{$N_{\rm lines}$} &
\colhead{$\log\epsilon$} &
\colhead{[X/Fe]\tablenotemark{a}} &
\colhead{$\sigma_{\rm statistical}$} &
\colhead{$\sigma_{\rm total}$} &
\colhead{$\sigma_{\rm neutrals}$} &
\colhead{$\sigma_{\rm ions}$} }
\startdata      
 Fe~\textsc{i}  & 106 &     4.19 &    -3.31 &  0.06 &  0.15 &  0.00 &  0.00  \\
 Fe~\textsc{ii} &  10 &     4.26 &    -3.24 &  0.07 &  0.15 &  0.00 &  0.00  \\
 Li~\textsc{i}  &   1 &     0.66 &   \nodata&  0.09 &  0.16 &  0.11 &  0.19  \\
 C~(CH)         &   1 &     6.22 &     1.03 &  0.15 &  0.25 &  0.19 &  0.19  \\
 N~(NH)         &   1 &     4.44 &    -0.15 &  0.30 &  0.36 &  0.32 &  0.32  \\
 O~\textsc{i}   &   2 &     6.60 &     1.21 &  0.18 &  0.22 &  0.19 &  0.25  \\
 Na~\textsc{i}  &   0 &  \nodata &   \nodata&\nodata&\nodata&\nodata&\nodata \\
 Mg~\textsc{i}  &   8 &     4.67 &     0.38 &  0.05 &  0.16 &  0.09 &  0.18  \\
\enddata
\tablenotetext{a}{[Fe/H] is indicated for Fe~\textsc{i} and Fe~\textsc{ii}}
\end{deluxetable*}
  % ''

\begin{deluxetable*}{cccccccc}
\tablecaption{Mean Abundances in CS~22878--101
\label{abundtab3}}
%\rotate
\tablewidth{0pt}
\tabletypesize{\scriptsize}
%\tabletypesize{\tiny}
\tablehead{
\colhead{Species} &
\colhead{$N_{\rm lines}$} &
\colhead{$\log\epsilon$} &
\colhead{[X/Fe]\tablenotemark{a}} &
\colhead{$\sigma_{\rm statistical}$} &
\colhead{$\sigma_{\rm total}$} &
\colhead{$\sigma_{\rm neutrals}$} &
\colhead{$\sigma_{\rm ions}$} }
\startdata      
 Fe~\textsc{i}  & 100 &     3.97 &    -3.53 &  0.06 &  0.15 &  0.00 &  0.00  \\
 Fe~\textsc{ii} &  11 &     4.20 &    -3.30 &  0.07 &  0.15 &  0.00 &  0.00  \\
 Li~\textsc{i}  &   1 & $<$ 0.43 &   \nodata&\nodata&\nodata&\nodata&\nodata \\
 C~(CH)         &   1 &     4.53 &    -0.60 &  0.15 &  0.25 &  0.19 &  0.19  \\
 N~(NH)         &   1 &     5.48 &     1.29 &  0.30 &  0.36 &  0.32 &  0.32  \\
 O~\textsc{i}   &   1 &     6.39 &     1.23 &  0.15 &  0.20 &  0.17 &  0.23  \\
 Na~\textsc{i}  &   0 &  \nodata &   \nodata&\nodata&\nodata&\nodata&\nodata \\
 Mg~\textsc{i}  &   8 &     4.67 &     0.60 &  0.06 &  0.16 &  0.09 &  0.17  \\
\enddata
\tablenotetext{a}{[Fe/H] is indicated for Fe~\textsc{i} and Fe~\textsc{ii}}
\end{deluxetable*}
  % ''

\begin{deluxetable*}{cccccccc}
\tablecaption{Mean Abundances in CS~22891--200
\label{abundtab4}}
%\rotate
\tablewidth{0pt}
\tabletypesize{\scriptsize}
%\tabletypesize{\tiny}
\tablehead{
\colhead{Species} &
\colhead{$N_{\rm lines}$} &
\colhead{$\log\epsilon$} &
\colhead{[X/Fe]\tablenotemark{a}} &
\colhead{$\sigma_{\rm statistical}$} &
\colhead{$\sigma_{\rm total}$} &
\colhead{$\sigma_{\rm neutrals}$} &
\colhead{$\sigma_{\rm ions}$} }
\startdata      
%\hline
%\multicolumn{8}{c}{CS~22891--200     } \\ % RG  T=4490  logg=0.50  vt=1.70
%\hline
%species        Nline logeps     [X/Fe]  e_stat  e_total e_I     e_II
 Fe~\textsc{i}  &  87 &     3.44 &    -4.06 &  0.06 &  0.15 &  0.00 &  0.00  \\
 Fe~\textsc{ii} &   9 &     3.62 &    -3.88 &  0.08 &  0.16 &  0.00 &  0.00  \\
 Li~\textsc{i}  &   1 & $<$ 0.13 &   \nodata&\nodata&\nodata&\nodata&\nodata \\
 C~(CH)         &   1 &     4.90 &     0.35 &  0.15 &  0.25 &  0.19 &  0.20  \\
 N~(NH)         &   1 &     5.15 &     1.20 &  0.30 &  0.36 &  0.32 &  0.33  \\
 O~\textsc{i}   &   4 & $<$ 6.30 & $<$ 1.67 &\nodata&\nodata&\nodata&\nodata \\
 Na~\textsc{i}  &   0 &  \nodata &   \nodata&\nodata&\nodata&\nodata&\nodata \\
 Mg~\textsc{i}  &   9 &     4.36 &     0.82 &  0.06 &  0.16 &  0.09 &  0.18  \\
\enddata
\tablenotetext{a}{[Fe/H] is indicated for Fe~\textsc{i} and Fe~\textsc{ii}}
\end{deluxetable*}
  % ''

\begin{deluxetable*}{cccccccc}
\tablecaption{Mean Abundances in CS~22893--010
\label{abundtab5}}
%\rotate
\tablewidth{0pt}
\tabletypesize{\scriptsize}
%\tabletypesize{\tiny}
\tablehead{
\colhead{Species} &
\colhead{$N_{\rm lines}$} &
\colhead{$\log\epsilon$} &
\colhead{[X/Fe]\tablenotemark{a}} &
\colhead{$\sigma_{\rm statistical}$} &
\colhead{$\sigma_{\rm total}$} &
\colhead{$\sigma_{\rm neutrals}$} &
\colhead{$\sigma_{\rm ions}$} }
\startdata      
%\hline
%\multicolumn{8}{c}{CS~22893--010     } \\ % RG  T=5150  logg=2.45  vt=1.35
%\hline
%species        Nline logeps     [X/Fe]  e_stat  e_total e_I     e_II
 Fe~\textsc{i}  & 112 &     4.38 &    -3.12 &  0.06 &  0.15 &  0.00 &  0.00  \\
 Fe~\textsc{ii} &   9 &     4.57 &    -2.93 &  0.07 &  0.16 &  0.00 &  0.00  \\
 Li~\textsc{i}  &   1 &     2.57 &   \nodata&  0.05 &  0.23 &  0.17 &  0.25  \\
 C~(CH)         &   1 &     5.62 &     0.13 &  0.15 &  0.25 &  0.19 &  0.19  \\
 N~(NH)         &   1 &     6.44 &     1.55 &  0.30 &  0.36 &  0.32 &  0.32  \\
 O~\textsc{i}   &   1 &     6.58 &     1.01 &  0.12 &  0.18 &  0.13 &  0.21  \\
 Na~\textsc{i}  &   2 &     4.70 &     1.58 &  0.11 &  0.17 &  0.12 &  0.20  \\
 Mg~\textsc{i}  &   7 &     5.04 &     0.56 &  0.05 &  0.15 &  0.08 &  0.18  \\
\enddata
\tablenotetext{a}{[Fe/H] is indicated for Fe~\textsc{i} and Fe~\textsc{ii}}
\end{deluxetable*}
  % ''

\begin{deluxetable*}{cccccccc}
\tablecaption{Mean Abundances in CS~22943--201
\label{abundtab7}}
%\rotate
\tablewidth{0pt}
\tabletypesize{\scriptsize}
%\tabletypesize{\tiny}
\tablehead{
\colhead{Species} &
\colhead{$N_{\rm lines}$} &
\colhead{$\log\epsilon$} &
\colhead{[X/Fe]\tablenotemark{a}} &
\colhead{$\sigma_{\rm statistical}$} &
\colhead{$\sigma_{\rm total}$} &
\colhead{$\sigma_{\rm neutrals}$} &
\colhead{$\sigma_{\rm ions}$} }
\startdata      
 Fe~\textsc{i}  &  72 &     4.82 &    -2.68 &  0.06 &  0.17 &  0.00 &  0.00  \\
 Fe~\textsc{ii} &   8 &     4.81 &    -2.69 &  0.06 &  0.15 &  0.00 &  0.00  \\
 Li~\textsc{i}  &   1 & $<$ 2.04 &   \nodata&\nodata&\nodata&\nodata&\nodata \\
 C~(CH)         &   1 &     7.63 &     1.89 &  0.15 &  0.25 &  0.19 &  0.19  \\
 N~(NH)         &   1 & $<$ 6.70 & $<$ 1.56 &\nodata&\nodata&\nodata&\nodata \\
 O~\textsc{i}   &   3 &     6.25 &     0.24 &  0.08 &  0.18 &  0.10 &  0.20  \\
 Na~\textsc{i}  &   0 &  \nodata &   \nodata&\nodata&\nodata&\nodata&\nodata \\
 Mg~\textsc{i}  &   7 &     5.54 &     0.63 &  0.05 &  0.17 &  0.08 &  0.20  \\
\enddata
\tablenotetext{a}{[Fe/H] is indicated for Fe~\textsc{i} and Fe~\textsc{ii}}
\end{deluxetable*}
  % ''

\begin{deluxetable*}{cccccccc}
\tablecaption{Mean Abundances in CS~22948--066
\label{abundtab8}}
%\rotate
\tablewidth{0pt}
\tabletypesize{\scriptsize}
%\tabletypesize{\tiny}
\tablehead{
\colhead{Species} &
\colhead{$N_{\rm lines}$} &
\colhead{$\log\epsilon$} &
\colhead{[X/Fe]\tablenotemark{a}} &
\colhead{$\sigma_{\rm statistical}$} &
\colhead{$\sigma_{\rm total}$} &
\colhead{$\sigma_{\rm neutrals}$} &
\colhead{$\sigma_{\rm ions}$} }
\startdata      
 Fe~\textsc{i}  &  99 &     4.00 &    -3.50 &  0.06 &  0.15 &  0.00 &  0.00  \\
 Fe~\textsc{ii} &  11 &     4.32 &    -3.18 &  0.07 &  0.16 &  0.00 &  0.00  \\
 Li~\textsc{i}  &   1 & $<$ 0.45 &   \nodata&\nodata&\nodata&\nodata&\nodata \\
 C~(CH)         &   1 &     4.50 &    -0.74 &  0.15 &  0.25 &  0.19 &  0.19  \\
 N~(NH)         &   1 &     5.69 &     1.05 &  0.30 &  0.36 &  0.32 &  0.32  \\
 O~\textsc{i}   &   3 & $<$ 6.57 & $<$ 1.38 &\nodata&\nodata&\nodata&\nodata \\
 Na~\textsc{i}  &   0 &  \nodata &   \nodata&\nodata&\nodata&\nodata&\nodata \\
 Mg~\textsc{i}  &   9 &     4.56 &     0.46 &  0.06 &  0.15 &  0.08 &  0.18  \\
\enddata
\tablenotetext{a}{[Fe/H] is indicated for Fe~\textsc{i} and Fe~\textsc{ii}}
\end{deluxetable*}
  % ''

\begin{deluxetable*}{cccccccc}
\tablecaption{Mean Abundances in CS~22949--037
\label{abundtab9}}
%\rotate
\tablewidth{0pt}
\tabletypesize{\scriptsize}
%\tabletypesize{\tiny}
\tablehead{
\colhead{Species} &
\colhead{$N_{\rm lines}$} &
\colhead{$\log\epsilon$} &
\colhead{[X/Fe]\tablenotemark{a}} &
\colhead{$\sigma_{\rm statistical}$} &
\colhead{$\sigma_{\rm total}$} &
\colhead{$\sigma_{\rm neutrals}$} &
\colhead{$\sigma_{\rm ions}$} }
\startdata      
%\hline
%\multicolumn{8}{c}{CS~22949--037     } \\ % RG  T=4630  logg=0.95  vt=1.70
%\hline
%species        Nline logeps     [X/Fe]  e_stat  e_total e_I     e_II
 Fe~\textsc{i}  &  58 &     3.12 &    -4.38 &  0.07 &  0.15 &  0.00 &  0.00  \\
 Fe~\textsc{ii} &   2 &     3.30 &    -4.20 &  0.07 &  0.16 &  0.00 &  0.00  \\
 Li~\textsc{i}  &   1 & $<$ 0.13 &   \nodata&\nodata&\nodata&\nodata&\nodata \\
 C~(CH)         &   1 &     5.21 &     0.99 &  0.15 &  0.25 &  0.19 &  0.19  \\
 N~(NH)         &   1 &     6.12 &     2.50 &  0.30 &  0.36 &  0.32 &  0.32  \\
 O~\textsc{i}   &   2 & $<$ 6.70 & $<$ 2.39 &\nodata&\nodata&\nodata&\nodata \\
 Na~\textsc{i}  &   0 &  \nodata &   \nodata&\nodata&\nodata&\nodata&\nodata \\
 Mg~\textsc{i}  &   7 &     4.78 &     1.56 &  0.05 &  0.15 &  0.09 &  0.18  \\
\enddata
\tablenotetext{a}{[Fe/H] is indicated for Fe~\textsc{i} and Fe~\textsc{ii}}
\end{deluxetable*}
  % ''

\begin{deluxetable*}{cccccccc}
\tablecaption{Mean Abundances in CS~22957--027
\label{abundtab10}}
%\rotate
\tablewidth{0pt}
\tabletypesize{\scriptsize}
%\tabletypesize{\tiny}
\tablehead{
\colhead{Species} &
\colhead{$N_{\rm lines}$} &
\colhead{$\log\epsilon$} &
\colhead{[X/Fe]\tablenotemark{a}} &
\colhead{$\sigma_{\rm statistical}$} &
\colhead{$\sigma_{\rm total}$} &
\colhead{$\sigma_{\rm neutrals}$} &
\colhead{$\sigma_{\rm ions}$} }
\startdata      
%\hline
%\multicolumn{8}{c}{CS~22957--027     } \\ % RG  T=5220  logg=2.65  vt=1.45
%\hline
%species        Nline logeps     [X/Fe]  e_stat  e_total e_I     e_II
 Fe~\textsc{i}  &  36 &     4.31 &    -3.19 &  0.06 &  0.14 &  0.00 &  0.00  \\
 Fe~\textsc{ii} &   3 &     4.50 &    -3.00 &  0.07 &  0.16 &  0.00 &  0.00  \\
 Li~\textsc{i}  &   1 & $<$ 0.86 &   \nodata&\nodata&\nodata&\nodata&\nodata \\
 C~(CH)         &   1 &     7.85 &     2.42 &  0.15 &  0.25 &  0.19 &  0.19  \\
 N~(NH)         &   1 &     6.33 &     1.50 &  0.30 &  0.36 &  0.32 &  0.32  \\
 O~\textsc{i}   &   3 & $<$ 6.70 & $<$ 1.20 &\nodata&\nodata&\nodata&\nodata \\
 Na~\textsc{i}  &   3 &     4.03 &     0.98 &  0.12 &  0.20 &  0.15 &  0.22  \\
 Mg~\textsc{i}  &   3 &     4.49 &     0.08 &  0.06 &  0.16 &  0.09 &  0.19  \\
\enddata
\tablenotetext{a}{[Fe/H] is indicated for Fe~\textsc{i} and Fe~\textsc{ii}}
\end{deluxetable*}
  % ''

\begin{deluxetable*}{cccccccc}
\tablecaption{Mean Abundances in CS~22958--042
\label{abundtab11}}
%\rotate
\tablewidth{0pt}
\tabletypesize{\scriptsize}
%\tabletypesize{\tiny}
\tablehead{
\colhead{Species} &
\colhead{$N_{\rm lines}$} &
\colhead{$\log\epsilon$} &
\colhead{[X/Fe]\tablenotemark{a}} &
\colhead{$\sigma_{\rm statistical}$} &
\colhead{$\sigma_{\rm total}$} &
\colhead{$\sigma_{\rm neutrals}$} &
\colhead{$\sigma_{\rm ions}$} }
\startdata      
%\hline
%\multicolumn{8}{c}{CS~22958--042     } \\ % SG  T=5760  logg=3.55  vt=0.95
%\hline
%species        Nline logeps     [X/Fe]  e_stat  e_total e_I     e_II
 Fe~\textsc{i}  &  19 &     4.10 &    -3.40 &  0.06 &  0.21 &  0.00 &  0.00  \\
 Fe~\textsc{ii} &   1 &     4.51 &    -2.99 &  0.11 &  0.16 &  0.00 &  0.00  \\
 Li~\textsc{i}  &   1 & $<$ 1.33 &   \nodata&\nodata&\nodata&\nodata&\nodata \\
 C~(CH)         &   1 &     7.59 &     2.15 &  0.15 &  0.25 &  0.19 &  0.21  \\
 N~(CN)         &   1 &     7.09 &     2.25 &  0.20 &  0.28 &  0.23 &  0.25  \\
 O~\textsc{i}   &   0 &  \nodata &   \nodata&\nodata&\nodata&\nodata&\nodata \\
 Na~\textsc{i}  &   2 & $<$ 4.73 & $<$ 1.89 &\nodata&\nodata&\nodata&\nodata \\
 Mg~\textsc{i}  &   3 &     4.95 &     0.74 &  0.05 &  0.23 &  0.08 &  0.23  \\
\enddata
\tablenotetext{a}{[Fe/H] is indicated for Fe~\textsc{i} and Fe~\textsc{ii}}
\end{deluxetable*}
  % ''

\begin{deluxetable*}{cccccccc}
\tablecaption{Mean Abundances in CS~22960--064
\label{abundtab12}}
%\rotate
\tablewidth{0pt}
\tabletypesize{\scriptsize}
%\tabletypesize{\tiny}
\tablehead{
\colhead{Species} &
\colhead{$N_{\rm lines}$} &
\colhead{$\log\epsilon$} &
\colhead{[X/Fe]\tablenotemark{a}} &
\colhead{$\sigma_{\rm statistical}$} &
\colhead{$\sigma_{\rm total}$} &
\colhead{$\sigma_{\rm neutrals}$} &
\colhead{$\sigma_{\rm ions}$} }
\startdata      
 Fe~\textsc{i}  & 102 &     4.73 &    -2.77 &  0.05 &  0.14 &  0.00 &  0.00  \\
 Fe~\textsc{ii} &  12 &     4.73 &    -2.77 &  0.07 &  0.15 &  0.00 &  0.00  \\
 Li~\textsc{i}  &   1 & $<$ 0.86 &   \nodata&\nodata&\nodata&\nodata&\nodata \\
 C~(CH)         &   1 &     5.80 &     0.14 &  0.15 &  0.25 &  0.19 &  0.19  \\
 N~(NH)         &   1 &     6.41 &     1.35 &  0.30 &  0.36 &  0.32 &  0.32  \\
 O~\textsc{i}   &   3 & $<$ 6.85 & $<$ 0.93 &\nodata&\nodata&\nodata&\nodata \\
 Na~\textsc{i}  &   2 &     3.80 &     0.33 &  0.10 &  0.30 &  0.24 &  0.29  \\
 Mg~\textsc{i}  &   5 &     5.27 &     0.44 &  0.05 &  0.15 &  0.08 &  0.17  \\
\enddata
\tablenotetext{a}{[Fe/H] is indicated for Fe~\textsc{i} and Fe~\textsc{ii}}
\end{deluxetable*}
  % ''

\begin{deluxetable*}{cccccccc}
\tablecaption{Mean Abundances in CS~29498--043
\label{abundtab13}}
%\rotate
\tablewidth{0pt}
\tabletypesize{\scriptsize}
%\tabletypesize{\tiny}
\tablehead{
\colhead{Species} &
\colhead{$N_{\rm lines}$} &
\colhead{$\log\epsilon$} &
\colhead{[X/Fe]\tablenotemark{a}} &
\colhead{$\sigma_{\rm statistical}$} &
\colhead{$\sigma_{\rm total}$} &
\colhead{$\sigma_{\rm neutrals}$} &
\colhead{$\sigma_{\rm ions}$} }
\startdata      
%\hline
%\multicolumn{8}{c}{CS~29498--043     } \\ % RG  T=4440  logg=0.50  vt=1.75
%\hline
%species        Nline logeps     [X/Fe]  e_stat  e_total e_I     e_II
 Fe~\textsc{i}  &  74 &     3.63 &    -3.87 &  0.06 &  0.15 &  0.00 &  0.00  \\
 Fe~\textsc{ii} &   6 &     3.65 &    -3.85 &  0.08 &  0.16 &  0.00 &  0.00  \\
 Li~\textsc{i}  &   1 & $<$-0.05 &   \nodata&\nodata&\nodata&\nodata&\nodata \\
 C~(CH)         &   1 &     7.31 &     2.72 &  0.15 &  0.25 &  0.19 &  0.20  \\
 N~(NH)         &   1 &     5.70 &     1.71 &  0.30 &  0.36 &  0.32 &  0.33  \\
 O~\textsc{i}   &   3 &     7.19 &     2.37 &  0.08 &  0.15 &  0.10 &  0.19  \\
 Na~\textsc{i}  &   4 &     3.40 &     1.03 &  0.17 &  0.23 &  0.19 &  0.25  \\
 Mg~\textsc{i}  &   5 &     5.51 &     1.78 &  0.05 &  0.18 &  0.11 &  0.18  \\
\enddata
\tablenotetext{a}{[Fe/H] is indicated for Fe~\textsc{i} and Fe~\textsc{ii}}
\end{deluxetable*}
  % ''

\begin{deluxetable*}{cccccccc}
\tablecaption{Mean Abundances in CS~29502--092
\label{abundtab14}}
%\rotate
\tablewidth{0pt}
\tabletypesize{\scriptsize}
%\tabletypesize{\tiny}
\tablehead{
\colhead{Species} &
\colhead{$N_{\rm lines}$} &
\colhead{$\log\epsilon$} &
\colhead{[X/Fe]\tablenotemark{a}} &
\colhead{$\sigma_{\rm statistical}$} &
\colhead{$\sigma_{\rm total}$} &
\colhead{$\sigma_{\rm neutrals}$} &
\colhead{$\sigma_{\rm ions}$} }
\startdata      
 Fe~\textsc{i}  &  90 &     4.20 &    -3.30 &  0.06 &  0.15 &  0.00 &  0.00  \\
 Fe~\textsc{ii} &   9 &     4.30 &    -3.20 &  0.07 &  0.15 &  0.00 &  0.00  \\
 Li~\textsc{i}  &   1 & $<$ 0.45 &   \nodata&\nodata&\nodata&\nodata&\nodata \\
 C~(CH)         &   1 &     6.19 &     0.96 &  0.15 &  0.25 &  0.19 &  0.29  \\
 N~(NH)         &   1 &     5.63 &     1.00 &  0.30 &  0.36 &  0.32 &  0.32  \\
 O~\textsc{i}   &   3 &     6.52 &     1.13 &  0.16 &  0.20 &  0.17 &  0.23  \\
 Na~\textsc{i}  &   1 &     2.92 &    -0.02 &  0.10 &  0.27 &  0.22 &  0.29  \\
 Mg~\textsc{i}  &   8 &     4.80 &     0.51 &  0.05 &  0.16 &  0.09 &  0.18  \\
\enddata
\tablenotetext{a}{[Fe/H] is indicated for Fe~\textsc{i} and Fe~\textsc{ii}}
\end{deluxetable*}
  % ''

\begin{deluxetable*}{cccccccc}
\tablecaption{Mean Abundances in CS~30314--067
\label{abundtab15}}
%\rotate
\tablewidth{0pt}
\tabletypesize{\scriptsize}
%\tabletypesize{\tiny}
\tablehead{
\colhead{Species} &
\colhead{$N_{\rm lines}$} &
\colhead{$\log\epsilon$} &
\colhead{[X/Fe]\tablenotemark{a}} &
\colhead{$\sigma_{\rm statistical}$} &
\colhead{$\sigma_{\rm total}$} &
\colhead{$\sigma_{\rm neutrals}$} &
\colhead{$\sigma_{\rm ions}$} }
\startdata      
%\hline
%\multicolumn{8}{c}{CS~30314--067     } \\ % RG  T=4320  logg=0.50  vt=1.85
%\hline
%species        Nline logeps     [X/Fe]  e_stat  e_total e_I     e_II
 Fe~\textsc{i}  &  91 &     4.19 &    -3.31 &  0.06 &  0.15 &  0.00 &  0.00  \\
 Fe~\textsc{ii} &   8 &     4.49 &    -3.01 &  0.06 &  0.15 &  0.00 &  0.00  \\
 Li~\textsc{i}  &   1 & $<$-0.41 &   \nodata&\nodata&\nodata&\nodata&\nodata \\
 C~(CH)         &   1 &     5.97 &     0.55 &  0.15 &  0.25 &  0.19 &  0.19  \\
 N~(NH)         &   1 &     6.00 &     1.18 &  0.30 &  0.36 &  0.32 &  0.32  \\
 O~\textsc{i}   &   3 &     6.49 &     1.11 &  0.21 &  0.25 &  0.22 &  0.27  \\
 Na~\textsc{i}  &   0 &  \nodata &   \nodata&\nodata&\nodata&\nodata&\nodata \\
 Mg~\textsc{i}  &   5 &     4.85 &     0.56 &  0.06 &  0.16 &  0.09 &  0.18  \\
\enddata
\tablenotetext{a}{[Fe/H] is indicated for Fe~\textsc{i} and Fe~\textsc{ii}}
\end{deluxetable*}
  % ''

\begin{deluxetable*}{cccccccc}
\tablecaption{Mean Abundances in CS~30492--001
\label{abundtab16}}
%\rotate
\tablewidth{0pt}
\tabletypesize{\scriptsize}
%\tabletypesize{\tiny}
\tablehead{
\colhead{Species} &
\colhead{$N_{\rm lines}$} &
\colhead{$\log\epsilon$} &
\colhead{[X/Fe]\tablenotemark{a}} &
\colhead{$\sigma_{\rm statistical}$} &
\colhead{$\sigma_{\rm total}$} &
\colhead{$\sigma_{\rm neutrals}$} &
\colhead{$\sigma_{\rm ions}$} }
\startdata      
 Fe~\textsc{i}  &  93 &     4.88 &    -2.62 &  0.06 &  0.21 &  0.00 &  0.00  \\
 Fe~\textsc{ii} &   7 &     5.15 &    -2.35 &  0.07 &  0.14 &  0.00 &  0.00  \\
 Li~\textsc{i}  &   1 &     2.02 &   \nodata&  0.06 &  0.20 &  0.09 &  0.20  \\
 C~(CH)         &   1 &     6.03 &    -0.05 &  0.15 &  0.25 &  0.19 &  0.19  \\
 N~(NH)         &   1 &     6.60 &     1.12 &  0.30 &  0.36 &  0.32 &  0.32  \\
 O~\textsc{i}   &   3 &     6.72 &     0.64 &  0.10 &  0.21 &  0.12 &  0.22  \\
 Na~\textsc{i}  &   1 &     3.47 &    -0.15 &  0.10 &  0.27 &  0.17 &  0.27  \\
 Mg~\textsc{i}  &   6 &     5.45 &     0.47 &  0.05 &  0.22 &  0.08 &  0.21  \\
\enddata
\tablenotetext{a}{[Fe/H] is indicated for Fe~\textsc{i} and Fe~\textsc{ii}}
\end{deluxetable*}
  % ''

\begin{deluxetable*}{cccccccc}
\tablecaption{Mean Abundances in HE~1012$-$1540
\label{abundtab17}}
%\rotate
\tablewidth{0pt}
\tabletypesize{\scriptsize}
%\tabletypesize{\tiny}
\tablehead{
\colhead{Species} &
\colhead{$N_{\rm lines}$} &
\colhead{$\log\epsilon$} &
\colhead{[X/Fe]\tablenotemark{a}} &
\colhead{$\sigma_{\rm statistical}$} &
\colhead{$\sigma_{\rm total}$} &
\colhead{$\sigma_{\rm neutrals}$} &
\colhead{$\sigma_{\rm ions}$} }
\startdata      
%\hline
%\multicolumn{8}{c}{HE~1012$-$1540    } \\ % RG  T=5230  logg=2.65  vt=1.70
%\hline
%species        Nline logeps     [X/Fe]  e_stat  e_total e_I     e_II
 Fe~\textsc{i}  &  48 &     3.33 &    -4.17 &  0.07 &  0.16 &  0.00 &  0.00  \\
 Fe~\textsc{ii} &   1 &     3.74 &    -3.76 &  0.14 &  0.20 &  0.00 &  0.00  \\
 Li~\textsc{i}  &   1 & $<$ 0.75 &   \nodata&\nodata&\nodata&\nodata&\nodata \\
 C~(CH)         &   1 &     6.66 &     1.99 &  0.15 &  0.25 &  0.19 &  0.23  \\
 N~(NH)         &   1 &     4.81 &     0.74 &  0.30 &  0.36 &  0.32 &  0.35  \\
 O~\textsc{i}   &   2 & $<$ 6.70 & $<$ 2.18 &\nodata&\nodata&\nodata&\nodata \\
 Na~\textsc{i}  &   2 &     3.72 &     1.65 &  0.10 &  0.29 &  0.23 &  0.32  \\
 Mg~\textsc{i}  &   3 &     5.24 &     1.80 &  0.05 &  0.15 &  0.09 &  0.22  \\
\enddata
\tablenotetext{a}{[Fe/H] is indicated for Fe~\textsc{i} and Fe~\textsc{ii}}
\end{deluxetable*}
  % ''


\begin{thebibliography}{}


\bibitem[Aoki et al.(2002a)]{aoki02a} Aoki, W., Norris, J.~E., 
Ryan, S.~G., Beers, T.~C., \& Ando, H.\ 2002a, \apj, 567, 1166 

\bibitem[Aoki et al.(2002b)]{aoki02b} Aoki, W., Norris, J.~E., 
Ryan, S.~G., Beers, T.~C., \& Ando, H.\ 2002b, \apjl, 576, L141 

\bibitem[Aoki et al.(2002c)]{aoki02c} Aoki, W., Ryan, S.~G., 
Norris, J.~E., et al.\ 2002c, \apj, 580, 1149 

\bibitem[Aoki et al.(2004)]{aoki04} Aoki, W., Norris, J.~E., 
Ryan, S.~G., et al.\ 2004, \apj, 608, 971 

\bibitem[Aoki et al.(2006)]{aoki06} Aoki, W., Frebel, A., 
Christlieb, N., et al.\ 2006, \apj, 639, 897 

\bibitem[Aoki et al.(2007)]{aoki07} Aoki, W., Beers, T.~C., 
Christlieb, N., Norris, J.~E., Ryan, S.~G., 
\& Tsangarides, S.\ 2007, \apj, 655, 492 

\bibitem[Asplund et al.(2009)]{asplund09} Asplund, M., Grevesse, N., 
Sauval, A.~J., \& Scott, P.\ 2009, \araa, 47, 481 

\bibitem[Barklem et al.(2000)]{barklem00} Barklem, P.~S., Piskunov, N., \& 
O'Mara, B.~J.\ 2000, \aaps, 142, 467 

\bibitem[Barklem \& Aspelund-Johansson(2005)]{barklem05b} Barklem, P.~S., \& 
Aspelund-Johansson, J.\ 2005, \aap, 435, 373 

\bibitem[Becker et al.(2012)]{becker12} Becker, G.~D., Sargent, 
W.~L.~W., Rauch, M., \& Carswell, R.~F.\ 2012, \apj, 744, 91 

\bibitem[Beers \& Christlieb(2005)]{beers05}
Beers, T.~C.\ \& Christlieb, N.\ 2005, \araa, 43, 531

\bibitem[Bernstein et al.(2003)]{bernstein03} Bernstein, R., 
Shectman, S.~A., Gunnels, S.~M., Mochnacki, S., 
\& Athey, A.~E.\ 2003, \procspie, 4841, 1694 

\bibitem[Bessell et al.(2004)]{bessell04} Bessell, M.~S., 
Christlieb, N., \& Gustafsson, B.\ 2004, \apjl, 612, L61 

\bibitem[Bisterzo et al.(2011)]{bisterzo11} Bisterzo, S., Gallino, 
R., Straniero, O., Cristallo, S., K\"{a}ppeler, F.\ 2011, \mnras, 418, 284 

\bibitem[Bonifacio et al.(1998)]{bonifacio98} Bonifacio, P., Molaro, P., 
Beers, T.~C., \& Vladilo, G.\ 1998, \aap, 332, 672 

\bibitem[Bonifacio et al.(2009)]{bonifacio09} Bonifacio, P., et al.\ 
2009, \aap, 501, 519 

\bibitem[Brown et al.(1989)]{brown89} Brown, J.~A., Sneden, C., 
Lambert, D.~L., \& Dutchover, E., Jr.\ 1989, \apjs, 71, 293 

\bibitem[Caffau et al.(2012)]{caffau12} Caffau, E., Bonifacio, P., 
Fran{\c c}ois, P., et al.\ 2012, \aap, 542, A51 

\bibitem[Carney et al.(2003)]{carney03} Carney, B.~W., Latham, 
D.~W., Stefanik, R.~P., Laird, J.~B., \& Morse, J.~A.\ 2003, \aj, 125, 293 

\bibitem[Carollo et al.(2012)]{carollo12} Carollo, D., Beers, 
T.~C., Bovy, J., et al.\ 2012, \apj, 744, 195 

\bibitem[Carretta et al.(2002)]{carretta02} Carretta, E., Gratton, 
R., Cohen, J.~G., Beers, T.~C., \& Christlieb, N.\ 2002, \aj, 124, 481 

\bibitem[Cayrel et al.(2004)]{cayrel04} Cayrel, R., Depagne, E., 
Spite, M., et al.\ 2004, \aap, 416, 1117 

\bibitem[Cescutti et al.(2013)]{cescutti13} Cescutti, G., Chiappini, C., 
Hirschi, R., Meynet, G., \& Frischknecht, U.\ 2013, \aap, 553, A51 

\bibitem[Christlieb et al.(2002)]{christlieb02} Christlieb, N., 
Bessell, M.~S., Beers, T.~C., et al.\ 2002, \nat, 419, 904 

\bibitem[Christlieb et al.(2004)]{christlieb04} Christlieb, N., 
Gustafsson, B., Korn, A.~J., et al.\ 2004, \apj, 603, 708 

\bibitem[Cohen et al.(2002)]{cohen02} Cohen, J.~G., Christlieb, 
N., Beers, T.~C., Gratton, R., \& Carretta, E.\ 2002, \aj, 124, 470 

\bibitem[Cohen et al.(2006)]{cohen06} Cohen, J.~G., McWilliam, 
A., Shectman, S., et al.\ 2006, \aj, 132, 137 

\bibitem[Cohen et al.(2008)]{cohen08} Cohen, J.~G., Christlieb, 
N., McWilliam, A., et al.\ 2008, \apj, 672, 320 

\bibitem[Cohen et al.(2013)]{cohen13} Cohen, J.~G., Christlieb, N., 
Thompson, I.~B., et al.\ 2013, \apj, 778, 56

\bibitem[Cooke et al.(2011a)]{cooke11a} Cooke, R., Pettini, M., 
Steidel, C.~C., Rudie, G.~C., \& Jorgenson, R.~A.\ 2011a, \mnras, 412, 1047 

\bibitem[Cooke et al.(2011b)]{cooke11b} Cooke, R., Pettini, M., 
Steidel, C.~C., Rudie, G.~C., \& Nissen, P.~E.\ 2011b, \mnras, 417, 1534 

\bibitem[Demarque et al.(2004)]{demarque04} Demarque, P., Woo, 
J.-H., Kim, Y.-C., \& Yi, S.~K.\ 2004, \apjs, 155, 667 

\bibitem[Depagne et al.(2002)]{depagne02} Depagne, E., Hill, V., 
Spite, M., et al.\ 2002, \aap, 390, 187 

\bibitem[Fabbian et al.(2009)]{fabbian09}
Fabbian, D., Asplund, M., Barklem, P.~S., Carlsson, M., Kiselman, D.\
2009, \aap, 500, 1221

\bibitem[Fran{\c c}ois et al.(2007)]{francois07} Fran{\c c}ois, P., 
Depagne, E., Hill, V., et al.\ 2007, \aap, 476, 935 

\bibitem[Frebel et al.(2005)]{frebel05} Frebel, A., Aoki, W., 
Christlieb, N., et al.\ 2005, \nat, 434, 871 

\bibitem[Frebel et al.(2006)]{frebel06} Frebel, A., Christlieb, 
N., Norris, J.~E., Aoki, W., \& Asplund, M.\ 2006, \apjl, 638, L17 

\bibitem[Frebel et al.(2008)]{frebel08} Frebel, A., Collet, R., 
Eriksson, K., Christlieb, N., Aoki, W.\ 2008, \apj, 684, 588

\bibitem[Frebel et al.(2010)]{frebel10} Frebel, A., Simon, 
J.~D., Geha, M., \& Willman, B.\ 2010, \apj, 708, 560 

\bibitem[Frebel \& Bromm(2012)]{frebel12} Frebel, A., \& Bromm, V.\ 
2012, \apj, 759, 115 

\bibitem[Frischknecht et al.(2012)]{frischknecht12} Frischknecht, U., 
Hirschi, R., \& Thielemann, F.-K.\ 2012, \aap, 538, L2 

\bibitem[Fryer et al.(2001)]{fryer01} Fryer, C.~L., Woosley, 
S.~E., \& Heger, A.\ 2001, \apj, 550, 372 

\bibitem[Fulbright(2002)]{fulbright02} Fulbright, J.~P.\ 2002, \aj, 123, 404 

\bibitem[Gallino et al.(1998)]{gallino98} Gallino, R., Arlandini, 
C., Busso, M., et al.\ 1998, \apj, 497, 388 

\bibitem[Gilmore et al.(2013)]{gilmore13} Gilmore, G., Norris, 
J.~E., Monaco, L., et al.\ 2013, \apj, 763, 61 

\bibitem[Giridhar et al.(2001)]{giridhar01} Giridhar, S., Lambert, 
D.~L., Gonzalez, G., \& Pandey, G.\ 2001, \pasp, 113, 519 

\bibitem[Gratton et al.(2003)]{gratton03} Gratton, R.~G., Carretta, E., 
Desidera, S., et al.\ 2003, \aap, 406, 131 

\bibitem[Gustafsson et al.(2008)]{gustafsson08} Gustafsson, B., 
Edvardsson, B., Eriksson, K., J{\o}rgensen, U.~G., Nordlund, {\AA}., \& 
Plez, B.\ 2008, \aap, 486, 951 

\bibitem[Heger \& Woosley(2002)]{heger02} Heger, A., \& Woosley, S.~E.\ 
2002, \apj, 567, 532 

\bibitem[Hollek et al.(2011)]{hollek11} Hollek, J.~K., Frebel, 
A., Roederer, I.~U., et al.\ 2011, \apj, 742, 54 

\bibitem[Honda et al.(2004)]{honda04a} Honda, S., Aoki, W., Ando, H., 
et al.\ 2004, \apjs, 152, 113 

\bibitem[Honda et al.(2006)]{honda06} Honda, S., Aoki, W., 
Ishimaru, Y., Wanajo, S., \& Ryan, S.~G.\ 2006, \apj, 643, 1180 

\bibitem[Ishigaki et al.(2010)]{ishigaki10} Ishigaki, M., Chiba, 
M., \& Aoki, W.\ 2010, \pasj, 62, 143 

\bibitem[Ishigaki et al.(2013)]{ishigaki13} Ishigaki, M.~N., Aoki, 
W., \& Chiba, M.\ 2013, \apj, 771, 67 

\bibitem[Ito et al.(2009)]{ito09} Ito, H., Aoki, W., Honda, 
S., \& Beers, T.~C.\ 2009, \apjl, 698, L37 

\bibitem[Ito et al.(2013)]{ito13} Ito, H., Aoki, W., Beers, T.~C., 
Tominaga, N., et al.\ 2013, \apj, 773, 33 

\bibitem[Johnson(2002)]{johnson02} Johnson, J.~A.\ 2002, \apjs, 139, 219 

\bibitem[Johnson et al.(2007)]{johnson07} Johnson, J.~A., Herwig, 
F., Beers, T.~C., \& Christlieb, N.\ 2007, \apj, 658, 1203 

\bibitem[Kelson(2003)]{kelson03} Kelson, D.~D.\ 2003, \pasp, 115, 688 

\bibitem[Kirby et al.(2012)]{kirby12} Kirby, E.~N., Fu, X., 
Guhathakurta, P., \& Deng, L.\ 2012, \apjl, 752, L16 

\bibitem[Koch et al.(2013)]{koch13} Koch, A., Feltzing, S., 
Ad{\'e}n, D., \& Matteucci, F.\ 2013, \aap, 554, A5 

\bibitem[Kurucz \& Bell(1995)]{kurucz95} Kurucz, R.~L., \& Bell, B.\ 
1995, Kurucz CD-ROM, Cambridge, MA: Smithsonian Astrophysical Observatory

\bibitem[Lai et al.(2004)]{lai04} Lai, D.~K., Bolte, M., 
Johnson, J.~A., \& Lucatello, S.\ 2004, \aj, 128, 2402 

\bibitem[Lai et al.(2008)]{lai08} Lai, D.~K., Bolte, M., 
Johnson, J.~A., Lucatello, S., Heger, A., 
\& Woosley, S.~E.\ 2008, \apj, 681, 1524 

\bibitem[Lai et al.(2011)]{lai11} Lai, D.~K., Lee, Y.~S., 
Bolte, M., et al.\ 2011, \apj, 738, 51 

\bibitem[Lambert \& Sawyer(1984)]{lambert84} Lambert, D.~L., \& 
Sawyer, S.~R.\ 1984, \apj, 283, 192 

\bibitem[Lebzelter et al.(2012)]{lebzelter12} Lebzelter, T., Uttenthaler, S., 
Busso, M., Schultheis, M., \& Aringer, B.\ 2012, \aap, 538, A36 

\bibitem[Lind et al.(2009)]{lind09} Lind, K., Asplund, M., \& Barklem, P.~S.\
2009, \aap, 503, 541

\bibitem[Lind et al.(2011)]{lind11} Lind, K., Asplund, M., Barklem, P.~S., \&
Belyaev, A.~K.\ 2011, \aap, 528, A103

\bibitem[Martell \& Shetrone(2013)]{martell13} Martell, S.~L., \& 
Shetrone, M.~D.\ 2013, \mnras, 430, 611 

\bibitem[McWilliam et al.(1995a)]{mcwilliam95a} McWilliam, A., 
Preston, G.~W., Sneden, C., \& Shectman, S.\ 1995a, \aj, 109, 2736 

\bibitem[McWilliam et al.(1995b)]{mcwilliam95b} McWilliam, A., 
Preston, G.~W., Sneden, C., \& Searle, L.\ 1995b, \aj, 109, 2757 

\bibitem[Meynet et al.(2006)]{meynet06} Meynet, G., Ekstr{\"o}m, S., 
\& Maeder, A.\ 2006, \aap, 447, 623 

\bibitem[Meynet et al.(2010)]{meynet10} Meynet, G., Hirschi, R., 
Ekstrom, S., et al.\ 2010, \aap, 521, A30 

\bibitem[Nissen \& Schuster(2010)]{nissen10} Nissen, P.~E., \& 
Schuster, W.~J.\ 2010, \aap, 511, L10 

\bibitem[Norris et al.(1997)]{norris97} Norris, J.~E., Ryan, 
S.~G., \& Beers, T.~C.\ 1997, \apjl, 489, L169 

\bibitem[Norris et al.(2001)]{norris01} Norris, J.~E., Ryan, 
S.~G., \& Beers, T.~C.\ 2001, \apj, 561, 1034 

\bibitem[Norris et al.(2002)]{norris02} Norris, J.~E., Ryan, S.~G., 
Beers, T.~C., Aoki, W., \& Ando, H.\ 2002, \apjl, 569, L107 

\bibitem[Norris et al.(2007)]{norris07} Norris, J.~E., 
Christlieb, N., Korn, A.~J., et al.\ 2007, \apj, 670, 774 

\bibitem[Norris et al.(2010a)]{norris10a} Norris, J.~E., Gilmore, 
G., Wyse, R.~F.~G., Yong, D., \& Frebel, A.\ 2010a, \apjl, 722, L104 

\bibitem[Norris et al.(2010b)]{norris10b} Norris, J.~E., Yong, D., 
Gilmore, G., \& Wyse, R.~F.~G.\ 2010b, \apj, 711, 350 

\bibitem[Norris et al.(2012)]{norris12} Norris, J.~E., 
Christlieb, N., Bessell, M.~S., et al.\ 2012, \apj, 753, 150 

\bibitem[Norris et al.(2013)]{norris13} Norris, J.~E., Yong, D., 
Bessell, M.~S., et al.\ 2013, \apj, 762, 28 

\bibitem[Pilachowski et al.(2000)]{pilachowski00} Pilachowski, C.~A., 
Sneden, C., Kraft, R.~P., Harmer, D., 
\& Willmarth, D.\ 2000, \aj, 119, 2895 

\bibitem[Piskunov \& Valenti(2002)]{piskunov02} Piskunov, N.~E., \& 
Valenti, J.~A.\ 2002, \aap, 385, 1095 

\bibitem[Preston \& Sneden(2001)]{preston01} Preston, G.~W., \& Sneden, C.\ 
2001, \aj, 122, 1545 

\bibitem[Primas et al.(1994)]{primas94} Primas, F., Molaro, P., \& 
Castelli, F.\ 1994, \aap, 290, 885 

\bibitem[Roederer et al.(2008)]{roederer08} Roederer, I.~U., 
Frebel, A., Shetrone, M.~D., et al.\ 2008, \apj, 679, 1549 

\bibitem[Roederer(2009)]{roederer09a} Roederer, I.~U.\ 2009, \aj, 137, 272 

\bibitem[Roederer et al.(2009)]{roederer09b} Roederer, I.~U., 
Kratz, K.-L., Frebel, A., et al.\ 2009, \apj, 698, 1963 

\bibitem[Roederer et al.(2010)]{roederer10} Roederer, I.~U., 
Cowan, J.~J., Karakas, A.~I., et al.\ 2010, \apj, 724, 975 

\bibitem[Roederer et al.(2012)]{roederer12b} Roederer, I.~U., Lawler, J.~E.,
Sobeck, J.~S., et al.\ 2012, \apjs, 203, 27

\bibitem[Roederer(2013)]{roederer13a} Roederer, I.~U.\ 2013, \aj, 145, 26

\bibitem[Roederer et al.(2014)]{roederer14} Roederer, I.~U., 
Preston, G.~W., Thompson, I.~B., Shectman, S.~A., Kelson, D., 
Sneden, C., 2014, \aj, submitted

\bibitem[Ruchti et al.(2011a)]{ruchti11a} Ruchti, G.~R., 
Fulbright, J.~P., Wyse, R.~F.~G., et al.\ 2011a, \apj, 737, 9 

\bibitem[Ruchti et al.(2011b)]{ruchti11b} Ruchti, G.~R., 
Fulbright, J.~P., Wyse, R.~F.~G., et al.\ 2011b, \apj, 743, 107 

\bibitem[Ryan et al.(2005)]{ryan05} Ryan, S.~G., Aoki, W., 
Norris, J.~E., \& Beers, T.~C.\ 2005, \apj, 635, 349 

\bibitem[Simon et al.(2010)]{simon10} Simon, J.~D., Frebel, A., 
McWilliam, A., Kirby, E.~N., \& Thompson, I.~B.\ 2010, \apj, 716, 446 

\bibitem[Sivarani et al.(2006)]{sivarani06} Sivarani, T., Beers, T.~C., 
Bonifacio, P., et al.\ 2006, \aap, 459, 125 

\bibitem[Sneden(1973)]{sneden73} Sneden, C.~A.\ 1973, 
Ph.D.~Thesis, Univ.\ of Texas at Austin

\bibitem[Sneden et al.(2003)]{sneden03} Sneden, C., Cowan, J.~J., 
Lawler, J.~E., et al.\ 2003, \apj, 591, 936 

\bibitem[Sneden et al.(2008)]{sneden08} Sneden, C., Cowan, J.~J., 
Gallino, R.\ 2008, \araa, 46, 241

\bibitem[Sneden et al.(2009)]{sneden09} Sneden, C., Lawler, J.~E., 
Cowan, J.~J., Ivans, I.~I., \& Den Hartog, E.~A.\ 2009, \apjs, 182, 80 

\bibitem[Sobeck et al.(2011)]{sobeck11} Sobeck, J.~S., Kraft, 
R.~P., Sneden, C., et al.\ 2011, \aj, 141, 175 

\bibitem[Spite et al.(2005)]{spite05} Spite, M., Cayrel, R., Plez, B., 
et al.\ 2005, \aap, 430, 655 

\bibitem[Stephens \& Boesgaard(2002)]{stephens02} Stephens, A., \& 
Boesgaard, A.~M.\ 2002, \aj, 123, 1647 

\bibitem[Takeda et al.(2002)]{takeda02} Takeda, Y., Zhao, G.,
Chen, Y.-Q., Qiu, H.-M., \& Takada-Hidai, M.\ 2002, \pasj, 54, 275

\bibitem[Thorburn(1994)]{thorburn94} Thorburn, J.~A.\ 1994, \apj, 421, 318 

\bibitem[Tominaga et al.(2007)]{tominaga07} Tominaga, N., Maeda, 
K., Umeda, H., et al.\ 2007, \apjl, 657, L77 

\bibitem[Tominaga et al.(2013)]{tominaga13} Tominaga, N., Iwamoto, 
N., \& Nomoto, K.\ 2013, \apj, submitted (arXiv:1309.6734)

\bibitem[Truran(1981)]{truran81} Truran, J.~W.\ 1981, \aap, 97, 391 

\bibitem[Tsangarides et al.(2004)]{tsangarides04} Tsangarides, S., 
Ryan, S.~G., \& Beers, T.~C.\ 2004, \memsai, 75, 772 

\bibitem[Tull et al.(1995)]{tull95} Tull, R.~G., MacQueen, 
P.~J., Sneden, C., \& Lambert, D.~L.\ 1995, \pasp, 107, 251 

\bibitem[Tumlinson(2006)]{tumlinson06} Tumlinson, J.\ 2006, \apj, 641, 1 

\bibitem[Umeda \& Nomoto(2002)]{umeda02} Umeda, H., \& Nomoto, K.\ 
2002, \apj, 565, 385 

\bibitem[Umeda \& Nomoto(2003)]{umeda03} Umeda, H., \& Nomoto, K.\ 
2003, \nat, 422, 871 

\bibitem[Umeda \& Nomoto(2005)]{umeda05} Umeda, H., \& Nomoto, K.\ 
2005, \apj, 619, 427 

\bibitem[Uns\"{o}ld(1955)]{unsold55} Uns\"{o}ld, A., Physik der
Sternatmosph\"{a}ren, Springer-Verlag, Berlin, 1955, p.\ 332--333

\bibitem[Yong et al.(2013)]{yong13a} Yong, D., Norris, J.~E., 
Bessell, M.~S., et al.\ 2013, \apj, 762, 26 


\end{thebibliography}
\end{document}